\documentclass[english,12pt,twoside]{article}
\usepackage[]{graphicx,rotating,times,pdfsync,epstopdf,amsfonts}
\usepackage{amsmath}
\usepackage{amssymb}
\usepackage[font=footnotesize]{caption}
\usepackage{empheq}
\usepackage{enumitem}

\newcommand{\be}{\begin{equation}}
\newcommand{\ee}{\end{equation}}
\newcommand{\ba}{\begin{eqnarray}}
\newcommand{\ea}{\end{eqnarray}}
\newcommand{\qqbar}{q\overline{q}}
\newcommand{\pbarp}{\overline{p}p}
\newcommand{\pbarn}{\overline{p}n}
\newcommand{\pbard}{\overline{p}d}
\newcommand{\nbarp}{\overline{n}p}
\newcommand{\nbarn}{\overline{n}n}
\newcommand{\NNbar}{N\overline{N}}
\newcommand{\pbar}{\overline{p}}
\newcommand{\nbar}{\overline{n}}
\newcommand{\qbar}{\overline{q}}
\newcommand{\ubar}{\overline{u}}
\newcommand{\dbar}{\overline{d}}
\newcommand{\sbar}{\overline{s}}

\newcommand{\uubar}{u\overline{u}}
\newcommand{\ddbar}{d\overline{d}}
\newcommand{\ssbar}{s\overline{s}}

\newcommand{\KKbar}{K\overline{K}}

\begin{document}

\begin{center}
{\bf\Large  Nucleon-antinucleon annihilation at LEAR{\let\thefootnote\relax\footnote{Invited talk  at the ECT* workshop on ``Antiproton-nucleus interactions and related phenomena'', \\ \hspace*{6mm}Trento 17--21, June 2019. E-mail: claude.amsler@cern.ch, http://amsler.web.cern.ch/amsler/}}}{}\\
\vspace*{6mm}
Claude AMSLER\vspace*{6mm}\\
Stefan Meyer Institute for Subatomic Physics, Austrian Academy of Sciences, \\
Boltzmanngasse 3, 1090 Vienna, Austria\\
\end{center}

\begin{abstract}
This report is a historical review of the salient results in  low energy antiproton-proton and  antineutron-proton annihilation obtained at the Low Energy Antiproton Ring (LEAR), which was operated at CERN between 1983 and 1996. The intention is to provide guidelines for future experiments at the CERN AD/ELENA complex and elsewhere. In the spirit of this workshop, hadron spectroscopy -- one of the cornerstones at LEAR -- is briefly mentioned, while emphasis is put on the annihilation mechanism on one and two nucleons, the final state multiplicity distributions and the contributions from quarks, in particular in annihilation  channels involving strangeness.
\end{abstract}

\subsection*{1. Nucleon-antinucleon bound states and resonances}
Nucleon-antinucleon annihilation at low energy proceeds through the emission of pions and kaons and is a tool to study meson resonances with masses below 2 GeV. Apart from $\qqbar$  mesons one can also produce exotic configurations such as tetraquark states ($qq\qbar\qbar$), meson-meson weakly bound ``molecules'',  $\qqbar g$ hybrid states with a valence gluon $g$, or glueballs (mesons made exclusively of gluons) \cite{LNP}. 

The existence of $\NNbar$ bound states and resonances was predicted a long time ago \cite{Shapiro}, based on the strongly attractive $\NNbar$ meson exchange potential. These  states were referred to as quasinuclear or baryonium states. Several candidates had been reported in the seventies by experiments at CERN, BNL and KEK, some of them being indisputably statistically significant (for a review see e.g. \cite{Mon80}). For example, a measurement of the low energy  $\pbarp$ annihilation and elastic cross sections on secondary extracted antiproton beams revealed a sharp resonance (the ``$S$-meson'') around 500 MeV/c ($m$ = 1940 MeV) \cite{Br77}, soon reported by  other experiments, see the inset in Figure \ref{Sigmas} below. This state was not confirmed  later by measurements performed at LEAR. Proton-antiproton resonances were also observed at 2020 and 2200 MeV in $\pi^-p\to p (\pbarp) \pi^-$ with 9 GeV/c pions \cite{Be77}. Even bound baryonium states $X$ were reported, in particular in radiative decays $\pbarp\to\gamma X$ \cite{Pa78}, none of which would survive over time.

\begin{figure}[htb]
\parbox{170mm}{\mbox{
\includegraphics[width=108mm]{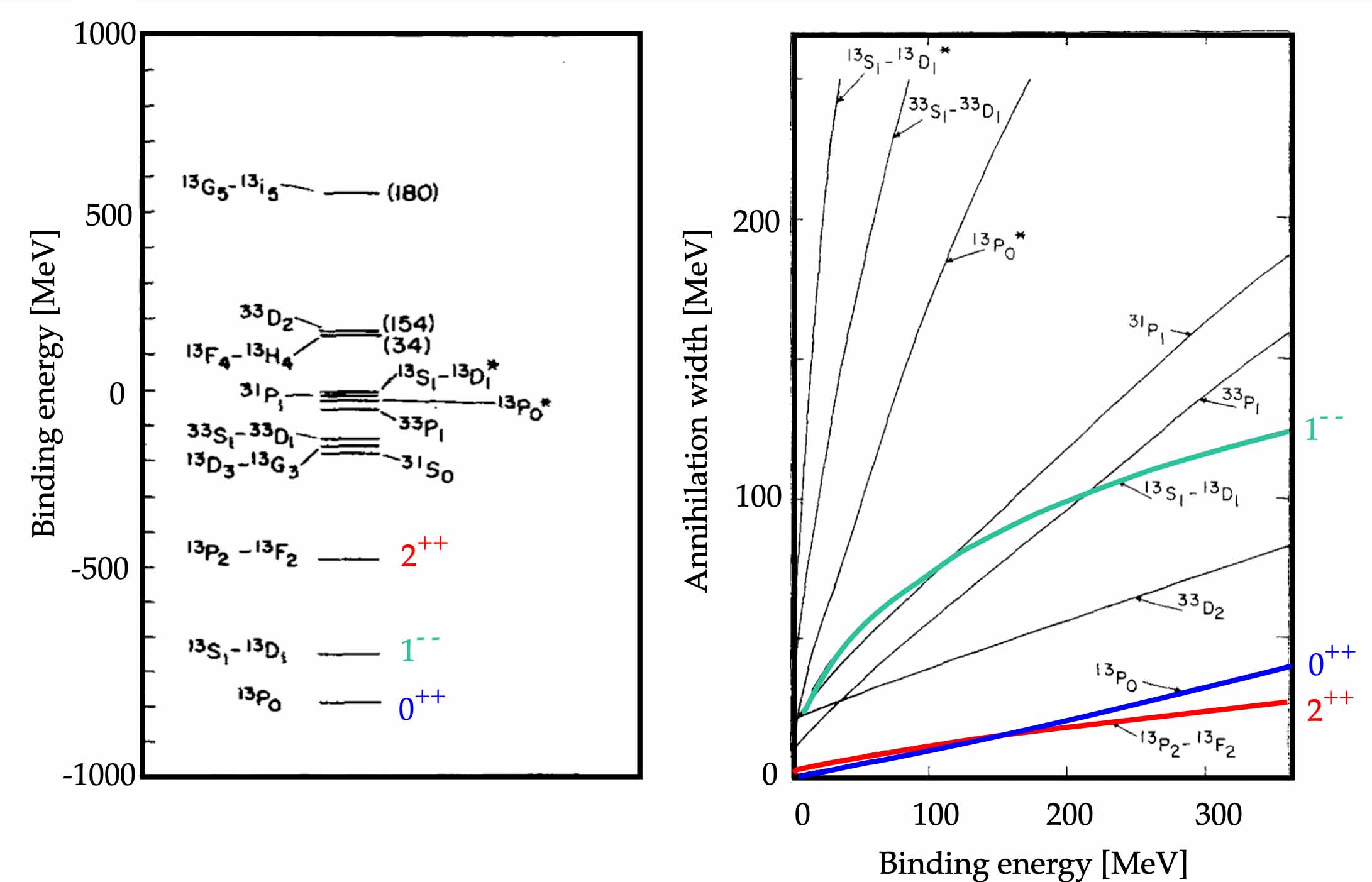}
}\centering}\hfill
\caption[]{Left: binding energies of baryonium states in the Paris  potential \cite{Bu79}. The most deeply bound states are isoscalar  ($i= 0$) and hence electrically neutral. Right: predicted  annihilation widths \cite{Do79}.
\label{DoverRichard}}
\end{figure}

The $\NNbar$ potential can be obtained from the $NN$ one by multiplying with the $G$-parity of the exchanged meson (a detailed discussion can be found e.g. in ref. \cite{Vogt}). Hence the short range repulsive $\omega$ exchange in $NN$ (leading to Pauli blocking) becomes attractive in $\NNbar$ since $G(\omega)$ = --1. On the other hand, the sign of the $\rho$ contribution does not change since $G(\rho)$= + 1, hence stays attractive  in the $\pbarp$ isospin $i$ = 0 state and repulsive in the $i$ =1 state. (In fact all meson exchange contributions to the $\NNbar$ central force are attractive for $i=0$.) Thus one expects more deeply bound states for $i$ = 0
and most potential models predict a sequence of deeply bound isoscalar baryonia  with quantum numbers $J^{PC}$ = $2^{++}$,  $1^{--}$ and $0^{++}$ ($P$ being the parity, $C$ the charge parity) the latter being the mostly bound state (Figure \ref{DoverRichard}, left). 

\begin{figure}[htb]
\parbox{85mm}{\mbox{
\includegraphics[width=80mm]{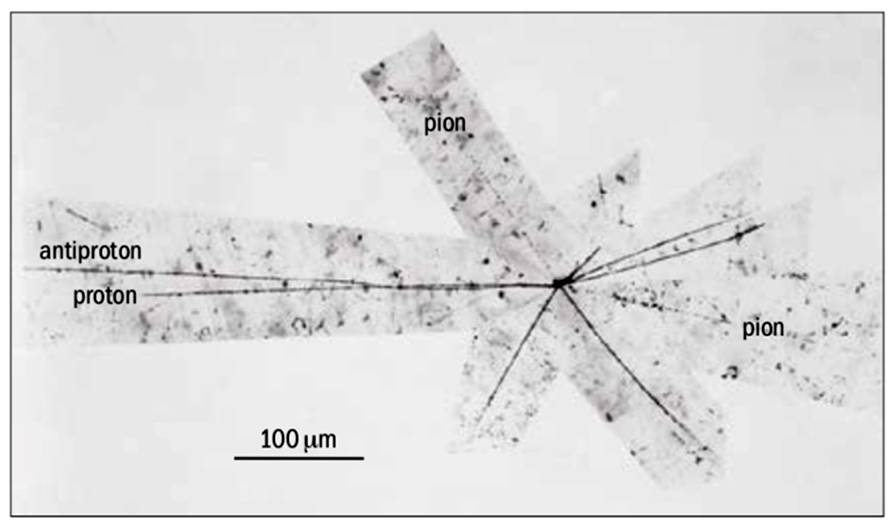}
}\centering}\hfill
\parbox{85mm}{\mbox{
\includegraphics[width=65mm]{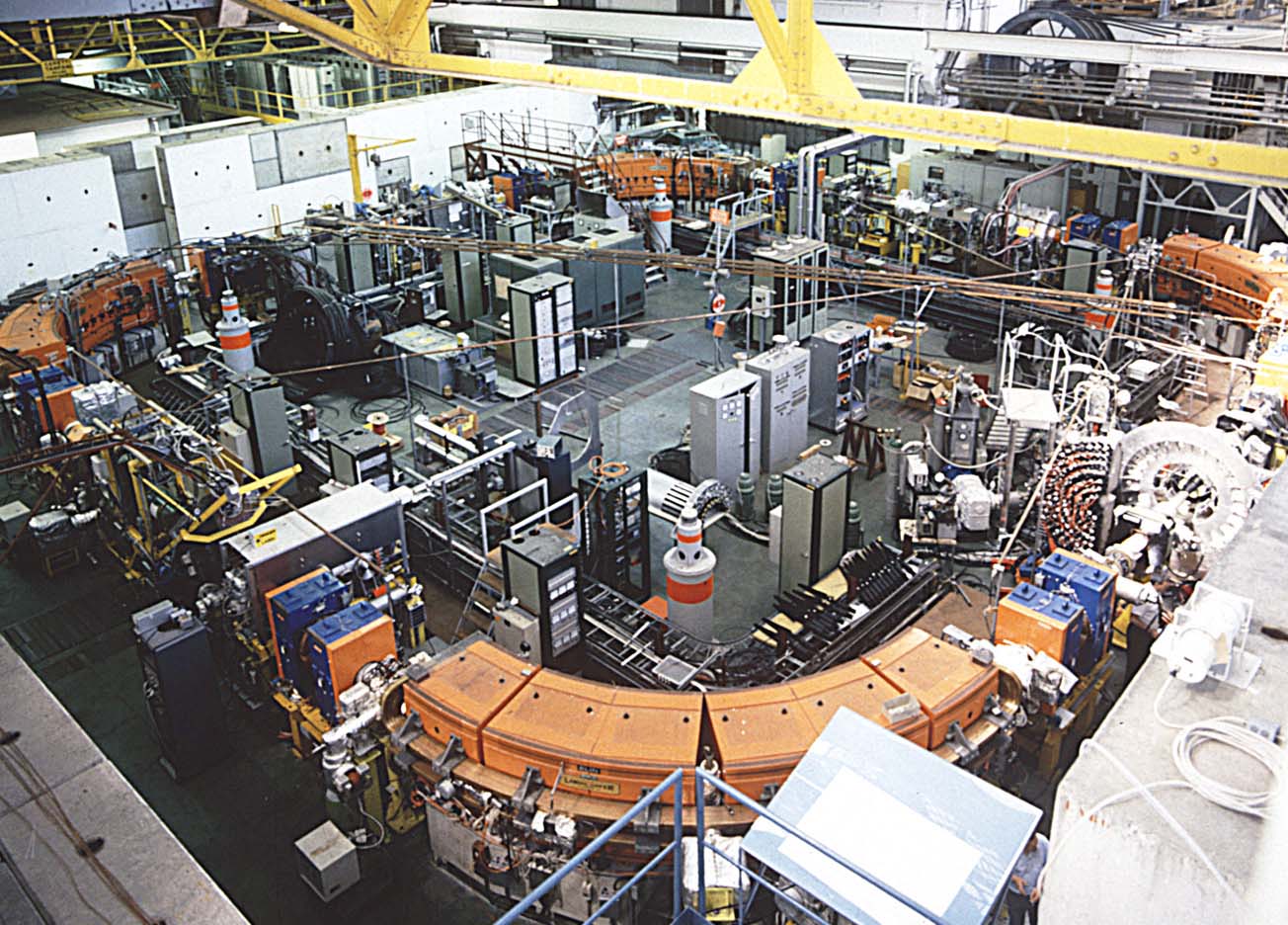}
}\centering} \caption[]{Left: one of the first annihilations observed in a nuclear emulsion \cite{Ch56}. Right: the LEAR storage ring in operation between 1983 and 1996 in the South Hall (photo CERN).
\label{LEAR}}
\end{figure}

Figure \ref{LEAR} (left) shows one of the first antiproton annihilations observed at the Bevatron in Berkeley \cite{Ch56}.
It was argued that annihilation had to be of very short range ($\sim$ 0.2 fm, of the order of the Compton wavelength of the exchanged nucleon) and that annihilation was therefore weak at $\NNbar$ distance separations of 1 fm, the predicted size of baryonium states from meson exchange potentials. Hence baryonium states ought to be narrow. However, due to their finite sizes the proton and antiproton already overlap at the much larger distance of typically 1 fm, thus decreasing the lifetime and correspondingly increasing the widths of baryonium states. The predicted annihilation widths (Figure \ref{DoverRichard}, right) are rather uncertain due to the lack of knowledge of the effective annihilation range. 

The construction of a low energy high intensity antiproton beam facility was mainly motivated by the predicted quasinuclear $\NNbar$ states. Thanks to the invention of stochastic cooling, intense 
and pure accelerator beams of low momentum antiprotons 
became feasible. The antiprotons were produced  around  3.5 GeV/c  by the CERN PS, stored in the Antiproton Accumulator  (and later in the Antiproton Collector), decelerated to 600 MeV/c and injected into the  Low Energy Antiproton Ring (LEAR), where they could be further decelerated to 60 MeV/c (2 MeV kinetic energy) or accelerated to 1940 MeV/c (Figure \ref{LEAR}, right). They were then slowly extracted and distributed to several experiments simultaneously with intensities of about 10$^6$ s$^{-1}$ and momentum bites $\Delta p/p$ better than 10$^{-3}$ during one hour long spills.  It is impressive to compare the high antiproton flux ($>$ 10$^6$ 
$\pbar$/s) with the rate of about 1 $\pbar$ every 15 minutes at the time of the antiproton discovery  \cite{Ch55}.
 
LEAR was decommissioned in 1996. I have counted 35 experiments at LEAR, about 15 of them dealing with annihilation\footnote{PS170 Precision measurements of the proton electromagnetic form factors in the time-like region\\
\hspace*{6mm}PS171 (ASTERIX) Study of proton-antiproton interactions at rest in a hydrogen gas target at LEAR \\
\hspace*{6mm}PS173 Measurement of antiproton-proton cross sections at low antiproton momenta\\
\hspace*{6mm}PS177 Study of the fission decay of heavy hypernuclei\\
\hspace*{6mm}PS179 Study of the interaction of low-energy antiprotons with $^2$H, $^3$He, $^4$He, Ne-Nuclei with a streamer chamber\\
\hspace*{6mm}PS182 Investigations on baryonium and other rare $\pbarp$ annihilation modes using high-resolution 
$\pi^0$ spectrometers\\
 \hspace*{6mm}PS183 Search for bound  $\NNbar$ states using a precision $\gamma$  and charged pion spectrometer at LEAR\\
\hspace*{6mm}PS184 Study of antiproton-nucleus interaction with the high resolution SPESII magnetic spectrometer\\
\hspace*{6mm}PS186 Nuclear excitations by antiprotons and antiprotonic atoms\\
\hspace*{6mm}PS187 A high statistics study of antiproton interactions with nuclei\\
\hspace*{6mm}PS197 (CRYSTAL BARREL) Meson spectroscopy at LEAR with a 4$\pi$ detector\\
\hspace*{6mm}PS201 (OBELIX) Study of $\pbarp$  and $\pbarn$ annihilations at LEAR with a large acceptance and high resolution detector\\
\hspace*{6mm}PS202 (JETSET) Physics at LEAR with an internal gas jet target and an advanced 
general purpose detector \\
\hspace*{6mm}PS203 Antiproton induced fission and fragmentation\\
\hspace*{6mm}PS208 Decay of hot  nuclei at low spins produced by antiproton-annihilation in heavy nuclei\\
}. 
The results from ASTERIX, CRYSTAL BARREL and OBELIX (on which this report will concentrate) totaled 350 journal publications. 

\subsection*{2. Global features of proton-antiproton annihilation at rest}
\begin{figure}[htb]
\begin{minipage}[c]{0.4\textwidth}
\begin{small}
\begin{tabular}{ c r c }
\hline
Prong & \multicolumn{2}{c}{[\%]}\\
\hline
\noalign{\vskip 1mm}
0  & 4.1&  $^{+0.2}_{-0.6}$ \\
\noalign{\vskip 1mm}
2  & 43.2 & $^{+0.9}_{-0.7}$ \\
\noalign{\vskip 1mm}
4  & 48.6 & $^{+0.9}_{-0.7}$ \\
\noalign{\vskip 1mm}
6  & 4.1 & $^{+0.2}_{-0.2}$ \\
\noalign{\vskip 1mm}
\hline
\end{tabular}
\end{small}
\centering
\end{minipage}
%don't skip line here
\begin{minipage}[c]{0.6\textwidth}
\includegraphics[width=80mm]{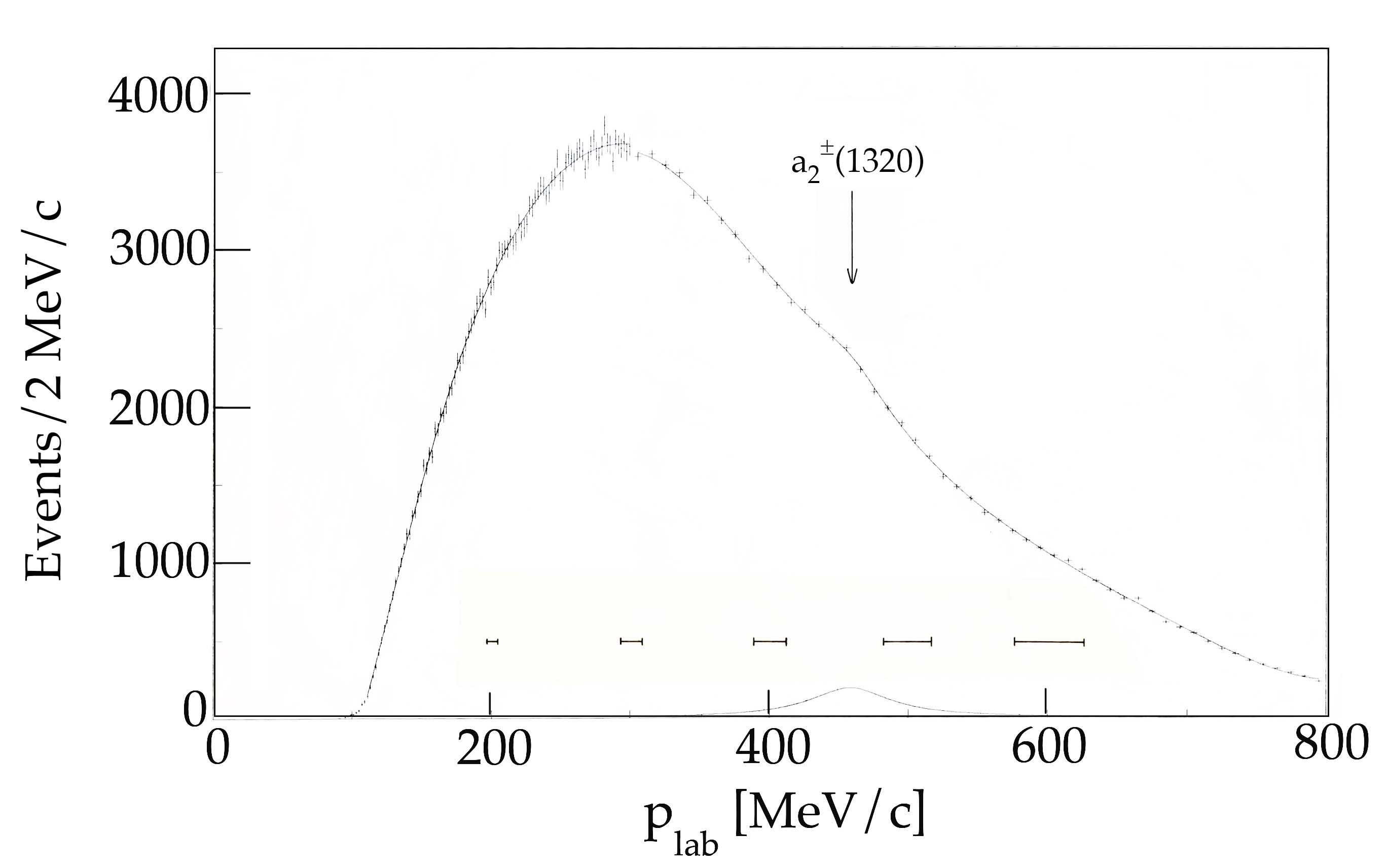}
\centering
\end{minipage}
\caption[]{Left: charged pion multiplicity measured in $\pbarp$ annihilation at rest with a hydrogen bubble chamber \cite{Gh74}. Right: inclusive charged pions momentum  spectrum in hydrogen gas \cite{Ah85}.  The average pion momentum is 350 MeV/c. The low energy cut is due to detector acceptance. The horizontal error bars show the experimental resolution. The curves are polynomial fits adding a Breit-Wigner for the 460 MeV/c bump. 
\label{Inclusive}}
\end{figure}

Annihilation at rest was studied in the sixties with bubble chambers \cite{Ar69} which analyzed 
final states involving charged mesons  ($\pi^{\pm}, K^{\pm}$, $K_S\rightarrow \pi^+\pi^-$), with at most one undetected $\pi^0$  due to the lack of $\gamma$ detection. The table in Figure \ref{Inclusive} gives the branching fractions for annihilation at rest into 0, 2, 4 and 6 pions (prongs), associated with any number of neutral pions \cite{Gh74}. Note that channels with two or more neutral pions, representing $\sim$60\% of all annihilations,  as well as 0 prong events ($\sim$4\%), have been studied for the first time at LEAR by CRYSTAL BARREL (discussed below). The branching fractions in Figure \ref{Inclusive} refer to pionic channels only and are normalized to 100\% (8 prong channels being neglected). Estimates for the contribution of channels with charged pions and more than one $\pi^0$ were made by including the intermediate resonances known at that time \cite{Gh74}. They are only in fair agreement with the multiplicity distribution measured at LEAR (in particular, they overestimate the five pion contribution). The contribution from channels involving $\eta$ mesons and kaons is estimated to be $\sim$7\% and  $\sim$6\% of all annihilations, respectively. 

In first approximation one can view the annihilation process as a hot fireball gas evaporating pions \cite{Ki81} with multiplicity  distributed statistically. In fireball models the total pion multiplicity $n = n_+ + n_- + n_0$ follows 
a Gaussian distribution \cite{Or73}. This is only a rough approach because  annihilation  proceeds via the excitation of intermediate resonances. For example, the annihilation channel $\pbarp\to 2\pi^+2\pi^-\pi^0$ proceeds through  the decays of intermediate states such as $\omega\rho^0$, $\omega f_2(1270)$ or $\rho^\pm\pi^\mp\pi^+\pi^-$. Figure \ref{Inclusive} (right) shows the inclusive momentum distribution of charged pions in $\pbarp$ annihilation at rest in gaseous hydrogen measured by ASTERIX \cite{Ah85}. The shape of the distribution appears to be statistical, apart from the bump at 460 MeV/c due to the $a_2(1320)^\pm$ recoiling against a $\pi^\mp$. 

Annihilation at rest occurs from protonium states (atomic $\pbarp$ states following $\pbar$ capture) and  further constraints arise from quantum number conservation. This will be described below, as well as  internal ``dynamical'' selection rules the origin of which is still open to discussion \cite{Kl05}.

\begin{figure}[htb]
\parbox{85mm}{\mbox{
\includegraphics[width=43mm]{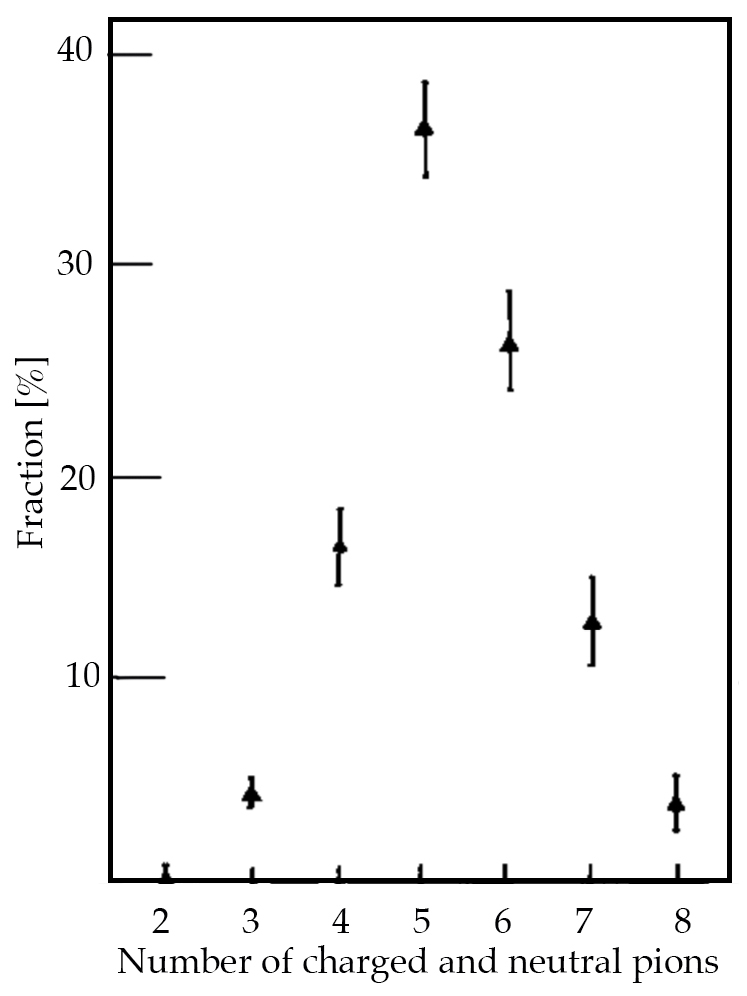}
}\centering}\hfill
\parbox{85mm}{\mbox{
\includegraphics[width=60mm]{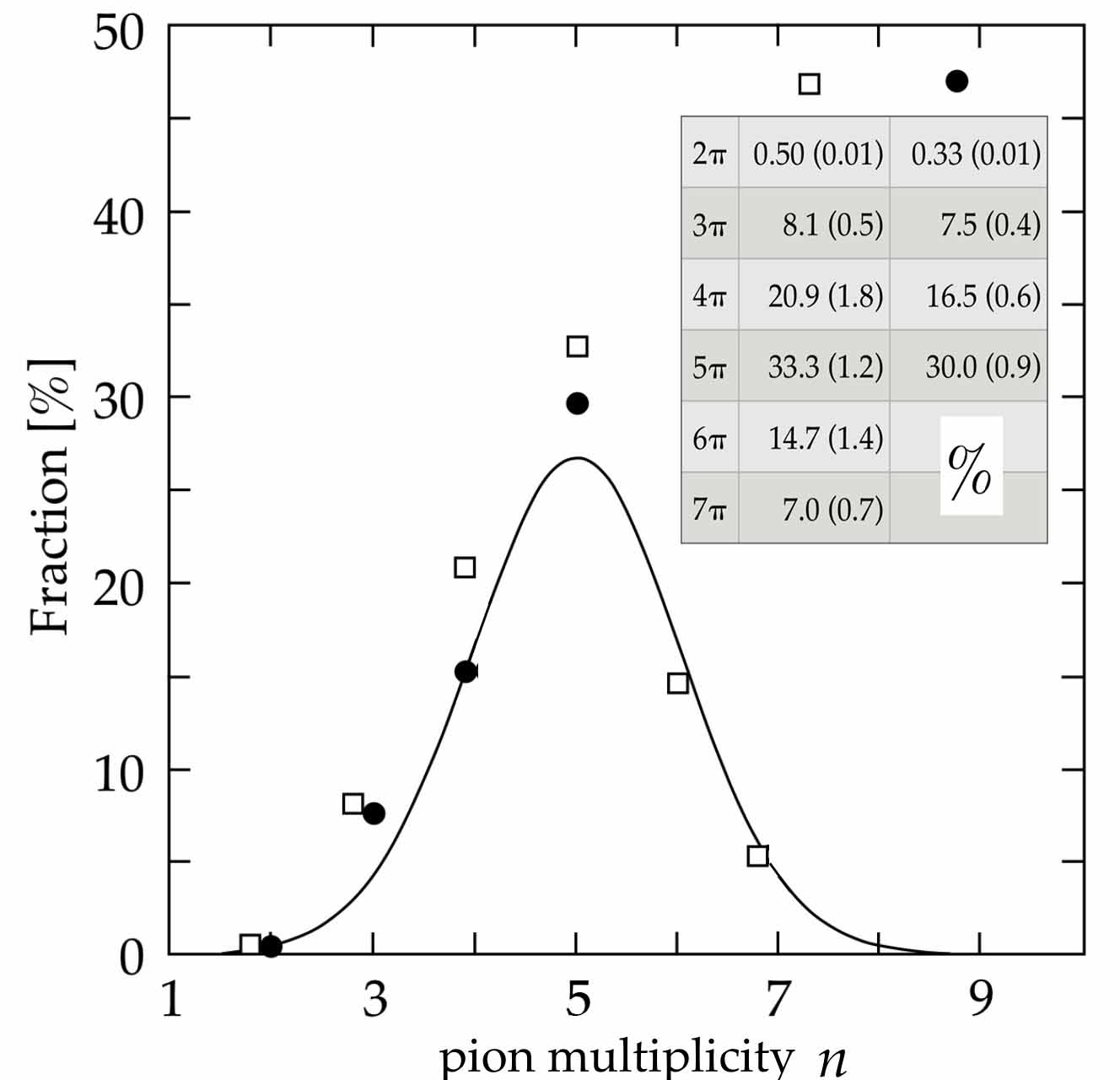}
}\centering} \caption[]{Left: total (charged + neutral) pion multiplicity at 1.6 GeV/c (normalized to 100\%) \cite{Fe77}. Right: branching fraction as a function of  multiplicity  for $\pbarp$ annihilation 
at rest in liquid hydrogen. Full 
circles: data from bubble chambers and CRYSTAL BARREL. Open squares: expected distribution using the factorial law.   The curve is a Gaussian fit assuming the average multiplicity $\langle n\rangle = 5$. The rms width is $\sigma = 1.04 \pm 0.01$.
\label{Multip}}
\end{figure}

Before LEAR, systematic measurement of both charged and neutral multiplicities at rest had not been performed, but data were available at 1.6 GeV/c, performed with CERN's Gargamelle bubble chamber (with which neutral currents were discovered). The chamber was filled with a heavy liquid (propane-freon) to convert the photons from $\pi^0$ decay into $e^+e^-$ pairs.
The multiplicity distribution is shown in Figure \ref{Multip} (left). The average 
charged pion multiplicity is 3.46 $\pm$ 0.04 and the  average $\pi^0$ multiplicity 1.92 $\pm$ 0.04. 

The Nearest Threshold Dominance Model describes  reasonably well the 
observed final state multiplicity up to $\pbar$ momenta of 3.5 GeV/c \cite{Va88}.  It assumes a branching ratio
\be
B \propto p\exp[-1.2\sqrt{s-(m_a + m_b)^2}],
\label{eq:phasespace}
\ee
for annihilation into the two possible  heaviest mesons with masses $m_a$ und $m_b$, 
where $p$ is the meson momentum in center-of-mass 
system with total energy $\sqrt{s}$ (all in GeV units). Thus annihilation into the heaviest possible 
meson pair is enhanced 
with respect to phase space $p$ by the exponential form factor. This is inspired by baryon exchange models which prefer small momentum transfers at the baryon-meson vertices. 

\begin{table}[htb]
\begin{scriptsize}
\begin{center}
\begin{tabular}{c | c  c c c c c | r}
\hline
$n$ & 0$\pi^0$ & 1$\pi^0$ & 2$\pi^0$ & 3$\pi^0$ & 4$\pi^0$ & 5$\pi^0$ & Total\ \ = 84.5\% \ \ \\
\hline
2$\pi$ & 0.33 $\pm$ 0.01&&  0.17 $\pm$ 0.01  && & & 0.50\\
& & &$\dagger$(6.9 $\pm$ 0.4) $\times$ 10$^{-2}$ & & & &\\
\hline
3$\pi$& & 6.9 $\pm$ 0.4  & &1.2 $\pm$ 0.1 & & & 8.1 $\pm$  0.5\\
&&&& $\dagger$0.62 $\pm$ 0.10 &&&  \\
\hline
4$\pi$ & 6.9 $\pm$ 0.6  && 13.8 $\pm$ 1.2 &&1.1 $\pm$ 0.1&& 20.9 $\pm$ 1.8\\
&&&  $\dagger$9.3 $\pm$ 0.2 && $\dagger$0.31$\pm$ 0.02&& \\
\hline
5$\pi$ &&19.6 $\pm$ 0.7  &&13.1 $\pm$ 0.5 && 0.65 $\pm$ 0.02 & 33.3 $\pm$ 1.2\\
&&&&$\dagger$9.7$\pm$0.6&& $\dagger$0.71 $\pm$ 0.14& \\
\hline
6$\pi$ & 2.1 $\pm$ 0.2 && 9.5 $\pm$ 0.9 && 3.1$\pm$ 0.3 &&14.7 $\pm$ 1.4\\
& &&  && & & \\
\hline
7$\pi$  && 1.9 $\pm$ 0.2 && 4.5 $\pm$ 0.5 && 0.57 $\pm$ 0.06 &7.0 $\pm$  0.7\\
& &&  && & & \\
\hline
\end{tabular}
\end{center}
\end{scriptsize}
\caption{Branching fractions in $\pbarp$ annihilation at rest in liquid hydrogen  (in \%) compared to data (whenever available) showing the neutral multiplicity. The columns $0\pi^0$ and $1\pi^0$ show bubble chamber data \cite{Ar69} from which predictions for $n(\pi^0)>1$ can be made with the factorial law (top rows). $\dagger$ CRYSTAL BARREL  data \cite{RMP98}.}
\label{tab:mult}
\end{table}

The branching ratios as a function of total pion multiplicity at rest in 
liquid hydrogen are shown in Figure \ref{Multip} (right). The missing 15.5\% are due to channels with kaons, $\eta$ mesons, $\omega\to \gamma\pi^0$, etc. Direct measurements are available for $n\leq 5$ from bubble chambers and CRYSTAL BARREL (black dots). For higher multiplicities one can resort to the statistical model \cite{Pa60} 
which predicts  the branching ratios to be distributed according to factorial law $\frac{1}{n_+! 
n_-! n_0!}$ for a given total multiplicity $n$: The open squares show the 
predictions derived from channels with charged pions \cite{Ar69}.  The agreement with data for $n\leq 5$ is only fair. The curve is a fit to data only (black dots) assuming $\langle n\rangle = 5$. The average number of charged pions is 3.0 $\pm$ 0.2, the average number of neutral ones 2.0 $\pm$ 0.2. 

Table \ref{tab:mult} shows the neutral multiplicity expected with the factorial law, compared to data from LEAR. Again the agreement with data is  fair.

\subsection*{3. The ASTERIX experiment}
\begin{figure}[htb]
\parbox{170mm}{\mbox{
\includegraphics[width=50mm]{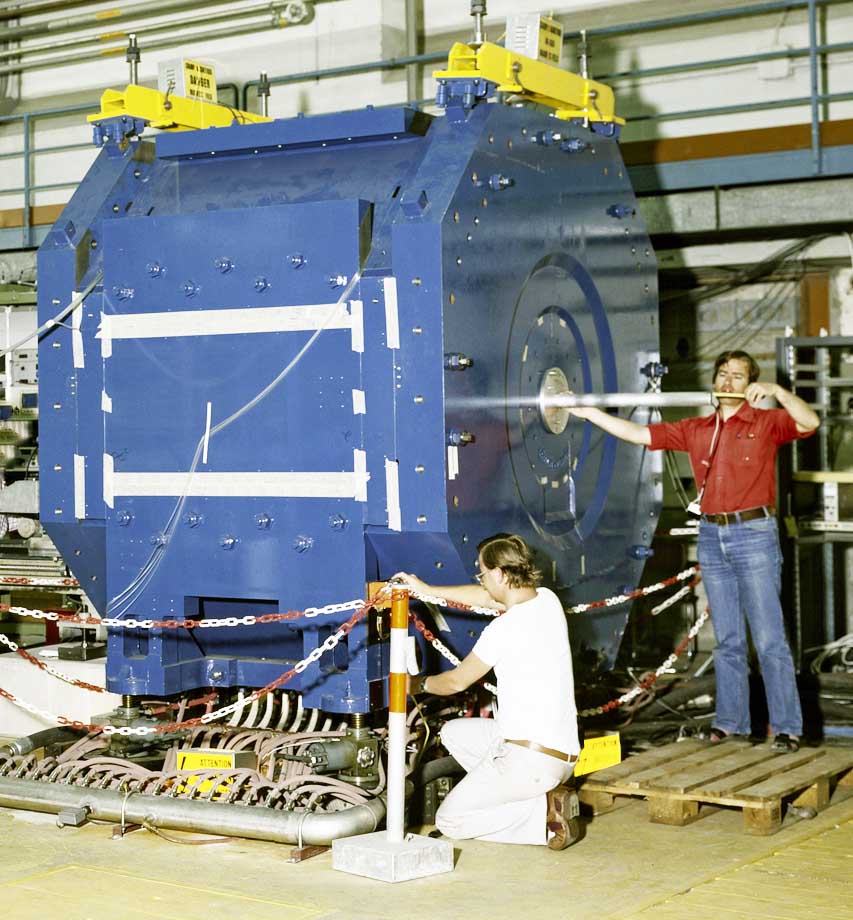}
}\centering}\hfill
\caption[]{The ASTERIX magnetic spectrometer in the South Hall in 1982 (photo CERN).
\label{ASTERIX-photo}}
\end{figure}

ASTERIX studied $\pbarp$ annihilation at rest in gaseous hydrogen at NTP. Figure \ref{ASTERIX-photo} shows the 0.8 T solenoidal magnet (upgraded from the former DM1 at LAL-Orsay). The 105 MeV/c antiprotons from LEAR entered the solenoid along the axis and stopped in  a cylindrical gas target. The gas target was surrounded by a drift chamber to detect x-rays, followed by multiwire proportional chambers. 

\begin{figure}[htb]
\parbox{85mm}{\mbox{
\includegraphics[width=70mm]{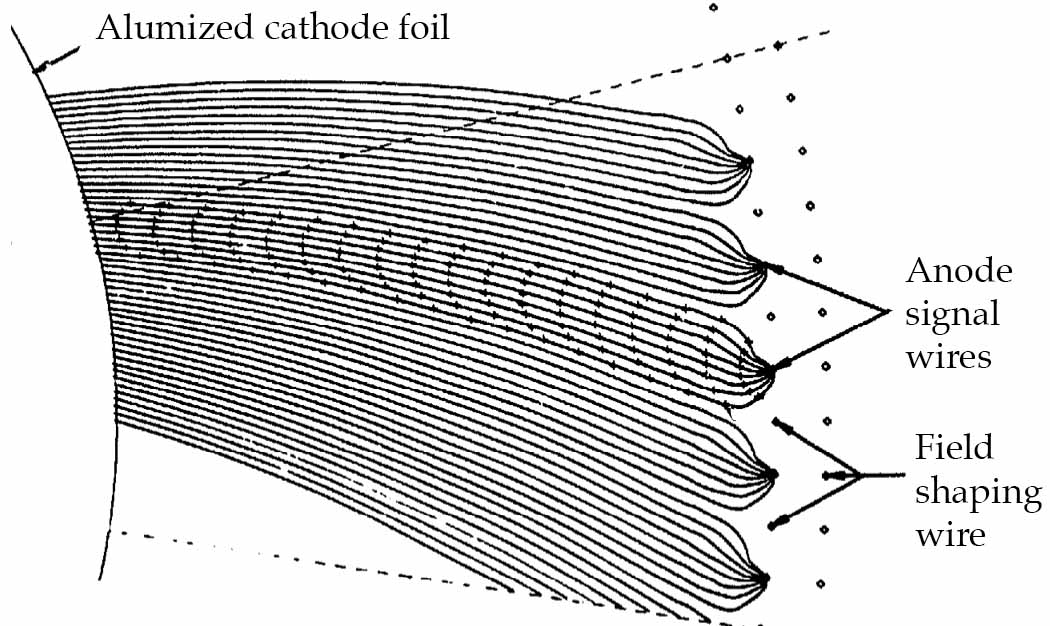}
}\centering}\hfill
\parbox{85mm}{\mbox{
\includegraphics[width=70mm]{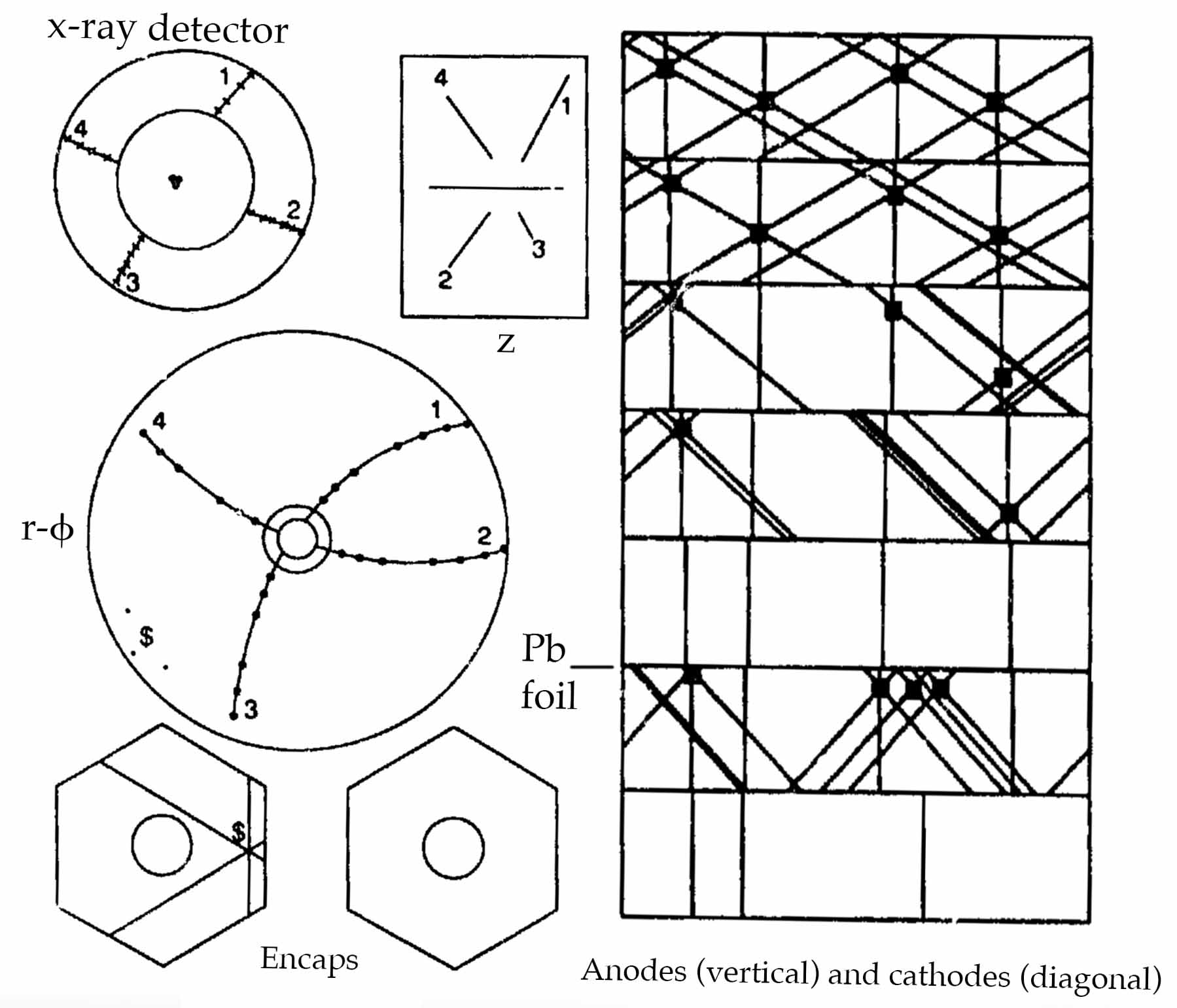}
}\centering} \caption[]{Left: the ASTERIX x-ray drift chamber. Right: a typical four prong event (after \cite{Ah90}).
\label{ASTERIXDet}}
\end{figure}

A drawing of the x-ray drift chamber  is shown in Figure \ref{ASTERIXDet} (left). The detection volume containing an argon/ethane mixture was separated from the target by a 6 $\mu$m thin aluminized mylar foil at a voltage of --10 kV. Drift electron from converted x-rays (with energies above 1 keV) emitted during $\pbar$ capture were detected on anode wires. The curved drift cell in the magnetic field was defined by five field shaping wires and one sense wire. The charged particles were tracked by seven multiwire proportional chambers, five of them with cathode wires or strips to supply coordinates along the detector axis. Limited $\gamma$ detection was achieved with a cylindrical  lead  conversion sheet inserted before the last two chambers and in front of  two end-cap multiwire proportional chambers. Figure \ref{ASTERIXDet} (right) shows a typical four prong event associated with $\gamma$ conversions detected in the endcap and in the last two chambers. Details of the apparatus can be found in \cite{Ah90}.

\subsubsection*{3.1 Protonium spectroscopy}
\begin{figure}[htb]
\parbox{85mm}{\mbox{
\includegraphics[width=70mm]{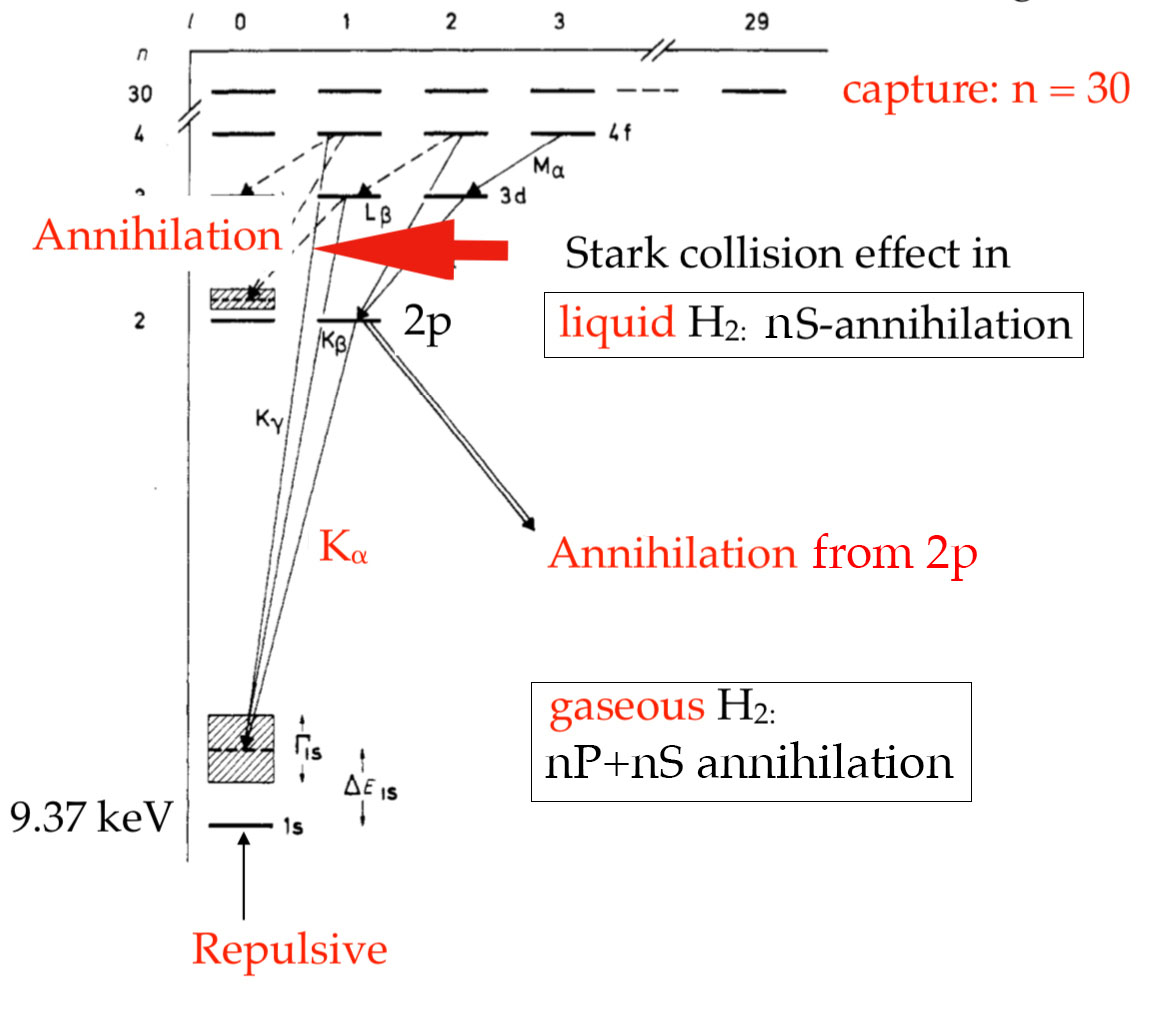}
}\centering}\hfill
\parbox{85mm}{\mbox{
\includegraphics[width=60mm]{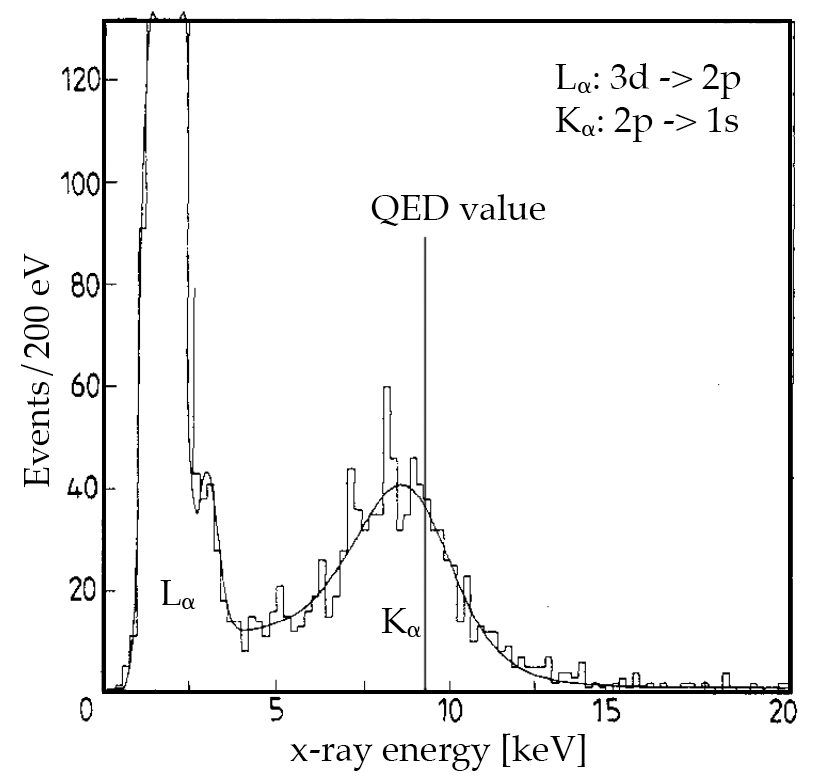}
}\centering} \caption[]{Left: x-ray cascade in liquid and in gas.  Right: x-ray spectrum from ASTERIX associated with 0 prong events and in coincidence with $L_\alpha$ lines \cite{Zi88}.
\label{Cascade}}
\end{figure}

The stopping antiproton is captured in the $n$ = 30 orbital of the antiprotonic atom by an hydrogen atom which ejects its shell electron  (for details see the presentations  at this workshop). The cascade down to the $P$ and $S$ states, strongly suppressed by collisions with neighbouring hydrogen atoms due to Stark mixing, depends on target density.  Therefore annihilation in liquid occurs mainly from high $nS$ orbitals, while in gaseous hydrogen at NTP the $nP$ levels are reached with a probability of about 50\% (see below), from which protonium annihilates with relative angular momentum $L(\pbarp)= 1$ (Figure \ref{Cascade}, left). The cascade time is typically 3 ps in liquid and 5 ns in gas at NTP \cite{Ba89}. By triggering on $L_\alpha$ lines one even obtains annihilation from a pure initial $P$-wave system, in contrast to $L(\pbarp)= 0$  in the liquid of bubble chambers. This was the main originality of ASTERIX which led to a new facet in annihilation studies and in hadron spectroscopy, as we will describe below. 

The $K_{\alpha}$ lines ($2p\to 1s$ transitions) were observed for the first time by ASTERIX with a yield of about $7\times 10^{-3}$/annihilation in gas at NTP, while the total yield of $L$ lines (transitions to $ 2p$) was measured to be about 0.13/annihilation   \cite{Ah85b}.   The main background to the x-ray spectrum was inner bremsstrahlung from charged pions being suddenly accelerated from the annihilation point. This background could be suppressed by triggering on 0 prong events and adding the $L_\alpha$ signal in coincidence (Figure \ref{Cascade}, right). The $K_\alpha$ line appears to be shifted from the QED prediction (9.37 keV) towards a lower value, indicating a repulsive shift from the strong interaction\footnote{The real part of the potential is attractive. The imaginary part due to annihilation reduces the $\pbarp$ wavefunction at short distances and therefore appears as repulsive.}.
The shift and width of the $1s$ protonium level have been measured more accurately by other LEAR experiments with the average values --0.72 $\pm$ 0.04 keV and 1.11 $\pm$ 0.07 keV, respectively, see \cite{Ba89} where comprehensive details on the x-ray cascade can be found.

\subsubsection*{3.2 Fraction of $\boldsymbol{S}$- and $\boldsymbol{P}$-wave annihilation at rest in liquid and gas}
The pairs of charge conjugated spinless mesons  $\pi^+\pi^-$, $K^+K^-$ and $K^0\overline{K}^0$ have the quantum numbers of parity $P$ and $C$-parity equal to = $(-1)^\ell$, where $\ell$ is the relative angular momentum, hence $J^{PC}$ = $0^{++}$, $1^{--}$, $2^{++}$, etc. Due to $P$ and $C$ conservations in strong interactions these are also the quantum numbers of the annihilating protonium levels, where now $P = (-1)^{L+1}$ and $C = (-1)^{L+S}$ with $L$ the orbital angular momentum and $S=0$ (singlet) or $S=1$ (triplet hyperfine state). Limiting ourselves to $s$ and $p$ orbitals, and with $J$ conservation, the protonium levels contributing to the meson pairs above are
\be
n^3P_0 (0^{++}), n^3S_1 (1^{--}), n^3P_2 (2^{++})
\ee
with the notation  $n^{2S+1}L_J$. On the other hand, $\pi^0\pi^0$ is excluded for $1^{--}$ because of Bose-Einstein symmetry and therefore its rate should be much reduced in $\pbarp$ annihilation in liquid. Before LEAR annihilation at rest into $\pi^0\pi^0$  had been hard to measure in liquid with  secondary antiproton beams, due to background from the stronger $3\pi^0$ channel and from $2\pi^0$ annihilation in flight. A reliable measurement, ($6.93 \pm 0.43) \times 10^{-4}$ \cite{Am92}, was eventually performed by CRYSTAL BARREL thanks to its excellent photon detection capability and the low energy narrow momentum bite of LEAR, leading to a sharp stopping peak (with a width $<$ 1mm at 200 MeV/c). 
The ratio of branching fractions
\be
f_p({\rm liq}) =  \frac{2 \times B(\pi^0\pi^0)_{{\rm  liq}}}{B(\pi^+\pi^-)_{\rm  2p}}
\ee
should give the fraction of $P$-wave annihilation in liquid (the factor 2 being a Clebsch-Gordan coefficient). With $B(\pi^+\pi^-)_{\rm 2p}$ = $(4.81 \pm 0.49)\times 10^{-3}$ from ASTERIX \cite{Do88b} one gets the surprisingly large result $f_p({\rm liq})\simeq$ 29\%. However, higher  $L$ orbitals ($n\geq 2$) also contribute to $\pi^0\pi^0$ and their populations need not be equal to that of the $2p$, and are density dependent. Hence a full atomic cascade calculation including Stark mixing and hadronic broadening by the strong interaction had to be  performed \cite{Ba96}, leading to the result
\be
\fbox{$f_p({\rm liq})=  (13 \pm 4)\%$} \ .
\ee

Let us now calculate the fraction of $P$-wave in gas at NTP. The channel $K^0\overline{K}^0$ appears as $K_S K_S$ (or $K_L K_L$) for $C=+1$ and $K_S K_L$ for $C=-1$. Hence $K_S K_S$ appears only in $P$-waves and $K_SK_L$ only in $S$-waves. From bubble chamber experiments one then obtains the branching ratio
\be
B(K^0\overline{K}^0)_S = \frac{B(K_SK_L)({\rm  liq})}{1-f_p({\rm liq})} = (8.7 \pm 0.6)\times 10^{-4} .
\label{eq:k0k0}
\ee
On the other hand, the measurement in gas gave  $(3.8 \pm 0.6)\times 10^{-4}$ \cite{Do88}, hence the $S$-fraction in gas is 0.437 $\pm$ 0.075 or
\be
\fbox{$f_p({\rm gas})=  (56 \pm 7)\%$} \ 
\ee
at NTP. This is in gratifying agreement with the cascade calculation (57 $\pm$ 6)\% \cite{Ba96}, where the $P$ fraction as a function of hydrogen density can also be found.

\begin{figure}[htb]
\parbox{170mm}{\mbox{
\includegraphics[width=130mm]{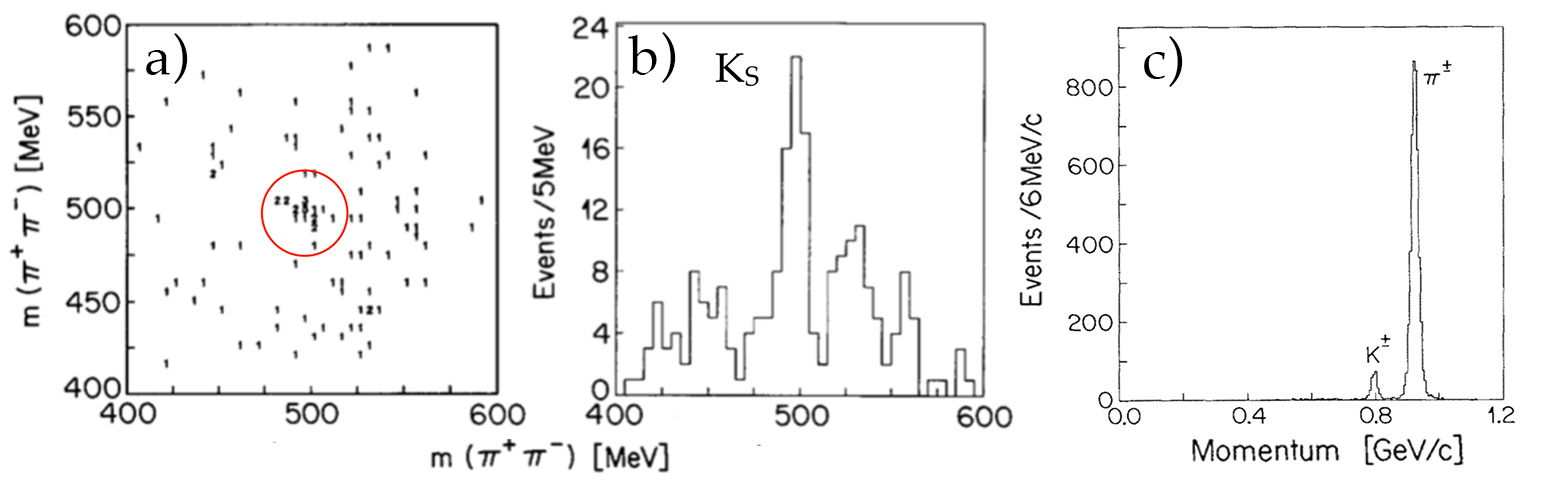}
}\centering}\hfill
\caption[]{a) and b) $\pi^+\pi^-$ invariant mass distribution of  $(\pi^+\pi^-)(\pi^+\pi^-)$ events with two separated vertices \cite{Do88}; \\c) momentum distribution in the two-body annihilation channels $\pi^+\pi^-$ and $K^+K^-$, in coincidence with $L$ x rays \cite{Do88b}.
\label{KKpp}}
\end{figure}

The branching ratio for $K_SK_S$ has been measured by ASTERIX. The rare signal is clearly seen in Figure \ref{KKpp}a,b which show  the distribution of two-pion invariant masses for $\pi^+\pi^-$  pairs with well separated vertices, measured in coincidence with $L$ x-rays. The branching ratio for $(K^0\overline{K}^0)_P$ is then, taking into account the unobserved $K_LK_L$,
\be
B(K^0\overline{K}^0)_P = (7.4 \pm 2.8)\times 10^{-5}.
\ee
Hence $K^0\overline{K}^0$ is suppressed from $P$-waves by an order of magnitude (compare with (\ref{eq:k0k0})). Thus the absence of $K_SK_S$ signal in bubble chambers - which was attributed to the strong $S$-wave dominance in liquid - is in reality due to the suppression from $P$-waves. Also $K^+K^-$ is suppressed from $P$-waves: The two-body $\pi^+\pi^-$ and $K^+K^-$ annihilations  appear as circular tracks traversing the detector. The momentum distribution is plotted in  Figure \ref{KKpp}c and  Table \ref{tab:pipiKK} lists the branching ratios for $\pbarp\to\pi^+\pi^-$ and $K^+K^-$ from $S$ and $P$ states.

\begin{table}[htb]
\begin{small}
\begin{center}
\begin{tabular}{c c c}
\hline
 & $S$ & $P$\\
\hline
$\pi^+\pi^-$ & $(3.19 \pm 0.20) \times 10^{-3}$ &$(4.81 \pm 0.49) \times 10^{-3}$\\
$K^+K^-$ & $(1.08 \pm 0.05) \times 10^{-3}$ &$(2.87 \pm 0.51) \times 10^{-4}$\\
\hline
\end{tabular}
\end{center}
\end{small}
\caption{Branching fractions for $\pbarp\to\pi^+\pi^-$ and $K^+K^-$ \cite{AMy}.}
\label{tab:pipiKK}
\end{table}

\begin{figure}[htb]
\parbox{170mm}{\mbox{
\includegraphics[width=100mm]{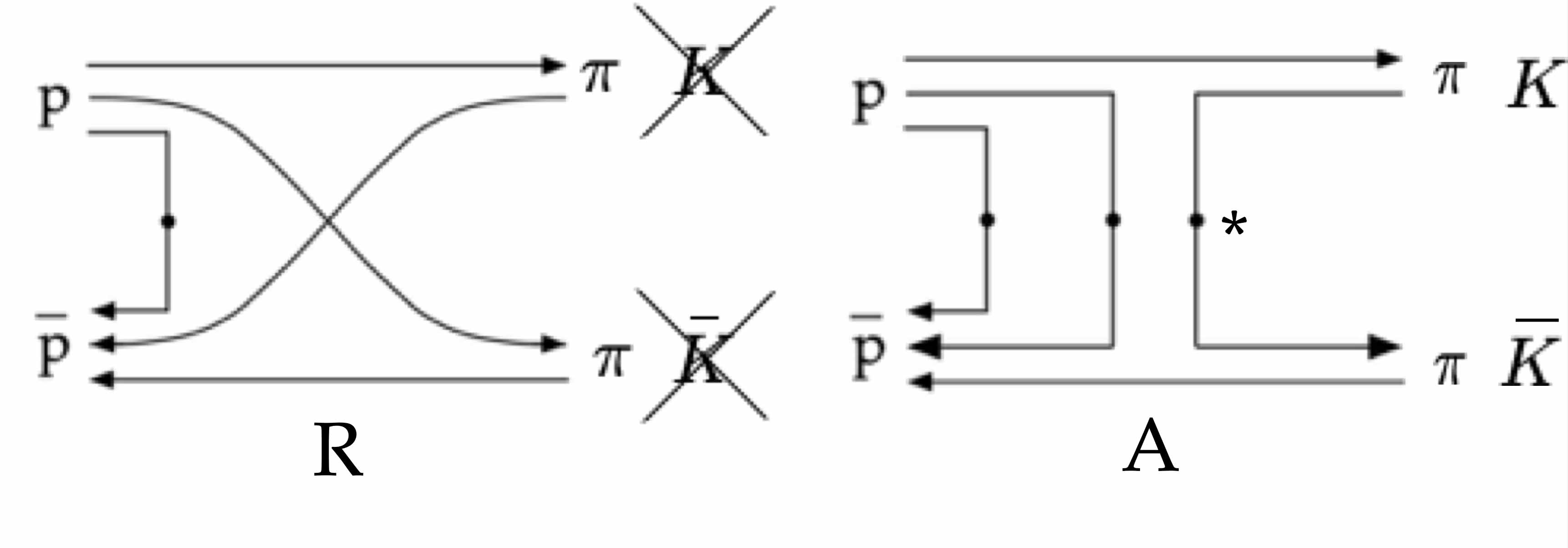}
}\centering}\hfill
\caption[]{Quark rearrangement (R) and quark annihilation (A) in two-body annihilations $\pi^+\pi^-$ and $K^+K^-$. The latter cannot proceed via R and appears to be suppressed in A from initial $P$-waves.
\label{RA}}
\end{figure}

The origin of the $K\overline{K}$ suppression from $P$-waves has been widely discussed in the literature and is still not settled. (For a discussion on dynamical selection rules in $\NNbar$ annihilation see \cite{Kl05,AMy}.) From a microscopic point of view this could be due to a missing contribution from the quark annihilation graph A (Figure \ref{RA}). In the $^3P_0$ model the $\qqbar$ pair (asterisk) generated from vacuum ($0^{++}$ = $^3P_0$) introduces a relative angular momentum $\ell = 1$ between the final state mesons which then violates parity conservation in the transition from the initial $P$ states with quantum numbers $^3P_0 (0^{++})$ or $^3P_2 (2^{++})$.

\subsubsection*{3.3 At last a baryonium state?}
\begin{figure}[htb]
\parbox{170mm}{\mbox{
\includegraphics[width=100mm]{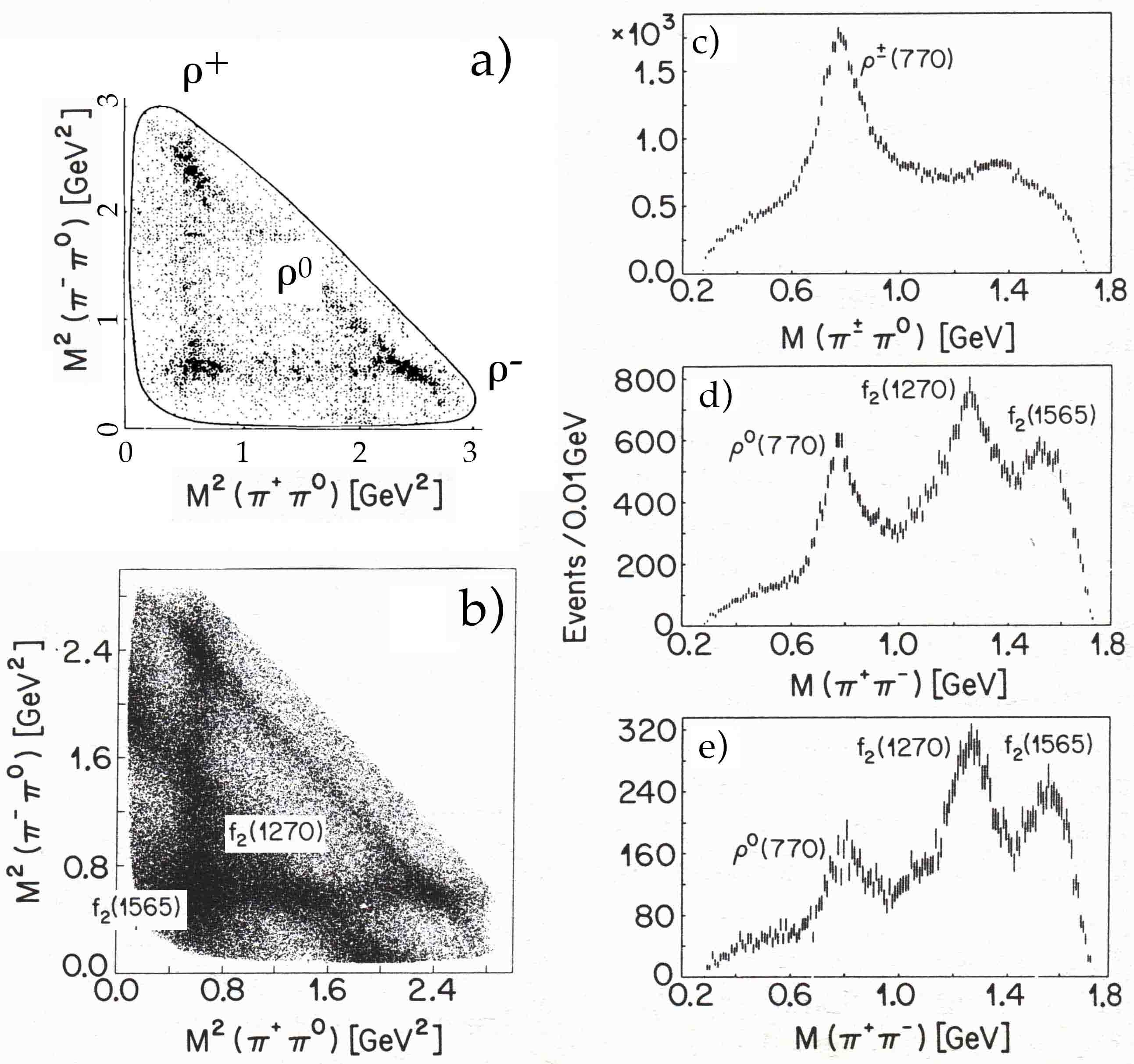}
}\centering}\hfill
\caption[]{a) $\pbarp\to\pi^+\pi^-\pi^0$ in liquid hydrogen \cite{Fo68},  in gas (b,c,d) and from the $2p$ state (e) \cite{Ma90}. 
\label{AX}}
\end{figure}

Proton-antiproton annihilation into $\pi^+\pi^-\pi^0$ has been studied earlier by bubble chamber experiments. The Dalitz plot (Figure \ref{AX}a) reveals the production of the intermediate states $\rho^+\pi^-$, $\rho^-\pi^+$ and $\rho^0\pi^0$ in equal intensities. $C$-conservation excludes  $\rho^0\pi^0$ from the $^1S_0 $ initial state and hence the annihilation occurs from $^3S_1$, as was confirmed by the angular distribution of the pions in the $\rho$ rest frames \cite{Fo68}.  The $^1S_0 $ suppression is another dynamical selection rule, the nature of which is not understood.

Figure \ref{AX}b shows the ASTERIX Dalitz plot in hydrogen gas, together with the invariant mass projections c) and d). A very strong $f_2(1270)\to\pi^+\pi^-$ signal appears (which is barely visible in Figure \ref{AX}a), together with a new peak at 1565 MeV \cite{Ma89} which is enhanced when triggering on $L$ x-rays (pure $2p$ annihilation, Figure \ref{AX}e). The amplitude analysis led to a new tensor meson, the $f_2(1565)$ (formerly called the $AX$) produced from $P$-waves with mass and width of 1565$\pm$10 MeV and 170$\pm$20 MeV, respectively \cite{Ma90}. Furthermore, this meson is isoscalar since no signal is observed in $\pi^\pm\pi^0$ (Figure \ref{AX}c). 

The $f_2(1565)$ (which has also been reported by OBELIX) is now well established and seems to be produced only in $\NNbar$ annihilations. Tensor mesons build one of the best known nonets comprising the $f_2(1270)$ ($\uubar + \ddbar$) and $f_2'(1525)$ ($\ssbar$) isoscalar mesons  \cite{PDG19}. Hence  the $f_2(1565)$ finds no room as a $\qqbar$ meson and must be of a different nature. We have seen that a $2^{++}$ isoscalar baryonium is predicted in this mass region (Figure \ref{DoverRichard}). The $f_2(1565)$ could be that state, produced in $\pbarp$ annihilation by shaking off a $\pi^0$, but with a larger width than predicted. Alternatively $f_2(1565)$ could be one of the  tetraquark states coupling to $\NNbar$ which was predicted a long time ago \cite{Ja77} (see also Table 16.2 in \cite{LNP}).

\subsection*{4. The CRYSTAL BARREL experiment}
\begin{figure}[htb]
\parbox{56mm}{\mbox{
\includegraphics[width=45mm]{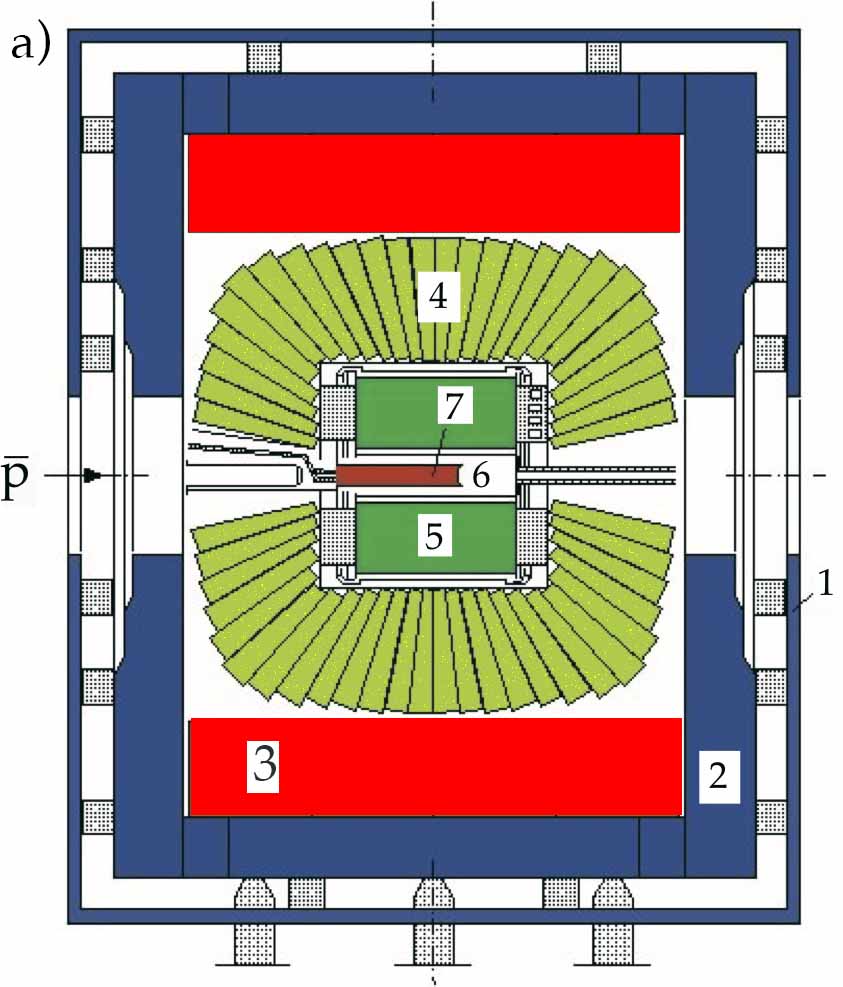}
}\centering}\hfill
\parbox{56mm}{\mbox{
\includegraphics[width=32mm]{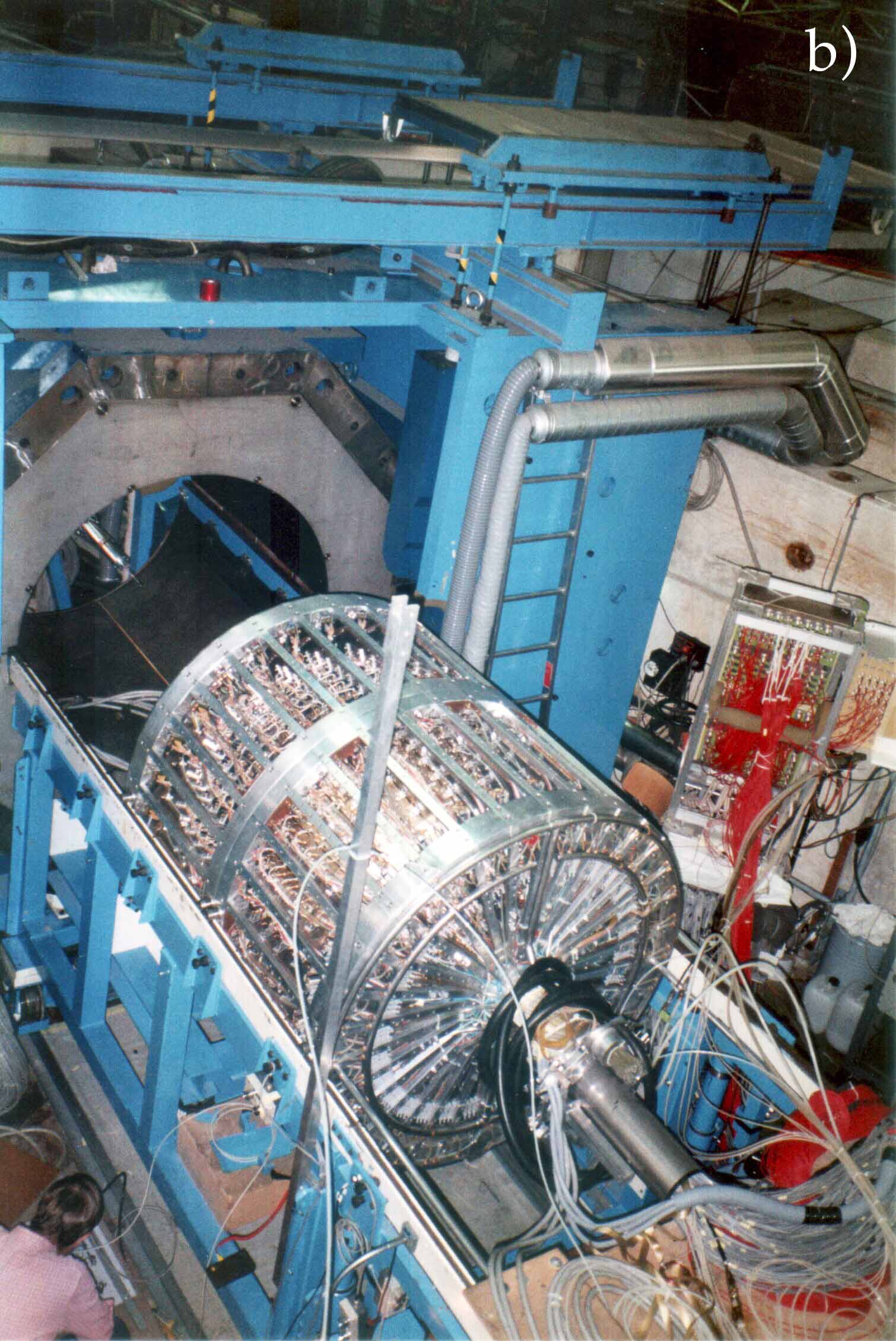}
}\centering}\hfill
\parbox{56mm}{\mbox{
\includegraphics[width=32mm]{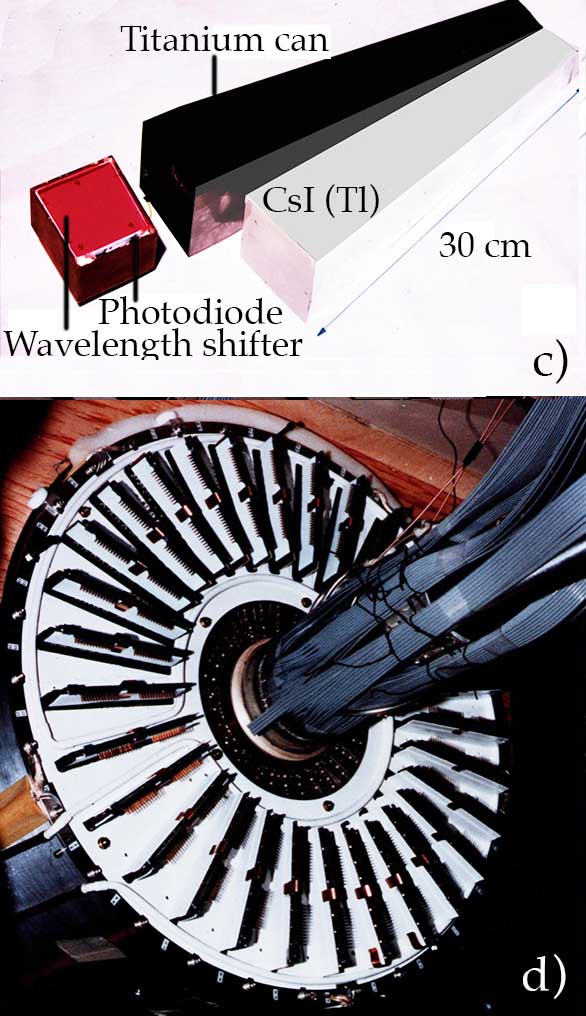}
}\centering}\hfill
\caption[]{a) Drawing of the Crystal Barrel detector with its magnet yoke \cite{Ak92}: (1,2), coil (3),  CsI(Tl) array (4), 
jet drift chamber (5), proportional wire chambers (6), and hydrogen target (7); b) CsI barrel
in front of the magnet (photo CERN); c) single crystal wrapped in teflon and aluminized mylar  and its titanium container; d)  jet drift chamber. 
\label{CB}}
\end{figure}

ASTERIX was succeeded by the CRYSTAL BARREL and OBELIX experiments which took data between 1989 and 1996. The  goal of the former was to study $\pbarp$ annihilation at rest with very high statistics, in particular to investigate the  unknown
channels with several neutral mesons (such as $\pi^0$, $\eta$, $\eta'$, $\omega$) leading to multiphoton final states. 
Figures \ref{CB}a,b show a drawing of the detector and a photograph. The 200 MeV/c antiprotons from LEAR entered the solenoidal 1.5 T magnet  along its axis and stopped in a  liquid hydrogen target. Photons were detected by a barrel-shaped assembly of 1380 CsI(Tl) 
crystals covering a solid angle of 97\% $\times 4\pi$, read out by photodiodes. The scintillation light was converted by a  wavelength shifter and detected by a photodiode (Figure \ref{CB}c). The momenta of the charged pions and kaons were measured by a jet drift chamber composed of 30 sectors, each with 23 sense wires, filled with CO$_{2}$/isobutane (Figure \ref{CB}d). Low energy ($<500$ MeV/c) kaons could be distinguished from pions by dE/dx sampling. Details on the equipment can be found in \cite{Ak92}.

Up to 10 $\gamma$'s with energies above 4 MeV could be detected and reconstructed with good efficiency. CRYSTAL BARREL measured  branching ratios for radiative annihilations ($\pbarp\to\gamma X$) \cite{RMP98} and annihilations into two or more mesons which will be discussed below. However, the emphasis was on the search for new mesons  with very high statistics, so that $T$-matrix analyses could be performed and resonances poles determined. Several mesons were discovered, $ f_0(1370)$, $\pi_1(1400)$, $\eta(1410)$, $ a_0(1450)$, $f_0(1500)$, $\eta_2(1645)$ \cite{PDG19}. I shall only mention briefly hadron spectroscopy here, since it is beyond the scope of this workshop (for a  detailed review of the physics results and a list of publications, see \cite{RMP98}). 

\subsubsection*{4.1 Annihilation into three pseudoscalar mesons}
Annihilation at rest into $\pi^0\pi^0\pi^0$, $\pi^0\pi^0\eta$ and $\pi^0\eta\eta$ has been studied with CRYSTAL BARREL  by detecting and reconstructing 6$\gamma$ events \cite{RMP98}. Note that for channels involving only neutral non-strange mesons (for which $C$-parity is defined),  conservation laws  restrict  the number of contributing partial waves between the initial $\pbarp$ orbitals and the mesonic final states. Also, contributions from the $\rho$ meson which dominates low energy annihilation (see e.g. Figure \ref{AX}) is not present since $\rho^0$ does not decay into $\pi^0\pi^0$. This considerably simplifies   spin-parity analyses.

\begin{figure}[htb]
\parbox{170mm}{\mbox{
\includegraphics[width=100mm]{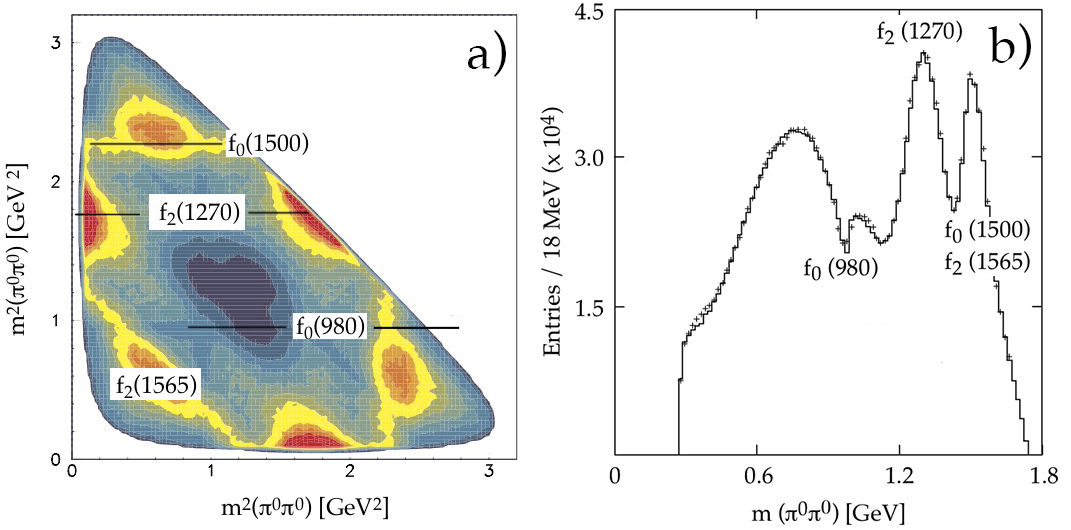}
}\centering}\hfill
\caption[]{a) $3\pi^0$ Dalitz plot confirming the ASTERIX $f_2(1565)$ and showing evidence for a new isoscalar spin 0 meson, the $f_0(1500)$ (6 entries per event, from \cite{RMP98}); b) $2\pi^0$ invariant mass projection (note the large number of entries in the $f_0(1500)$ peak).
\label{f01500}}
\end{figure}

Figure \ref{f01500} shows the Dalitz plot of the $3\pi^0$ final state and its $2\pi^0$ mass projection. Signals from the $f_2(1565)/AX$ and $f_2(1270)$ are clearly seen with the charateristic density distributions of spin 2 mesons. The scalar $f_0(980)$ appears as a dip. Surprising was the observation of a relatively narrow ($\simeq$110 MeV)  homogeneous band at 1500 MeV pointing to the existence of new isoscalar scalar meson, the $f_0(1500)$. This state has been proposed as a glueball \cite{CF96} mixed with the very broad $f_0(1370)$ (required by the Dalitz plot analysis) and the established $f_0(1710)$ (which is not seen here and would  lie at the edge of phase space).  For details on the 1500 MeV region the reader is invited to consult the reviews on ``Scalar mesons below 2 GeV'' and on ``Non-$\qqbar$ mesons'' in the Review of Particle Physics \cite{PDG19}. 

\begin{figure}[htb]
\parbox{170mm}{\mbox{
\includegraphics[width=90mm]{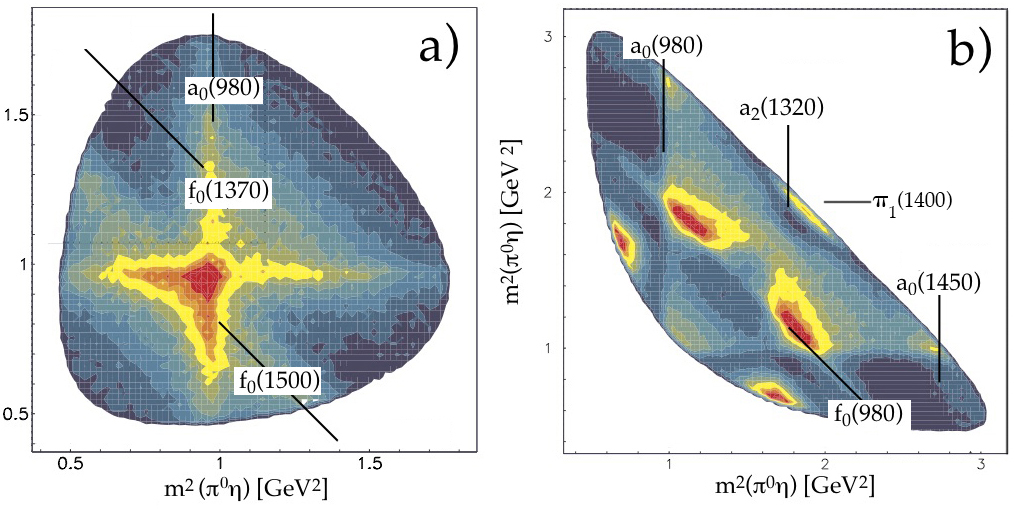}
}\centering}\hfill
\caption[]{$\pi^0\eta\eta$ (a) and $\pi^0\pi^0\eta$ Dalitz plot (b) \cite{RMP98}.
\label{a01450}}
\end{figure}

The $\pi^0\eta\eta$ and  $\pi^0\pi^0\eta$ Dalitz plots from CRYSTAL BARREL are reproduced in Figure \ref{a01450}. The  $\pi^0\pi^0\eta$ data require the presence of a new isovector scalar meson, the $a_0(1450)$ decaying into $\eta\pi^0$, as well as a new structure, the $\pi_1(1400)$ with ``exotic'' quantum numbers $1^{-+}$ \cite{Ca94} which are forbidden for a $\qqbar$ meson. The three Dalitz plots have been fitted jointly in a coupled channel analysis \cite{Ca95}.

\subsubsection*{4.2 Annihilation into two neutral mesons}
\begin{figure}[htb]
\begin{minipage}[c]{0.5\textwidth}
\includegraphics[width=80mm]{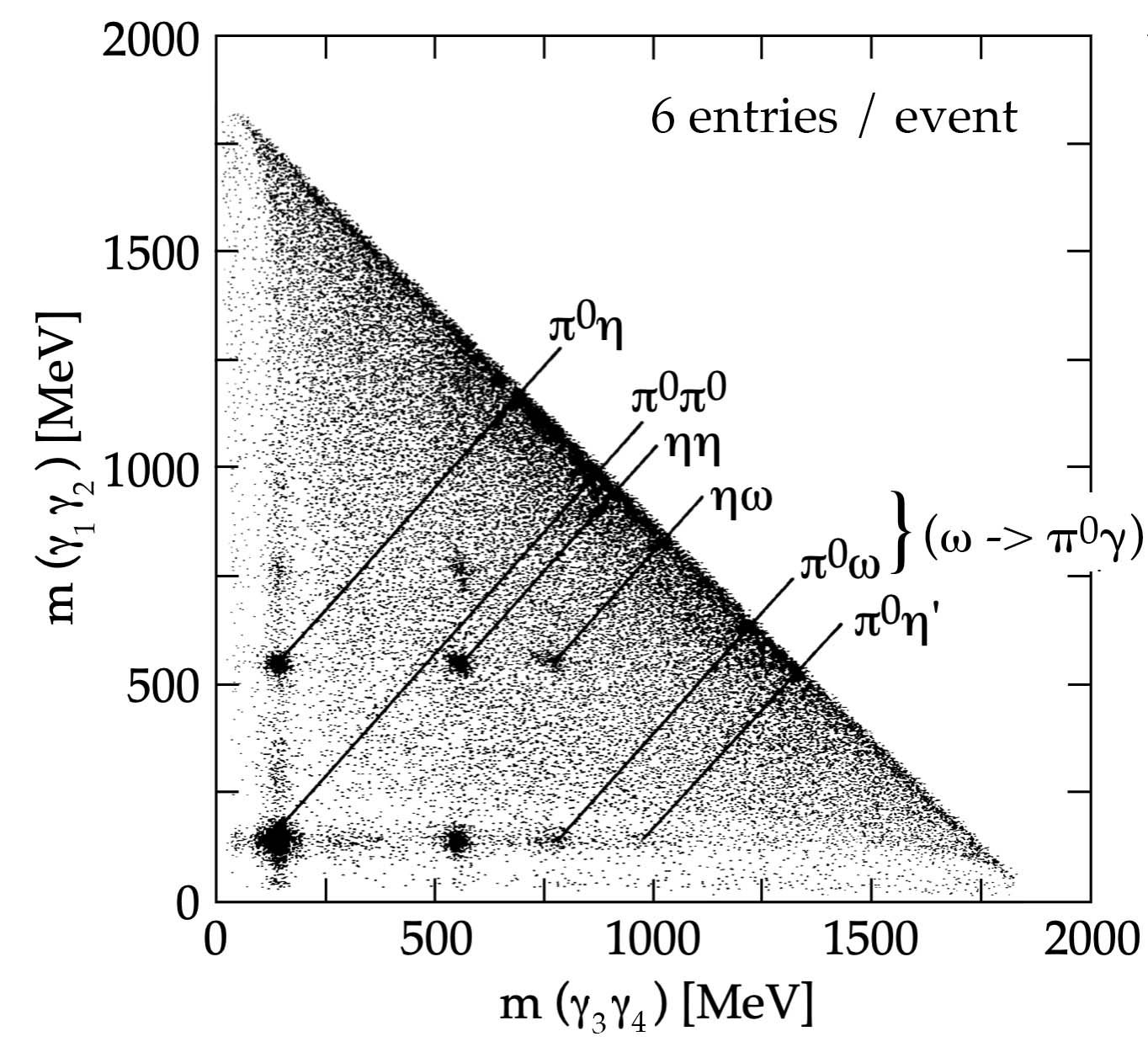}
\centering
\end{minipage}
\begin{minipage}[c]{0.5\textwidth}
%\begin{small}
\begin{tabular}{l c c}
\hline
Channel & \multicolumn{2}{c}{Branching ratio $B$}  \\
\hline
$\pi^0\pi^0$ & 6.93 $\pm$ 0.43 & $\times 10^{-4}$\\ 
$\pi^0\eta$ & 2.12 $\pm$ 0.12 & $\times 10^{-4}$\\ 
$\pi^0\eta'$ & 1.23 $\pm$ 0.13 & $\times 10^{-4}$\\ 
$\eta\eta$ & 1.64 $\pm$ 0.10 & $\times 10^{-4}$\\ 
$\pi^0\omega$ & 5.73 $\pm$ 0.47 & $\times 10^{-3}$\\
$\eta\omega$ & 1.51 $\pm$ 0.12 & $\times 10^{-2}$\\  
\hline
\end{tabular}
%\end{small}
\centering
\end{minipage}
\caption[]{$2\gamma$ mass distribution in $4\gamma$ events and two-body branching ratios (including all meson decay modes) \cite{RMP98,Ca93}.
\label{Fourgam}}
\end{figure}

CRYSTAL BARREL has measured the branching fractions for annihilations into two neutral mesons. Figure \ref{Fourgam} shows a mass scatterplot of events with four detected photons. The accumulations of events are due to two-body annihilations with branching ratios given in the table. The $\eta\omega$ and $\pi^0\omega$ events stem from $\omega\to\pi^0\gamma$ with a missing (undetected) soft $\gamma$ and the dark diagonal band to wrong combinations (three possible combinations per event and symmetrized scatterplot). 

The SU(3) wavefunctions of isoscalar mesons can be decomposed into  light ($u$, $d$) and  strange ($s$) quark contents. For the $i=0$ pseudoscalar mesons
\ba
\eta &=& \frac{1}{\sqrt{2}}(\uubar + \ddbar) \sin(\theta_i-\theta_P) - \ssbar\cos(\theta_i-\theta_P),\nonumber\\
\eta' &=& \frac{1}{\sqrt{2}}(\uubar + \ddbar) \cos(\theta_i-\theta_P) + \ssbar\sin(\theta_i-\theta_P),
\label{eq:mix}
\ea
where $\theta_i = 35.3^\circ$ is the ideal mixing angle and $\theta_P$ the pseudoscalar nonet mixing angle which can be estimated from the masses of the $0^{-+}$ nonet mesons $\pi$, $K$, $\eta$, $\eta'$: $\theta_P$ = $-24.5^\circ$ or $-11.3^\circ$, depending on whether one uses the linear or the quadratic mass formula \cite{LNP,PDG19}. For a nonet mixing angle of $35.3^\circ$ the light and heavy components would decouple and the two isoscalars would be pure $\ssbar$ and pure $\uubar + \ddbar$, as is nearly the case for the vector mesons $\phi$ and $\omega$ discussed below. The $\pi^0 (\rho^0)$ and $\omega$ wavefunctions are given by
\be 
\pi^0 = \frac{1}{\sqrt{2}}(\ddbar-\uubar)\ \ {\rm and} \ \ \omega = \frac{1}{\sqrt{2}}(\ddbar + \uubar).
\label{eq:omrho}
\ee

If we now assume that the nucleon and antinucleon do not contain any $s$ nor $\sbar$ quark,  an $\ssbar$ pair cannot be produced by the
graphs shown in Figure \ref{RA} and we may ignore the $\ssbar$ components in eqn. (\ref{eq:mix}) when calculating the annihilation branching ratios. Thus, following \cite{Ge85} one gets for example the ratio of phase space corrected branching ratios
\be
\frac{\tilde{B}(\eta\eta)}{\tilde{B}(\eta\eta')} = \frac{1}{2}\bigg(\frac{\sin^2\Delta}{\sin\Delta\cos\Delta}\bigg)^2 = \frac{\tan^2\Delta}{2},
\ee
where the factor $\frac{1}{2}$ takes into account the two identical mesons in the numerator and $\Delta = \theta_i-\theta_P$. This prediction can be used to obtain $\theta_P$ from the measured branching ratios $B$ divided by the phase space factor, e.g. using the prescription  (\ref{eq:phasespace}). Table \ref{tab:twobodyBR} lists the pseudoscalar mixing angles derived from some of the measured branching ratios \cite{RMP98}. They are in agreement with expectations, which suggests that quarks dynamics is indeed relevant to the annihilation mechanism. 

However, if we limit the quark contributions to the annihilation graph A which does not contribute to the pairs of components ($\uubar$, $\uubar$) nor ($\ddbar$, $\ddbar$), then the results for $\theta_P$ are not as consistent, in particular the ones involving $2\pi^0$. Furthermore, one would expect from A dominance the branching ratios for $\rho^0\rho^0$ and $\omega\omega$ to be equal, see eqn. (\ref{eq:omrho}). This disagrees with measurements, presumably due to the contribution from rearrangement R: $B(\rho^0\rho^0)$ = $(1.2 \pm 1.2) \times 10^{-3}$ and $B(\omega\omega)$ = $(3.32 \pm 0.34) \times 10^{-2}$ (phase space factors are equal).

\begin{table}[htb]
\begin{center}
\begin{tabular}{c c c  }
\hline
Ratio  & Prediction & $\theta_P$ \\
& ($\Delta = \theta_i-\theta_P$) & [$^{\circ}$]\\
\hline
\noalign{\vskip 1mm} 
$\frac{\tilde{B}(\eta\eta)}{\tilde{B}(\eta\eta')}$  & 
$\frac{1}{2}\tan^2\Delta$ & --17.7 $\pm$ 1.9\\
\noalign{\vskip 1mm}
$\frac{\tilde{B}(\pi^0\eta)}{\tilde{B}(\pi^0\eta')}$   & 
$\tan^2\Delta$ & 
--18.1 $\pm$ 1.6\\
\noalign{\vskip 1mm}  
$\frac{\tilde{B}(\omega\eta)}{\tilde{B}(\omega\eta')}$  & 
$\tan^2\Delta$ 
& --21.1 $\pm$1.5\\
\noalign{\vskip 1mm} 
$\frac{\tilde{B}(\eta\rho^0)}{\tilde{B}(\eta'\rho^0)}$  & 
$\tan^2\Delta$ & 
--25.4 $^{+}_{-}$ $^{5.0}_{2.9}$\\
\noalign{\vskip 1mm} 
\hline
\noalign{\vskip 1mm} 
$\frac{\tilde{B}(\eta\eta)}{\tilde{B}(\pi^0\pi^0)}$  & 
$\sin^4\Delta$ & --6.2 $^{+}_{-}$ $^{0.6}_{1.1}$\\
\noalign{\vskip 1mm} 
$\frac{\tilde{B}(\eta\eta')}{\tilde{B}(\pi^0\pi^0)}$  & 
$2\sin^2\Delta\cos^2\Delta$ & 14.6 $\pm$ 1.8\\
& & or --34.0 $\pm$ 1.8\\
\noalign{\vskip 1mm} 
\hline
\end{tabular}
\end{center}
\caption[]{Pseudoscalar mixing angle $\theta_P$ predicted from
the ratios of measured phase space corrected two-body branching ratios $\tilde{B}$. The bottom rows assume that only graph A contributes.}
\label{tab:twobodyBR}
\end{table}

\subsection*{5 The OBELIX experiment}

\begin{figure}[htb]
\parbox{100mm}{\mbox{
\includegraphics[width=90mm]{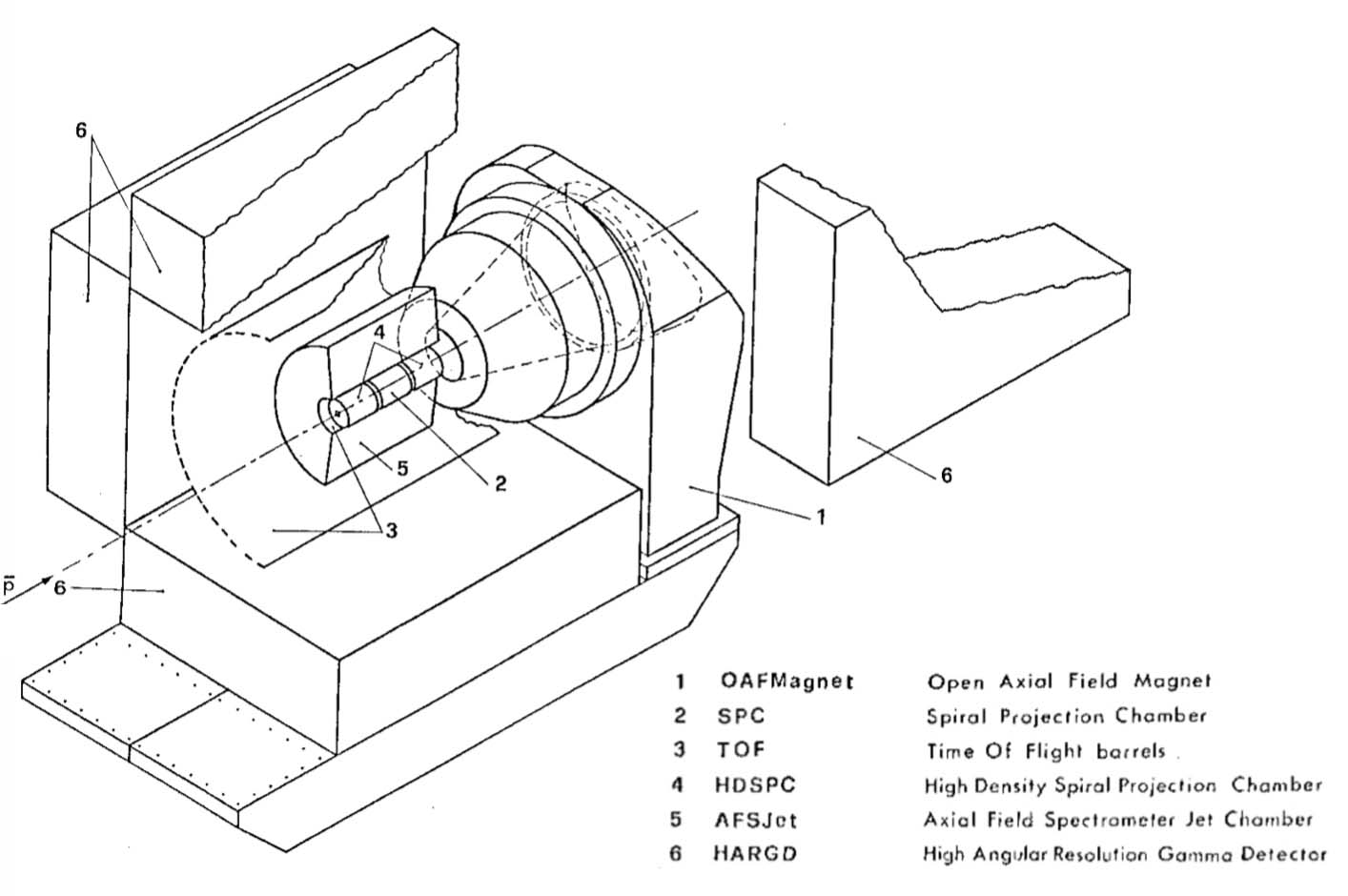}
}\centering}\hfill
\parbox{70mm}{\mbox{
\includegraphics[width=56mm]{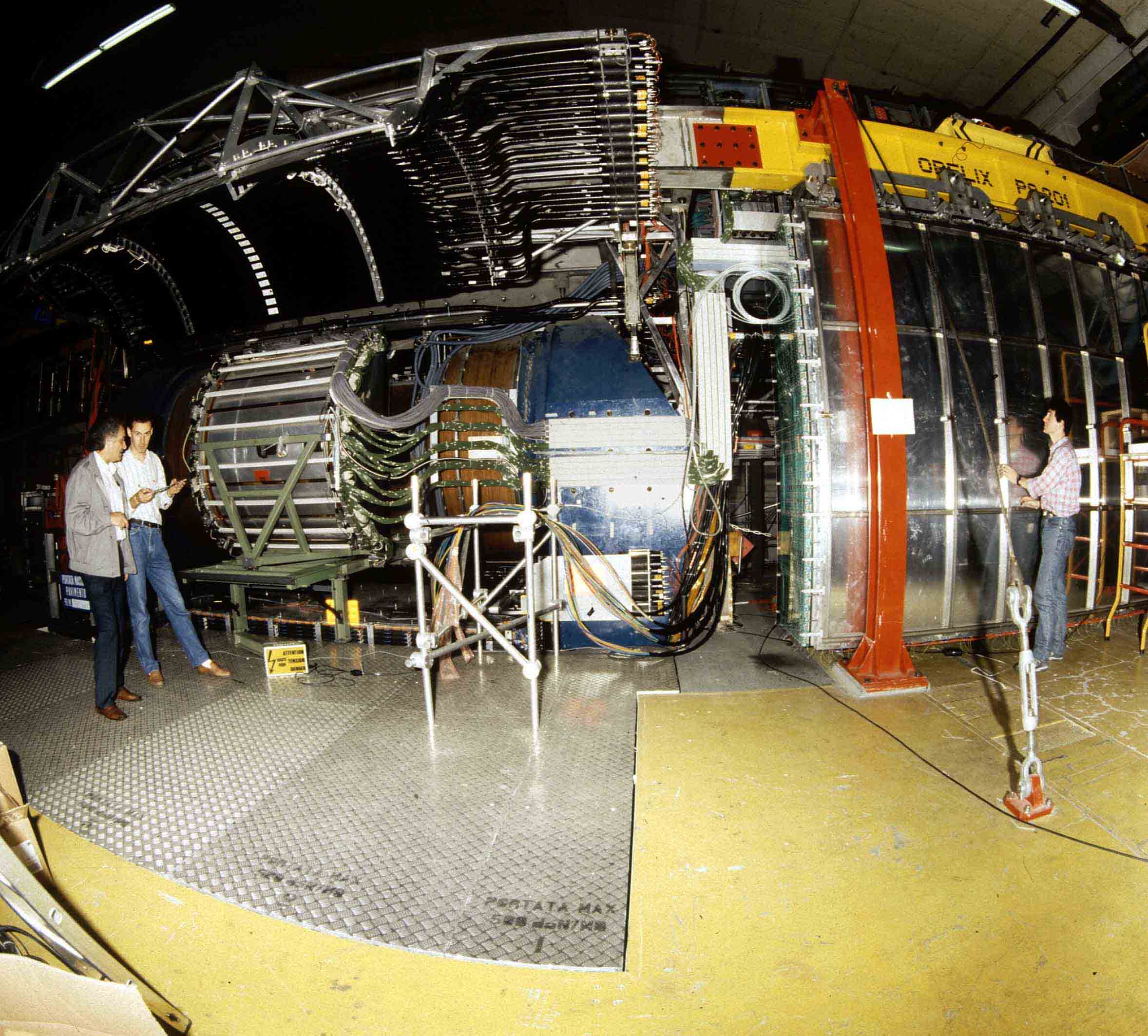}
}\centering} \caption[]{Sketch of the OBELIX apparatus  \cite{Br03} and photograph of the open detector (photo CERN).
\label{Obelix}}
\end{figure}

The OBELIX collaboration employed the large Open Axial Field Magnet which was operated earlier at the CERN Intersecting Storage Rings (Figure \ref{Obelix}). The magnetic field between the two poles separated by a distance of 1.5 m reached a maximum value of 0.6 T. The apparatus was equipped with four subdetectors located around the axis of the magnet. The time-of-flight detector consisted of two co\-axial scintillator barrels. A jet drift chamber inserted between the two scintillator barrels tracked charged particles which could be identified through dE/dx sampling.  The antiproton momentum was 200 MeV/c for measurements in liquid  and 105 MeV/c in gas at NTP and at lower densities. Photon detection was provided by a sandwich of lead plates and limited streamer tubes.

\begin{figure}[htb]
\parbox{105mm}{\mbox{
\includegraphics[width=90mm]{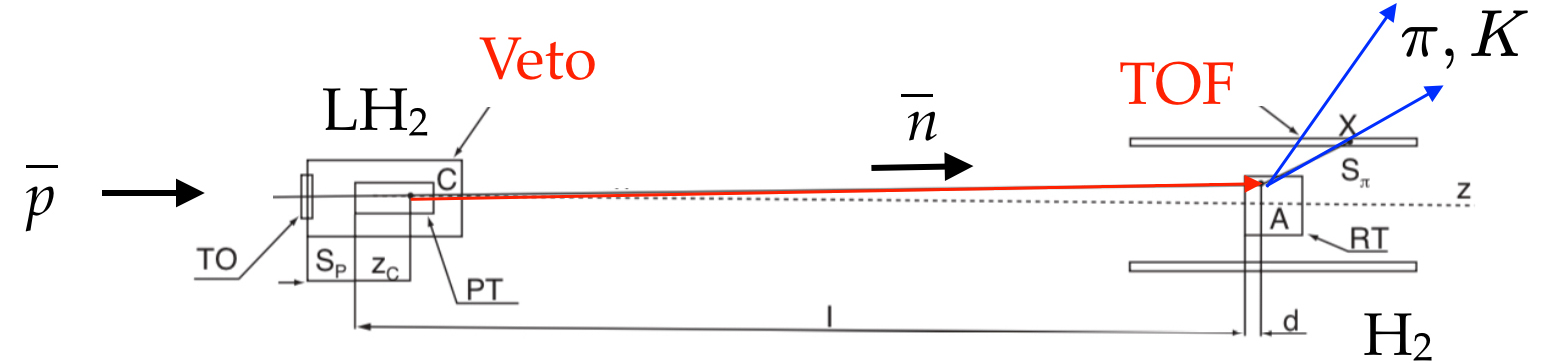}
}\centering}\hfill
\parbox{65mm}{\mbox{
\includegraphics[width=60mm]{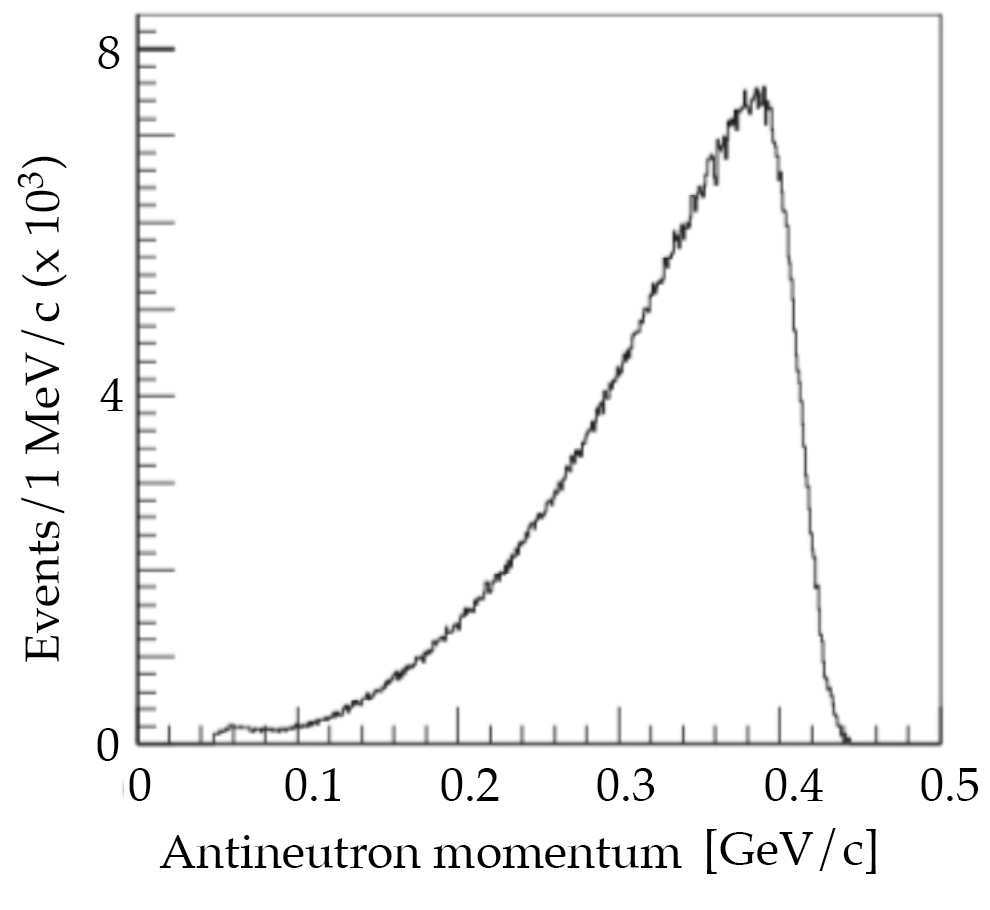}
}\centering} \caption[]{Left: antineutron production by charge exchange $\pbarp\to\nbarn$, followed by $\nbar$ annihilation in hydrogen. Right: momentum distribution of antineutrons produced by 412 MeV/c antiprotons \cite{Br03}.
\label{NeutronBeam}}
\end{figure}

One of the most original features of OBELIX was its antineutron beam facility (Figure \ref{NeutronBeam}, left). The antiprotons provided the start signal for the time-of-flight measurement.  While slowing down some of them converted into antineutrons through the charge exchange reaction $\pbarp\to \nbarn$ (threshold of 98 MeV/c). Charged particle from antiproton annihilation were removed by a veto box. Antineutron-proton interactions were studied with the antineutrons produced in the forward direction and reaching the second hydrogen target located at the center of the OBELIX detector. The antineutron momentum was determined by time-of-flight.

The main advantages of antineutrons over antiprotons is the absence of Coulomb interaction which complicates measurements of strong interaction cross sections. Furthermore, there is no energy loss nor range straggling which allows lower energy ranges. Antiproton-neutron interaction, (equivalent to antineutron-proton by $C$-invariance) has been studied earlier in deuterium bubble chambers \cite{Ban85}, but with the complication introduced by the spectator proton\footnote{The charged multiplicities in $\nbarp$ and $\pbarn$ have been measured in bubble chambers between 700 and 760 MeV/c and are found (as expected) to be equal:  1 prong $\simeq$14\%, 3 prong $\simeq$61\%, 5 prong $\simeq$24\%, 7 prong $\sim$1\%   \cite{Ban85} .}. Also, $\nbarp$ is a pure isospin 1 state, which simplifies the amplitude analyses when searching for resonances. The main drawback is the low antineutron flux (typically 40 $\nbar$ / 10$^6$ $\pbar$ in OBELIX) and the antineutron flux as a function of energy (Figure \ref{NeutronBeam}, right) has to be calculated by Monte Carlo simulations.

\begin{figure}[htb]
\parbox{85mm}{\mbox{
\includegraphics[width=66mm]{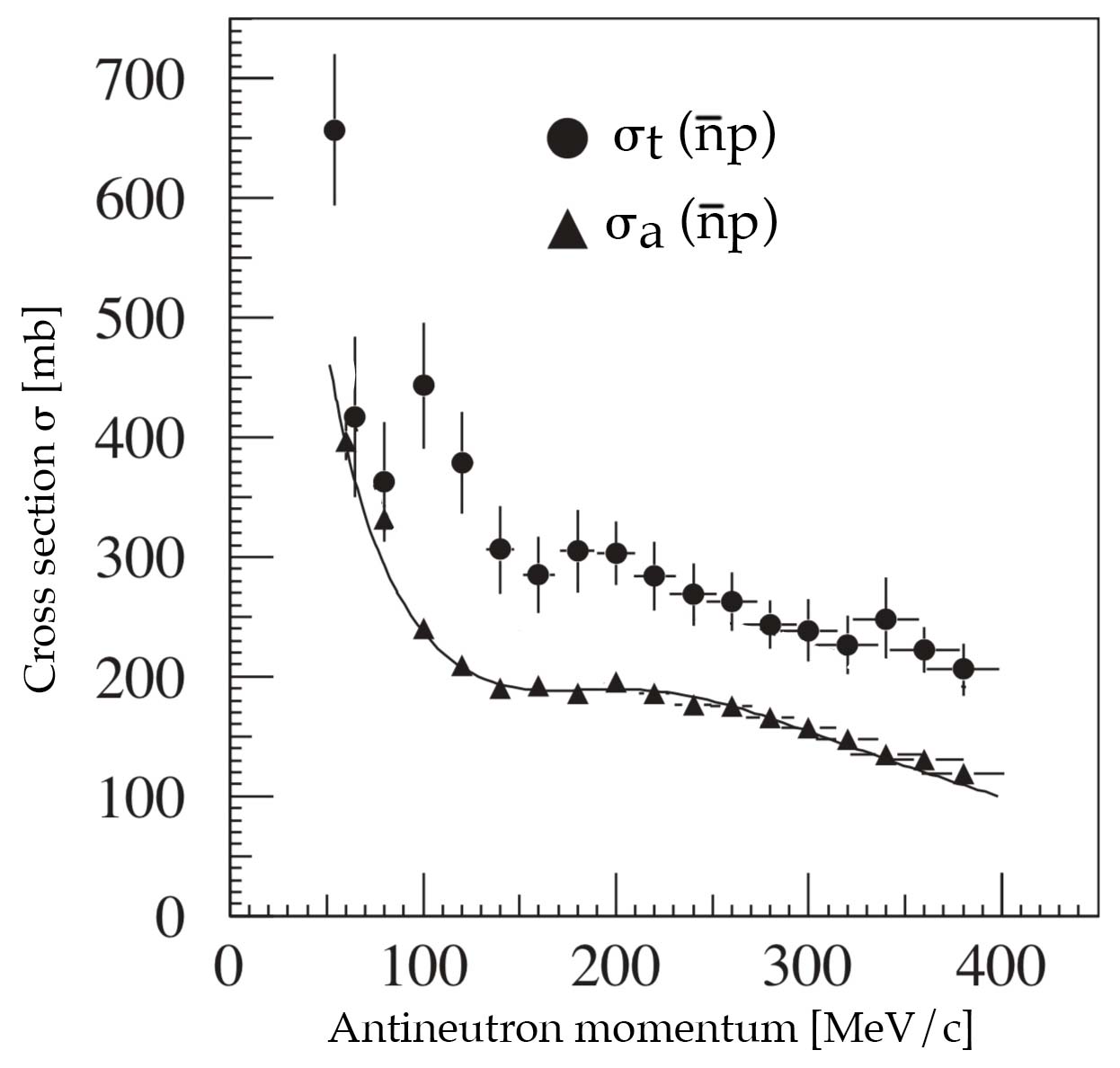}
}\centering}\hfill
\parbox{85mm}{\mbox{
\includegraphics[width=70mm]{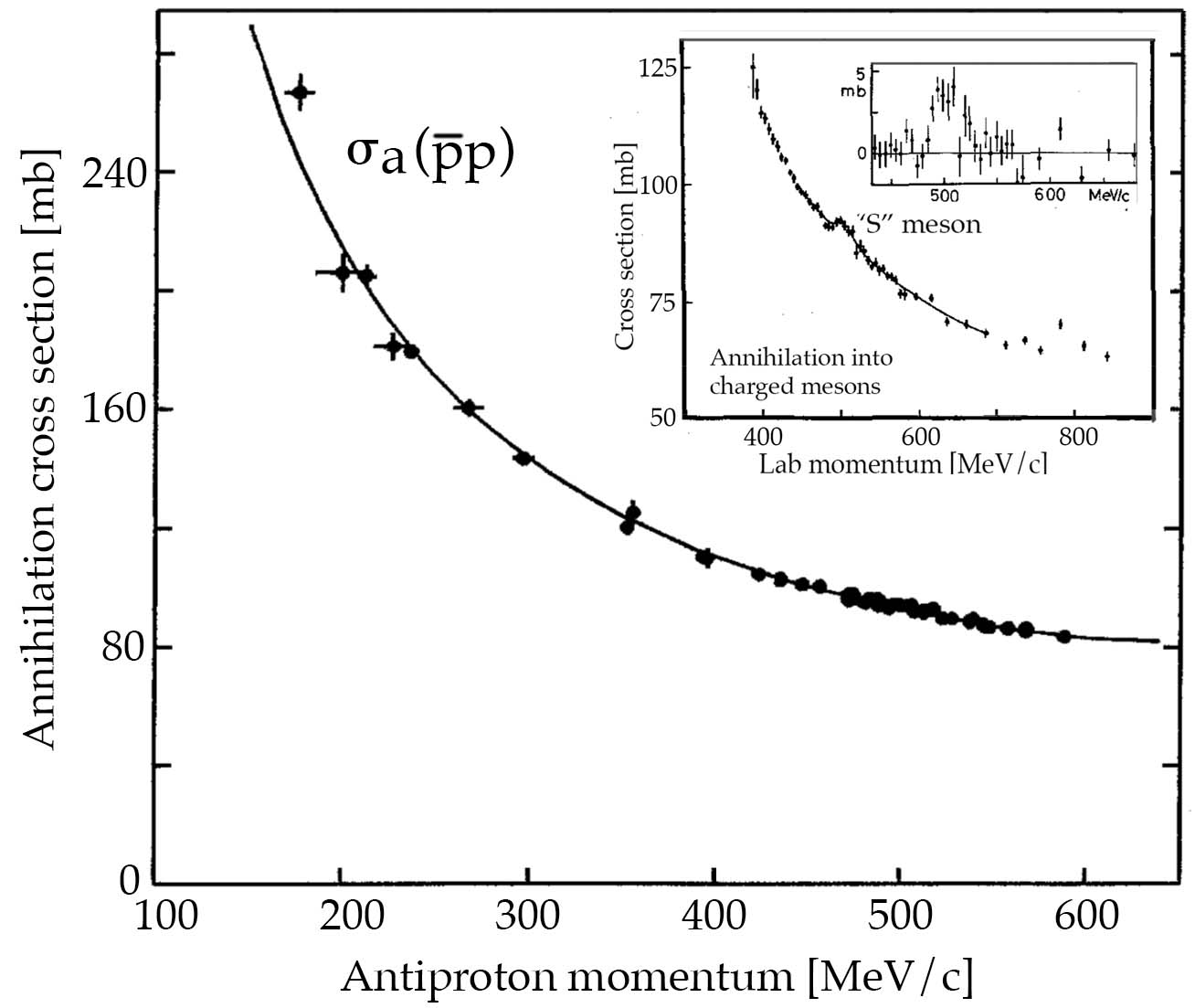}
}\centering} \caption[]{Left:  $\nbarp$ total cross section $\sigma_t$ and  $\nbarp$ annihilation  cross section $\sigma_a$ \cite{Br03}. Right: $\pbarp$  annihilation cross section $\sigma_a$ into prongs \cite{Br90}. Inset: the spurious ``$S$-meson'' reported in the annihilation cross section measured on a secondary extracted $\pbar$ beam before the LEAR era (from \cite{Br77}). 
\label{Sigmas}}
\end{figure}

Figure \ref{Sigmas} (left)  shows the $\nbarp$ total and annihilation  cross sections from OBELIX. The $P$-wave is already strong at low momenta (rapid onset with momentum leading to the plateau around 200 MeV/c). This is due to the strongly attractive potential which pulls the wavefunction with large impact parameters into smaller distances. The shape of $\sigma_a$ is otherwise structureless down to 50 MeV/c. The total cross section $\sigma_t$ shows a dip  around 80 MeV/c, not seen in the annihilation cross section, which is ascribed to the interference between the passing and interacting waves (Ramsauer-Townsend effect). Figure \ref{Sigmas} (right) shows the structureless $\pbarp$ cross sections measured by PS173 \cite{Br90}. At low momenta $\sigma_a(\pbarp)>\sigma_a(\nbarp)$ due to the strong attraction in the $i=0$ $\pbarp$ state, while above $\sim$400 MeV/c the two cross sections become nearly equal.

\begin{figure}[htb]
\parbox{170mm}{\mbox{
\includegraphics[width=160mm]{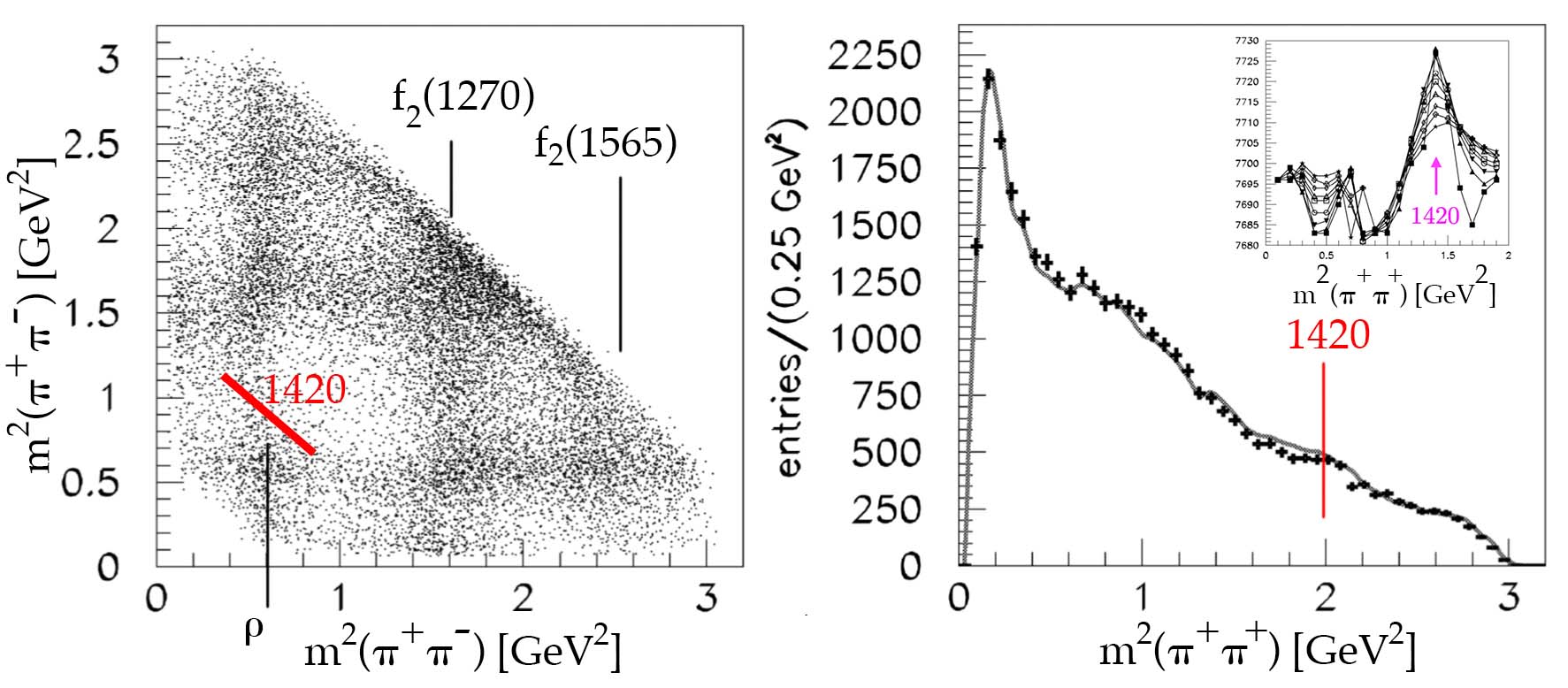}
}\centering}\hfill
\caption[]{Dalitz plot and ($\pi^+\pi^+$)-invariant mass distribution of the reaction $\nbarp\to\pi^+\pi^+\pi^-$. The inset shows the log-likelihood distribution for a ($\pi^+\pi^+$)-resonance with spin 0 at 1420 MeV, as a function of mass and for various widths (after  \cite{Fi00}).
\label{Isospin2}}
\end{figure}

OBELIX has searched for doubly charged mesons ($i=2$) in a channel with low pion multiplicity. Figure \ref{Isospin2} shows the Dalitz plot of the annihilation channel $\nbarp\to\pi^+\pi^+\pi^-$ collected with antineutrons between 50 and 405 MeV/c. Apart from the $\rho$, $f_2(1270)$ and $f_2(1565)/AX$ mesons the analysis required the presence of a resonance decaying into $\pi^+\pi^+$. The log-likelihood as a function of mass, displayed in the inset of Figure \ref{Isospin2}, illustrates the quality of the fit as a function of width. The mass is 1420$\pm$20 MeV, the width 100$\pm$10 MeV, and the $J^{P}$ determined to be $0^{+}$. If confirmed this signal could be due to a scalar tetraquark meson, e.g. a member of the $0^+$ 36-plet \cite{Ja77}.

Further results from OBELIX will be recalled in the next section.

\subsection*{6. Violation of the OZI rule}

\begin{figure}[htb]
\parbox{75mm}{\mbox{
\includegraphics[width=50mm]{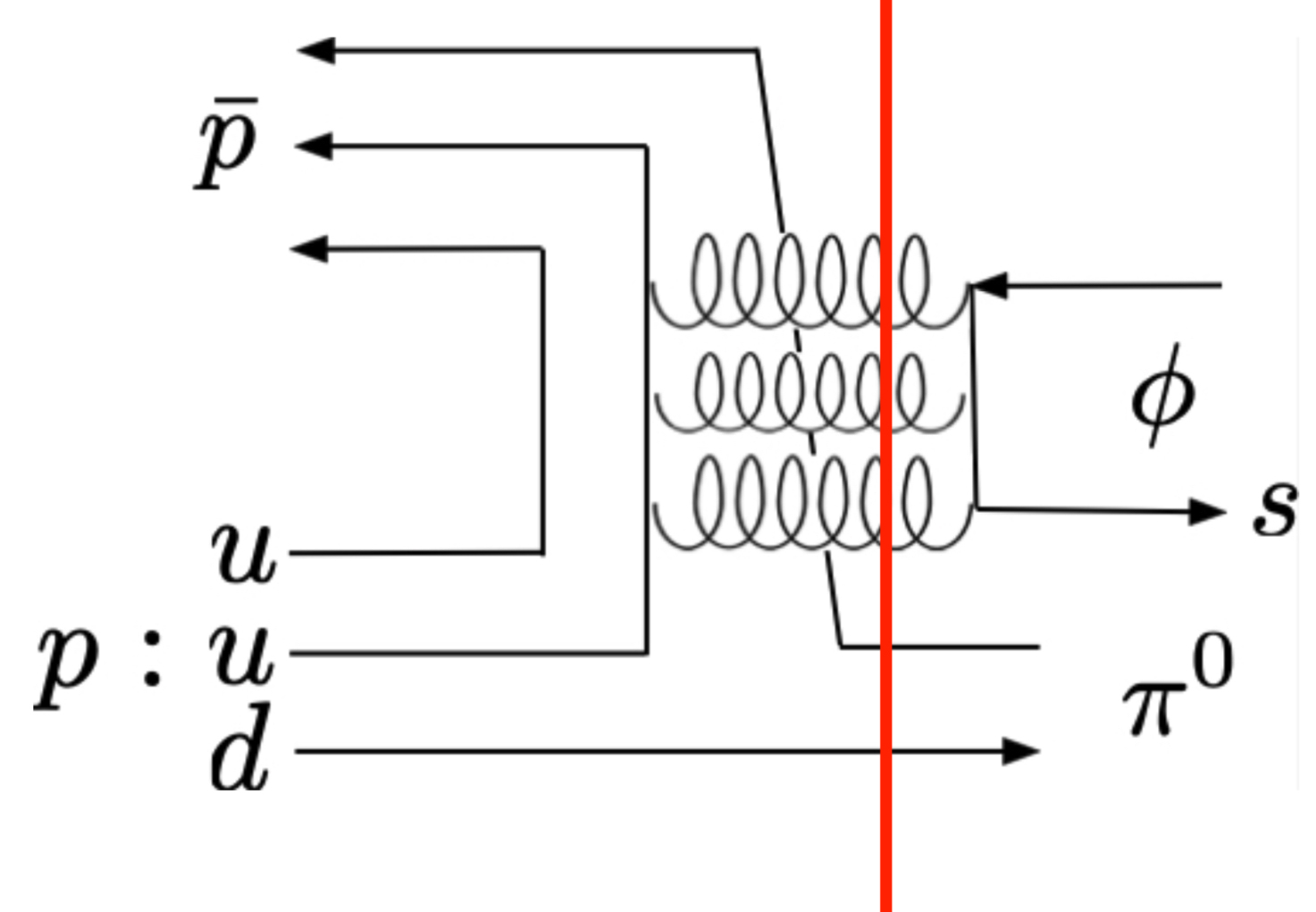}
}\centering}\hfill
\parbox{95mm}{\mbox{
\includegraphics[width=74mm]{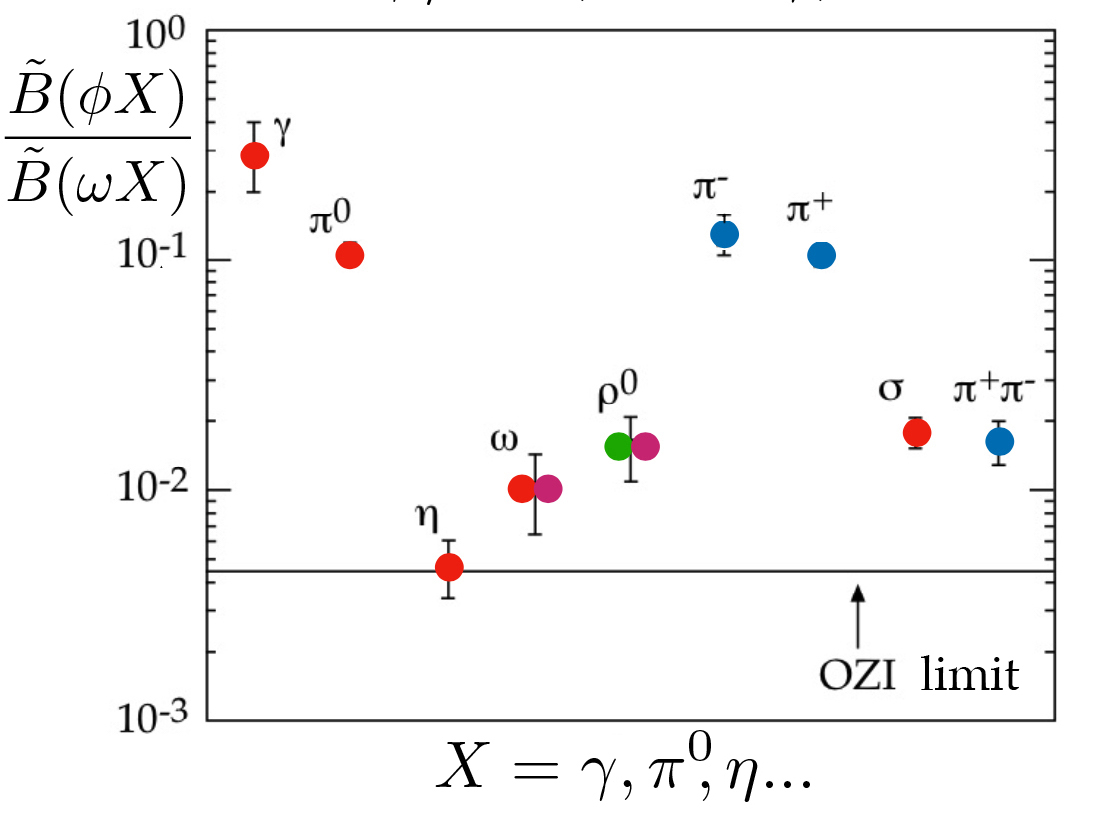}
}\centering} \caption[]{Left: in the limit $\theta_V=\theta_i$ the Feynman diagram for $\phi$ production in $\NNbar$ annihilation is disconnected and hence suppressed. Right: ratio of phase space corrected annihilation rates  into $\phi X$ and $\omega X$ in liquid hydrogen. The red dots are from CRYSTAL BARREL, the blue dots from OBELIX ($\phi \pi^+$ from $\nbarp$ and $\phi\pi^-$ from $\pbarn$ in deuterium), the violet from a bubble chamber, and the green one from ASTERIX extrapolated to liquid. For references see \cite{RMP98}. 
\label{OZI}}
\end{figure}

For the vector meson nonet ($\rho$, $K^*$, $\phi$ and $\omega$) the SU(3) wavefunctions of the two isoscalar mesons read, following (\ref{eq:mix}), 
\ba
\phi &=& \frac{1}{\sqrt{2}}(\uubar + \ddbar) \sin(\theta_i-\theta_V) - \ssbar\cos(\theta_i-\theta_V),\nonumber\\
\omega&=& \frac{1}{\sqrt{2}}(\uubar + \ddbar) \cos(\theta_i-\theta_V) + \ssbar\sin(\theta_i-\theta_V),
\label{eq:mix2}
\ea
with the mixing angle $\theta_V = 36.5^\circ$ (linear mass formula) or $39.2^\circ$ (quadratic mass formula) \cite{LNP,PDG19}. Thus $\phi$ production should be strongly suppressed since $\theta_V\simeq\theta_i$, hence $\phi\simeq-\ssbar$ and the disconnected graph in $\pbarp\to\phi\pi^0$ (Figure \ref{OZI}, left) violates the Okubo-Zweig-Iizuka (OZI) rule. Ignoring the $\ssbar$ components in (\ref{eq:mix2}) one gets, apart from phase space factors,  the prediction 
\be
\frac{\tilde{B}(\phi X)}{\tilde{B}(\omega X)} = \tan^2(\theta_i-\theta_V) = 4.7 \times 10^{-3}
\ee
with the quadratic mass formula and the even smaller ratio $4.4 \times 10^{-4}$ from the linear one. 

\begin{figure}[htb]
\parbox{85mm}{\mbox{
\includegraphics[width=60mm]{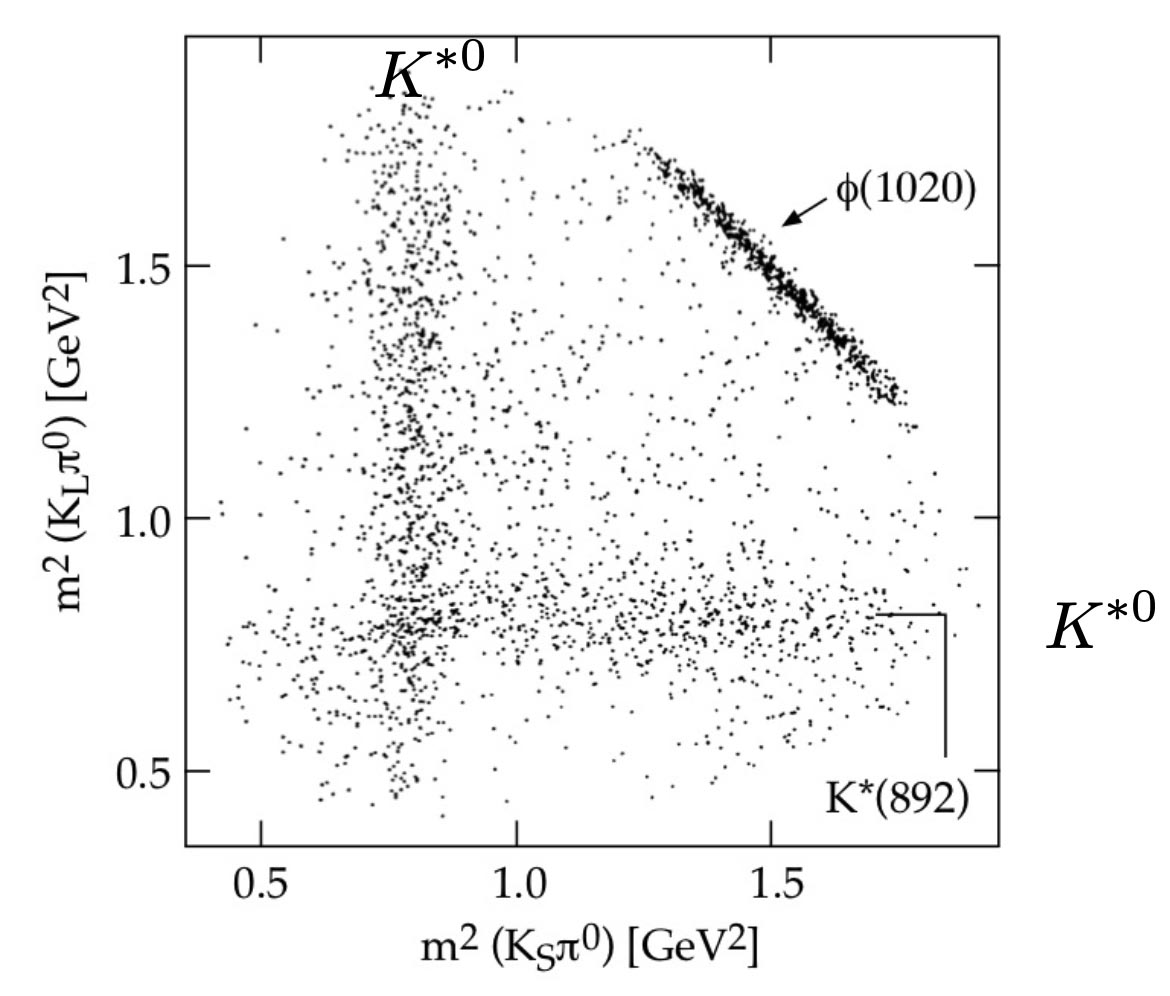}
}\centering}\hfill
\parbox{85mm}{\mbox{
\includegraphics[width=53mm]{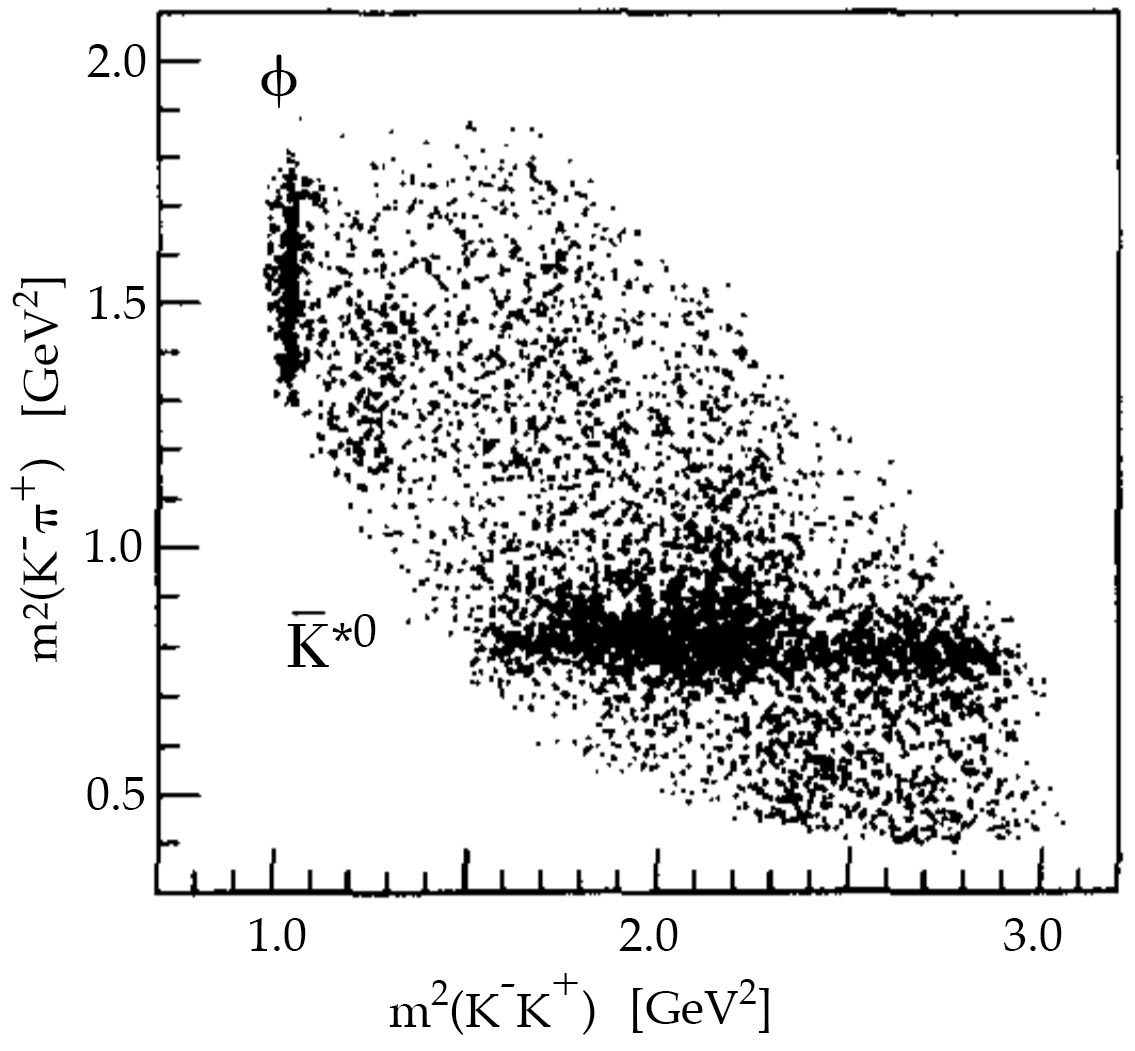}
}\centering} \caption[]{Dalitz plots for $\pbarp\to K_SK_L\pi^0$ \cite{Am93} (left) and $\nbarp\to K^+K^-\pi^+$  (right) \cite{Br03,Fi99}.
\label{KKpi}}
\end{figure}

The annihilation branching ratios into $\phi X$ and $\omega X$ have been measured at rest or with low energy antineutrons at LEAR and in bubble chambers.  The experimental results in liquid hydrogen (corrected by the phase space factor (\ref{eq:phasespace}))  strongly violate the OZI rule with almost any associated meson $X$ (Figure \ref{OZI}, right).
Figure \ref{KKpi} (left) shows the $K_SK_L\pi^0$ Dalitz plot measured by CRYSTAL BARREL with $K_S\to \pi^0\pi^0$ ($7\gamma$ final state). A strong $\phi\to K_SK_L$ signal appears together with the production of $K^*\to K\pi$  \cite{Am93}. The annihilation rate is then compared to $\pbarp\to\omega\pi^0$, with $\omega\to \pi^0\gamma$ or $\omega\to \pi^+\pi^-\pi^0$. 

A  strong $\phi$ production also appears in the $\nbarp\to K^+K^-\pi^+$ Dalitz plot from OBELIX (Figure \ref{KKpi}, right) \cite{Br03,Fi99} which is then compared to the $\omega\pi^+$ signal in $\nbarp\to (\pi^+\pi^-\pi^0)\pi^+$.

\begin{figure}[htb]
\parbox{56mm}{\mbox{
\includegraphics[width=48mm]{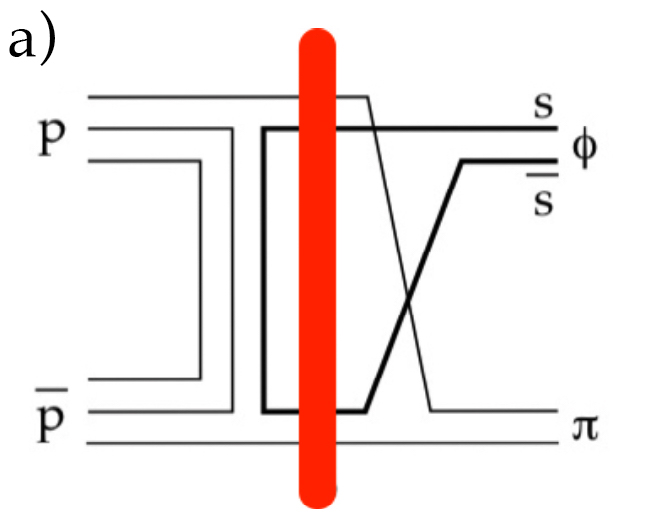}
}\centering}\hfill
\parbox{56mm}{\mbox{
\includegraphics[width=48mm]{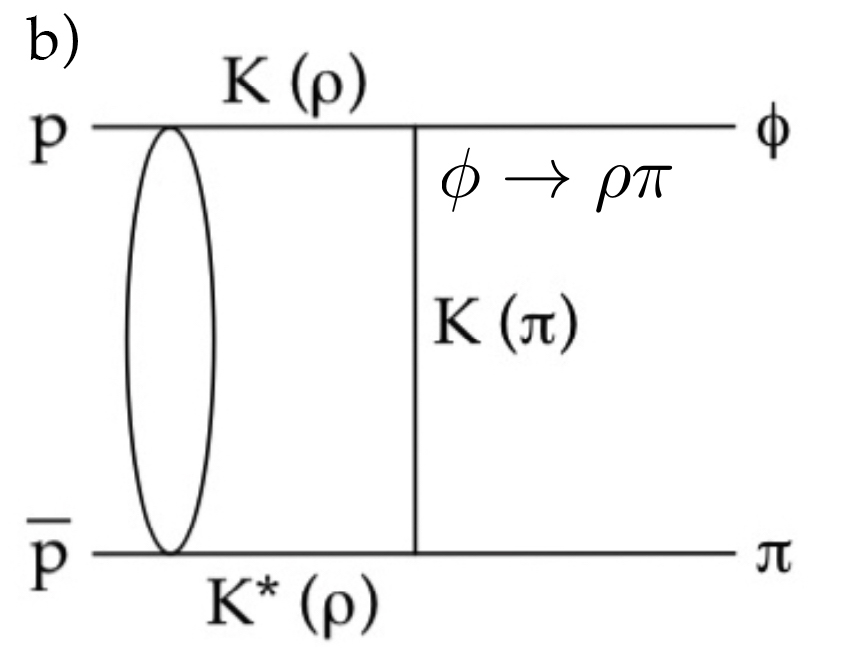}
}\centering}
\parbox{56mm}{\mbox{
\includegraphics[width=48mm]{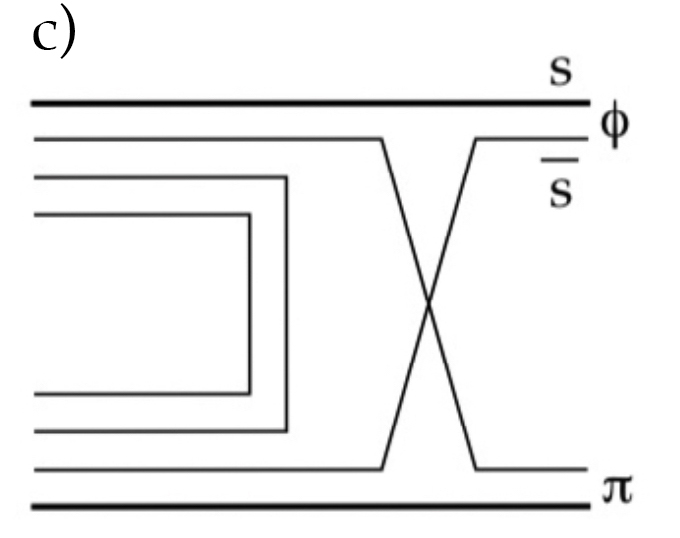}
}\centering} 
\caption[]{Mechanisms leading to enhanced $\phi$ production: a) tetraquark, b) rescattering, c) $\ssbar$ sea pairs.
\label{Tetra}}
\end{figure}

Several explanations have been proposed for the $\phi$ enhancement: $\phi\pi^0$ is produced from the $i=1$ $^3S_1$ ($1^{--}$) orbital in liquid, which could transit through the formation of nearby vector tetraquark mesons with quark content $sq\sbar\qbar$ (Figure \ref{Tetra}a) \cite{DoFi89}. Such a candidate state, the $C(1480)$ was reported in Serpukhov in the reaction $\pi^-p\to Cn$ with $C\to\phi\pi^0$, while the $\omega\pi^0$ decay was not observed \cite{Bi87}. A  $1^{--}$ enhancement is also reported around 1500 MeV by BABAR in $e^+e^-\to \phi\pi^0$ \cite{Au08}. However, the $C(1480)$ still needs to be confirmed. The well established isosinglet $\phi(2170)\to\phi\pi\pi$ is another tetraquark candidate which could enhance $\phi$ production in $i = 0$ channels such as $\phi\omega$ or $\phi\sigma$.

Rescattering and constructive interference between the intermediate $\overline{K}K^*$ and $\rho\rho$ were proposed as an alternative explanation (Figure \ref{Tetra}b) \cite{Go96} (the $\phi$ couples to $\rho\pi$ through its $u\ubar + d\dbar$ component). The very large OZI violation in $\phi\gamma$ was explained by the constructive interference between $\phi\rho$ and $\phi\omega$, while $\omega\rho$ and $\omega\omega$ interfere destructively. The $\rho$ and $\omega$ then couple to $\gamma$ via vector dominance (VDM) \cite{Lo94}.

\begin{figure}[htb]
\parbox{170mm}{\mbox{
\includegraphics[width=90mm]{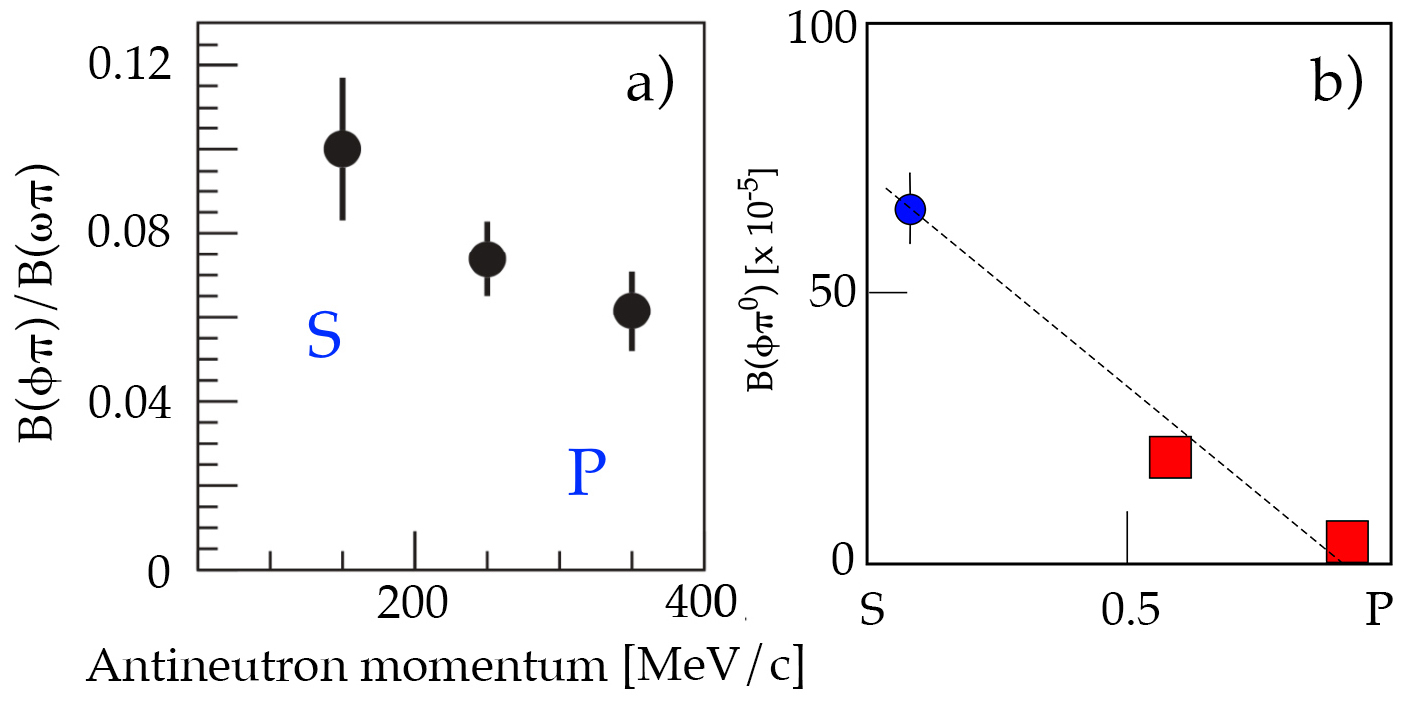}
}\centering}\hfill
\caption[]{Ratio $\phi\pi^+$ over $\omega\pi^+$ from OBELIX as a function of antineutron momentum \cite{Br03}; b) branching ratio for $\phi\pi^0$ in liquid from CRYSTAL BARREL (blue circle), in gas and from $2p$ from ASTERIX (red boxes, adapted from \cite{Rei91}).
\label{phipiSP}}
\end{figure}

Another palatable explanation resorts to sea $\ssbar$ pairs in the nucleon and antinucleon \cite{El95}.
In deep inelastic muon scattering the $s$ and $\sbar$ spins are known to be antiparallel to that of nucleon spin. The spin of the $\ssbar$ pair in  Figure \ref{Tetra}c would be antiparallel  to that of the $\pbarp$, hence in the spin triplet $^3S_1$ state, as required for the spin-1 $\phi$ meson. On the other hand $\phi\pi$ is also produced from $P$ states ($^1P_1$), but with proton and antiproton in the spin singlet state. Therefore $\phi$ production should be reduced from $P$ states. The ratio $\phi\pi/\phi\omega$ indeed decreases with increasing antineutron momentum, due to the onset of $P$-waves, as observed by OBELIX with antineutrons (Figure \ref{phipiSP}a). A similar suppression of $\phi\pi^0$ has been reported by ASTERIX in gaseous hydrogen (Figure \ref{phipiSP}b).

\begin{figure}[htb]
\parbox{170mm}{\mbox{
\includegraphics[width=160mm]{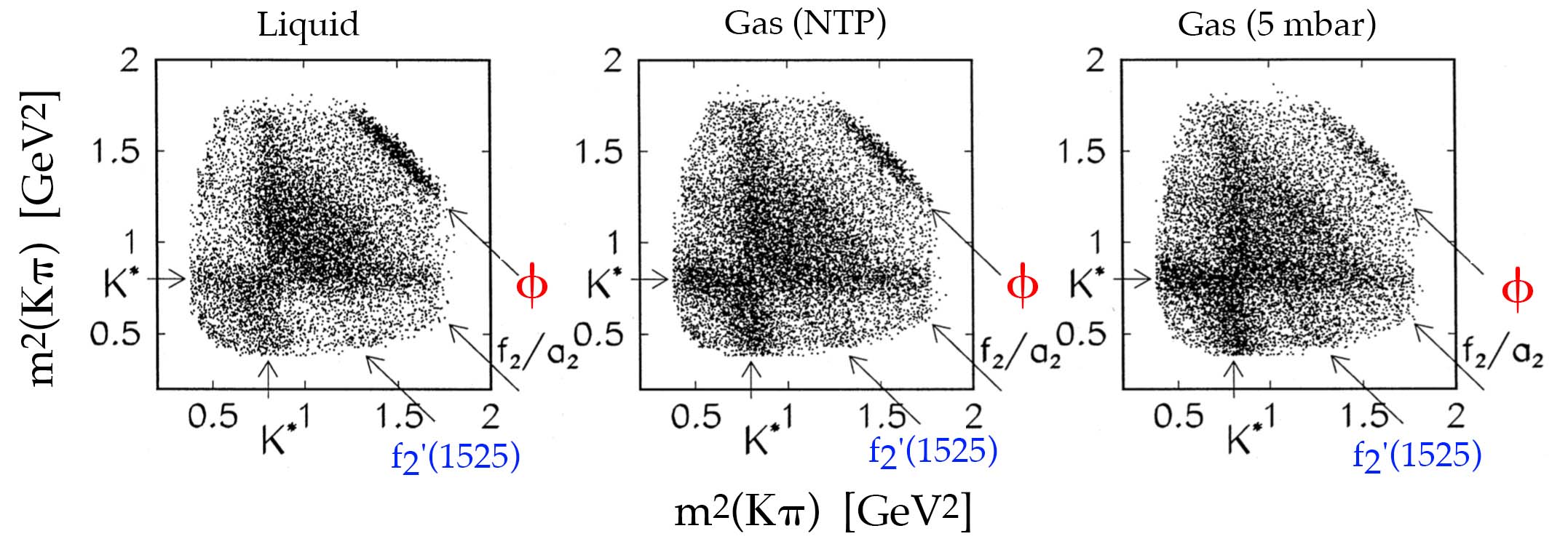}
}\centering}\hfill
\caption[]{$\pbarp\to K^+K^-\pi^0$ Dalitz plot  with stopping  $\pbar$ in liquid, gas at NTP and  at 5 mbar (see the text) \cite{Al98}.
\label{f2prime}}
\end{figure}

Let us now consider the ($2^{++}$) tensor nonet with the $a_2(1320), K^*_2(1430), f_2'(1525)$ and $f_2(1270)$, the latter two isoscalars being nearly ideally mixed (hence $f_2'(1525)\simeq \ssbar$). With sea quarks and antiquarks the spin triplet $f_2'(1525)\to\KKbar$  should be copiously produced at rest from the $^3P_1$ and $^3P_2$ states, associated with the emission of a $\pi^0$ ($f_2'(1525)\pi^0$ is also possible from the spin singlet $^1S_0$ with final state angular momentum $\ell = 2$). Its increasing production  has indeed been observed  by OBELIX in $\pbarp\to K^+K^-\pi^0$ at rest when reducing the hydrogen density  \cite{Al98}. Figure \ref{f2prime} shows the Dalitz plot  in liquid, gas at NTP and at 5 mbar. The diagonal bands clearly reveal an enhancement from $f_2'(1525)$ with decreasing pressure, and the corresponding decreasing $\phi$ production. The branching fractions into $\KKbar$ of the $f_2'(1525)$ and of the (mainly $\uubar+\ddbar$) $f_2(1270$ are well known. Hence the ratio of production rates can be calculated, yielding  $B(f_2'(1525)\pi^0)/B(f_2(1270)\pi^0)$ = 0.149 $\pm$ 0.020 from $P$-orbitals  \cite{Al98} (phase space corrections would lead to an even larger ratio). This ratio exceeds the the OZI prediction from the linear or quadratic mass formulae  of typically 1\% \cite{LNP} by more than one order of magnitude.

\subsection*{7. Pontecorvo reactions}
Unusual annihilation processes involving more than one nucleon were proposed in 1956 by Bruno Pontecorvo \cite{Pontecorvo56}. The two-body final states in $\pbard$ annihilation listed in Table \ref{tab:Pontecorvo} have been measured by CRYSTAL BARREL and OBELIX. The rates are very small, lying between 10$^{-6}$ and 10$^{-5}$. 

\begin{table}[htb]
\begin{center}
\begin{tabular}{ l r l r}
\hline
Reaction & \multicolumn{2}{c}{Branching ratio}& Ref.\\
\hline
$\bar{p}d \to \pi^-p$ & 1.46 $\pm 0.08 $ & $\times 10^{-5}$ & $\dagger$\cite{De99}\\
$\bar{p}d \to \pi^0 n$ & 7.02 $\pm 0.72 $ &$\times 10^{-6}$ & \cite{Am95} \\
$\bar{p}d \to \eta n $& 3.19 $\pm 0.48 $ &$\times 10^{-6}$ &\cite{Am95}\\
$\bar{p}d \to \omega n$ & 22.8 $\pm 4.1$ & $\times 10^{-6}$ &\cite{Am95}\\
$\bar{p}d \to \eta' n$ & 8.2$\pm 3.4 $ &$\times 10^{-6}$ & \cite{Am95}\\
$\bar{p}d \to \phi n$ & 3.56 $\pm 0.25 $ &$\times 10^{-6}$ & $\dagger$\cite{Go02}\\
$\bar{p}d \to \rho^-p$ & 2.9 $\pm 0.6 $ &$\times 10^{-5}$ & $\dagger$\cite{Ab94}\\
$\bar{p}d \to \pi^-\Delta^+(\to\pi^0p)$ & 1.01 $\pm 0.08 $ &$\times 10^{-5}$ & $\dagger$\cite{De99}\\
$\bar{p}d \to \pi^0\Delta^0(\to\pi^-p)$ & 1.12 $\pm 0.20 $ &$\times 10^{-5}$ & $\dagger$\cite{De99}\\
$\bar{p}d \to \pi^0\Delta^0$ & 2.21 $\pm 0.24 $ &$\times 10^{-5}$ & \cite{Am95b}\\
$\bar{p}d \to \Sigma^0 K^0$ & 2.35 $\pm 0.45 $ &$\times 10^{-6}$ & \cite{Ab99}\\
$\bar{p}d \to \Lambda K^0 $& 2.15 $\pm 0.45 $ &$\times 10^{-6}$ & \cite{Ab99}\\
\hline
\end{tabular}
\end{center}
\caption{Branching ratios of Pontecorvo reactions measured at LEAR in liquid deuterium and in gas   ($\dagger$).}
\label{tab:Pontecorvo}
\end{table}

\begin{figure}[htb]
\parbox{56mm}{\mbox{
\includegraphics[width=49mm]{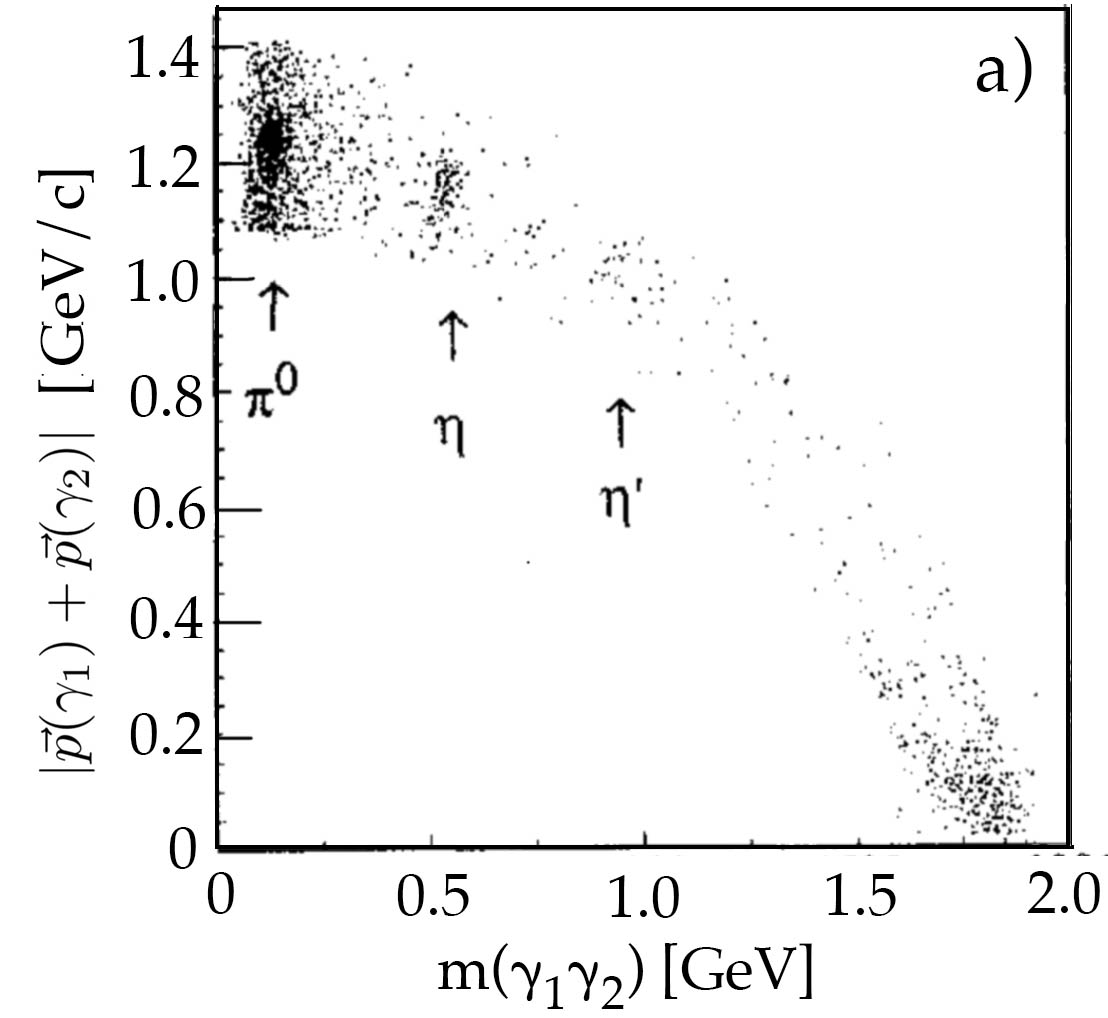}
}\centering}\hfill
\parbox{56mm}{\mbox{
\includegraphics[width=49mm]{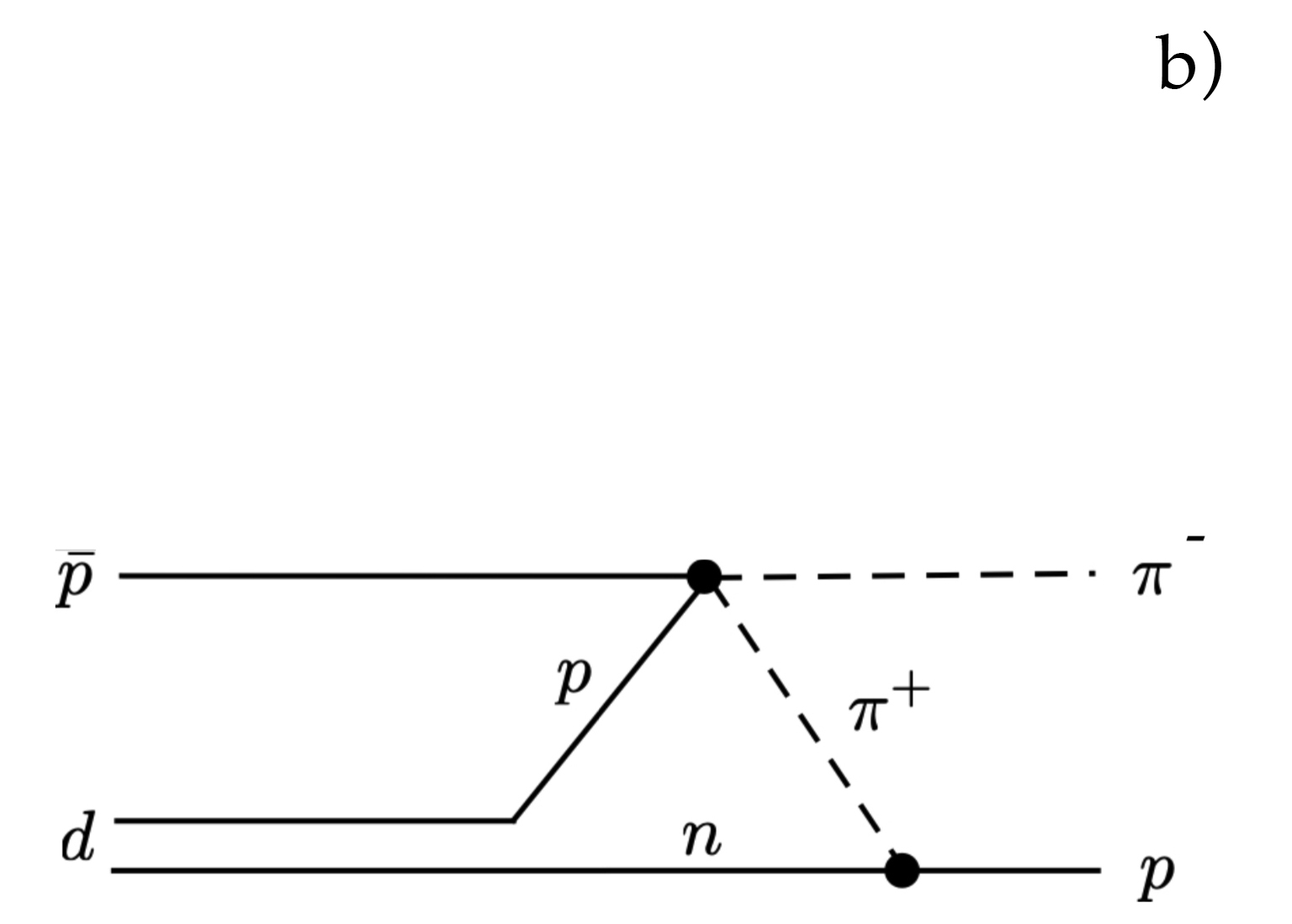}
}\centering}
\parbox{56mm}{\mbox{
\includegraphics[width=40mm]{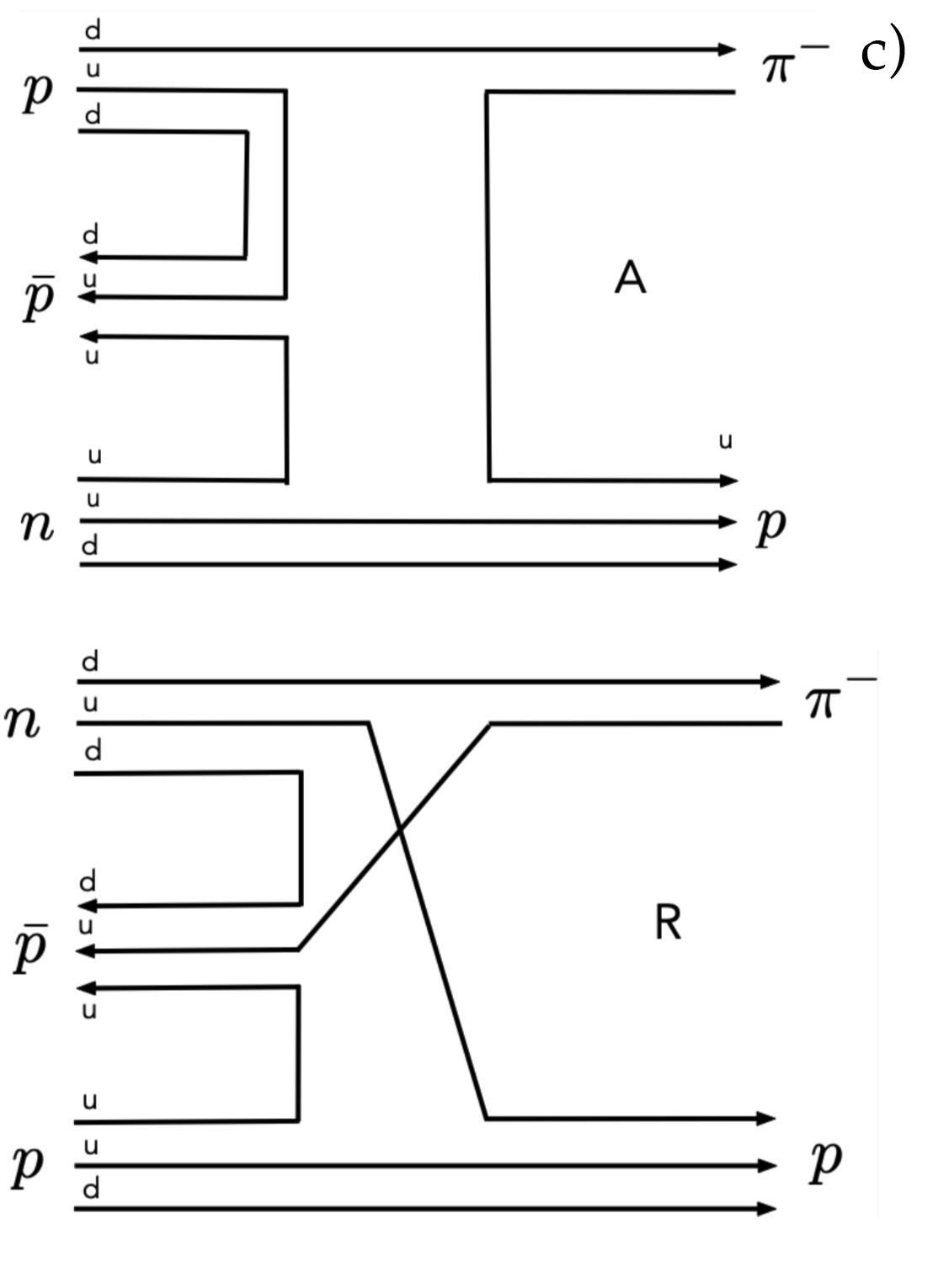}
}\centering}
\caption[]{a) Scatterplot for $\pbard\to 2\gamma n$. The clustering in the bottom corner is due to background  \cite{Am95}; b) rescattering diagram for $\pbard\to\pi^-p$;  c) annihilation and rearrangement graphs.
\label{pi0neutron}}
\end{figure}

Figure \ref{pi0neutron} shows as a first illustration  data collected by CRYSTAL BARREL  in $\pbard\to 2\gamma n$ with 200 MeV/c antiprotons stopping in liquid deuterium. Plotted is the $2\gamma$ momentum vs. the $2\gamma$ invariant mass. The neutron was not detected but the kinematics could be fully reconstructed by assuming  a missing back-to-back neutron with the momentum of the $2\gamma$ pair.  The total energy was required to lie within 300 MeV of 3 nucleon masses. Accumulation of events corresponding to $\pi^0n$, $\eta n$ and $\eta'n$ are observed. A 0-prong trigger was used to enhance the reactions of interest.  The branching fractions were measured by collecting a sample of unbiased annihilations (without 0-prong trigger) and by normalizing to the more frequent $\pi^0\pi^0n$ channel. The results for $\pi^0n$, $\eta n$ and $\eta'n$ are given in Table\ref{tab:Pontecorvo}. The rate for the $\pi^-p$ channel is derived by doubling the one for $\pi^0n$, hence $(1.41 \pm 0.14)\times 10^{-5}$, in agreement with the measurement by OBELIX in gas \cite{De99}. 

The reaction $\pbard\to\pi^-p$ involves both nucleons in deuterium. Annihilation rates from the rescattering graph illustrated in Figure \ref{pi0neutron}b  lie in the somewhat higher range $4 - 12 \times 10^{-5}$, depending on  the choice of $\NNbar$ potential and its short range uncertainties \cite{Os89}. In the analysis \cite{Ko89}, which predicts rates between $10^{-3}$ and  $10^{-6}$,  the reaction is sensitive to the deuteron wave function at small distances  and also to the meson form factors. A promising approach, which will be discussed below, is that of the fireball model \cite{Cu84,Cu89} in which a highly excited $3\overline{q}6q$ bag decays into the observed final state through the quark annihilation and rearrangement diagrams depicted in Figure \ref{pi0neutron}c.

\begin{figure}[htb]
\parbox{85mm}{\mbox{
\includegraphics[width=66mm]{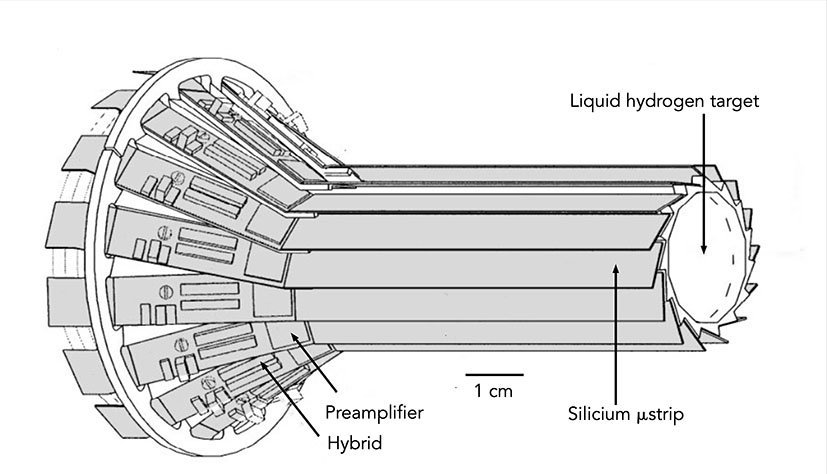}
}\centering}\hfill
\parbox{85mm}{\mbox{
\includegraphics[width=46mm]{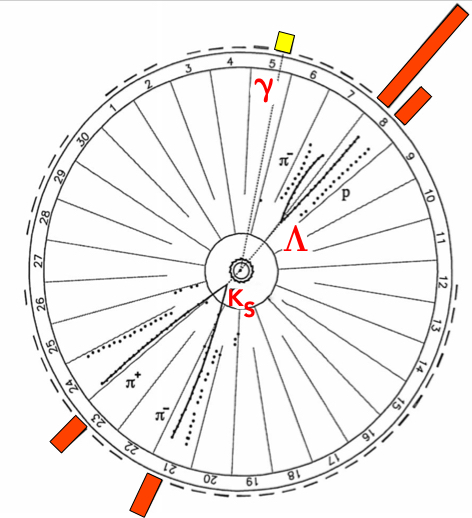}
}\centering} \caption[]{Left: the microstrip detector to trigger on hyperon decays was made of 15 single-sided silicon strips with 50$\mu$m pitch \cite{Do98}. Right: reconstructed $\Sigma^0 K_S\to \gamma\Lambda K_S$ event \cite{Ab99}.
\label{Microstrip}}
\end{figure}

\begin{figure}[htb]
\parbox{170mm}{\mbox{
\includegraphics[width=100mm]{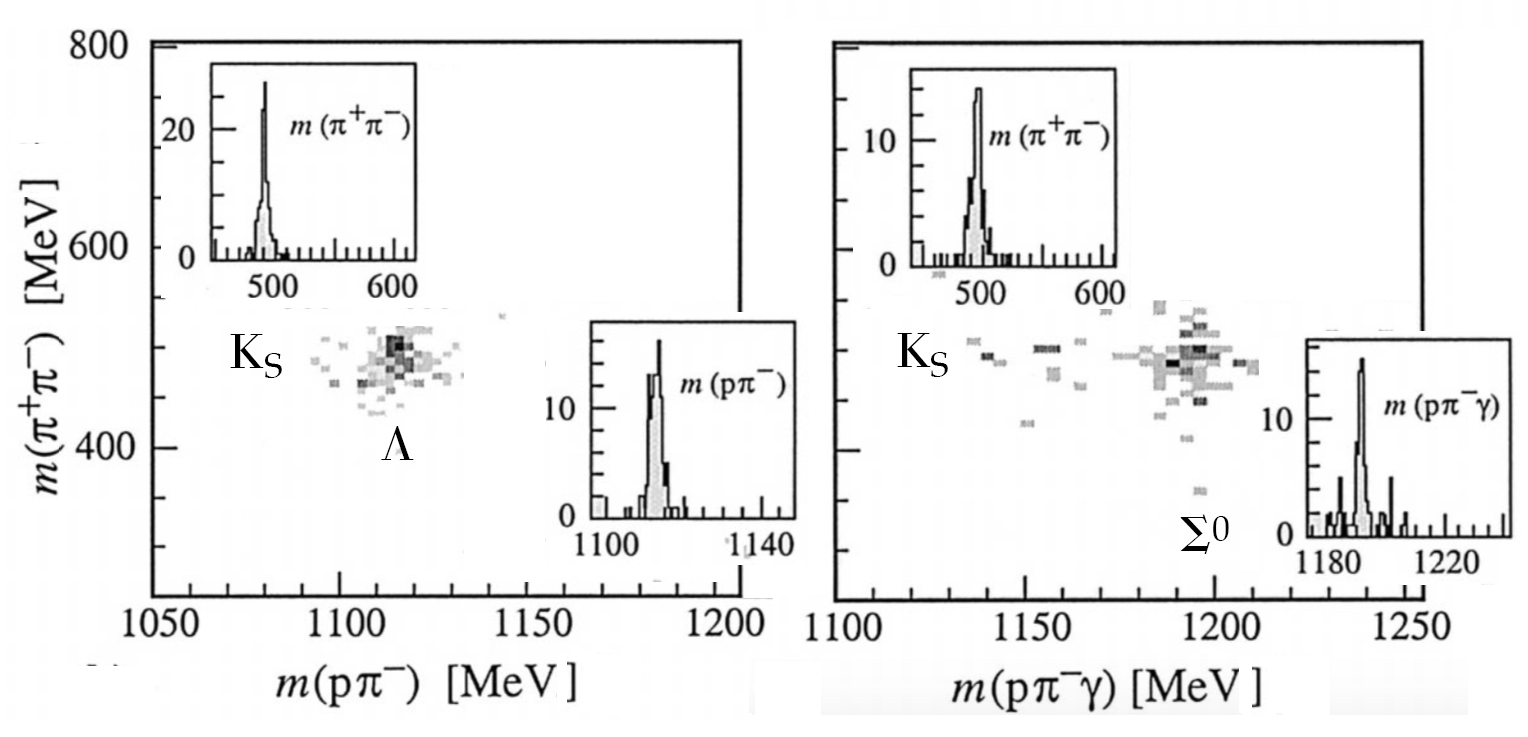}
}\centering}\hfill
\caption[]{Scatterplot and mass projections for $\pbard\to \Lambda (\to p \pi^-) K_S(\to \pi^+\pi^-)$ (left) and $\pbard\to \Sigma^0 (\to p\pi^-\gamma) K_S(\to \pi^+\pi^-) $ (right) \cite{Ab99}.
\label{LambdaSigma}}
\end{figure}

As a second example let us discuss the Pontecorvo processes $\bar{p}d \to \Lambda K^0$  and $\bar{p}d \to \Sigma^0 K^0$ measured by  CRYSTAL BARREL \cite{Ab99}. With $\Lambda\to p\pi^-$ and $\Sigma^0\to\Lambda\gamma$ these reactions lead to 2 or 4 prongs, depending on whether the $K^0$ decays as an unobserved $K_L$ or a $K_S\to\pi^+\pi^-$. The multiplicity increase was measured between a microstrip detector  surrounding the target and the first layers of the jet drift chamber (Figure \ref{Microstrip}, left).  A multiplicity increase between 0 and 2 or 4 prongs was required online to trigger on $\Lambda K^0$ pairs. The $\Lambda$ and $K_S$ decay vertices were reconstructed offline. Figure \ref{Microstrip} (right) shows a typical $\Lambda K_S$ event associated with a $\gamma$ from $\Sigma^0\to \Lambda\gamma$ decay. Signals from $\Lambda K_S$ and $\Sigma^0 K_S$ are presented in Figure \ref{LambdaSigma}. 

\begin{figure}[htb]
\parbox{85mm}{\mbox{
\includegraphics[width=70mm]{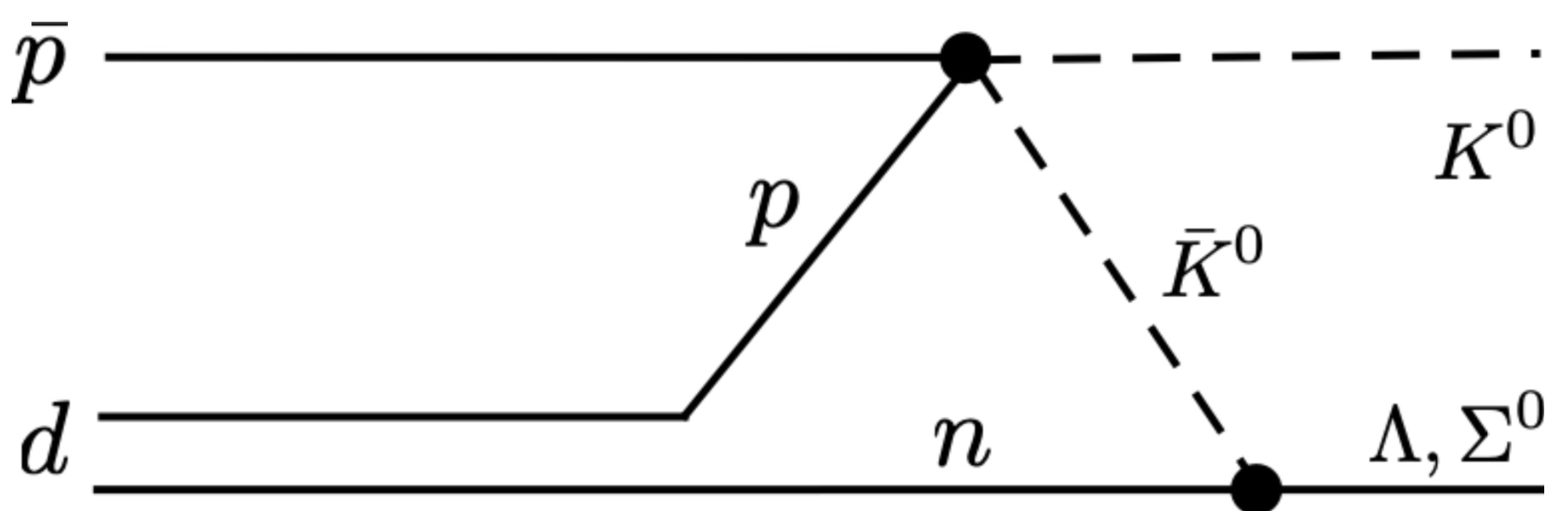}
}\centering}\hfill
\parbox{85mm}{\mbox{
\includegraphics[width=70mm]{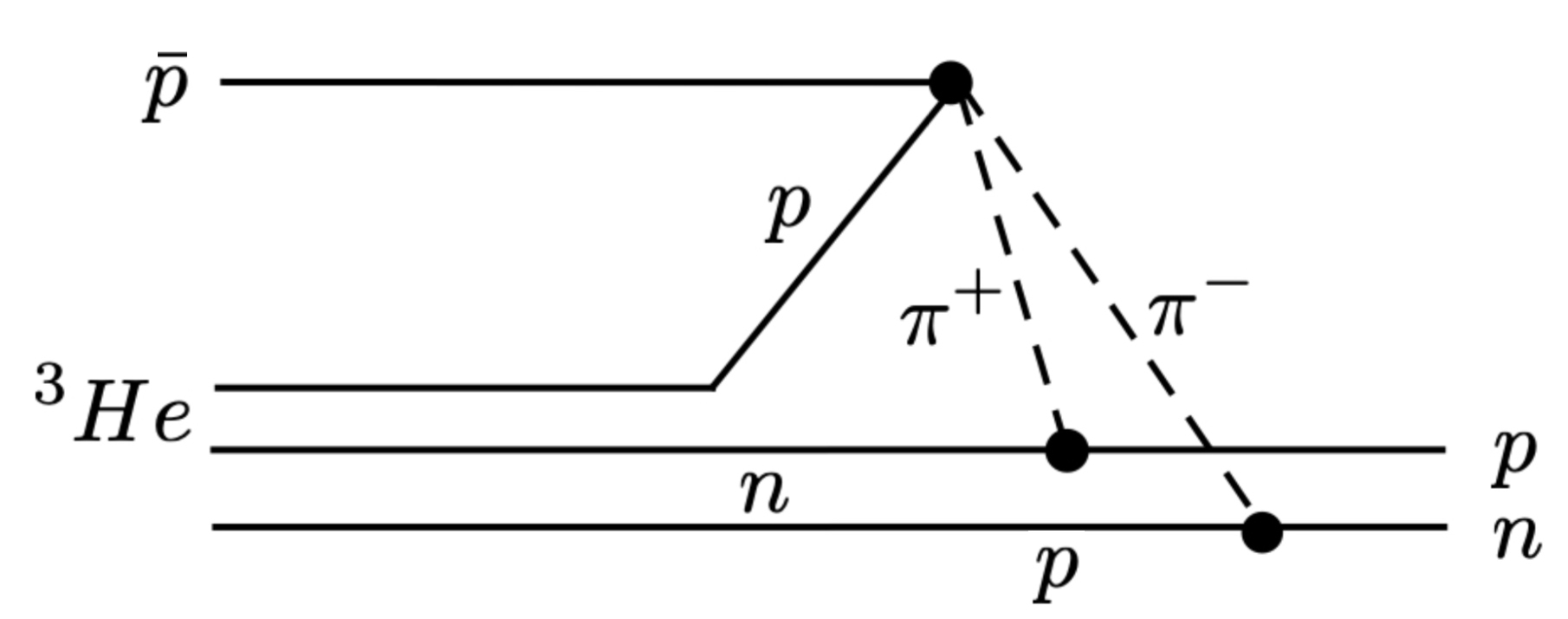}
}\centering} \caption[]{Rescattering diagram for $\bar{p}d \to \Lambda K^0, \Sigma^0 K^0$ (left) and for $\pbar^3$He$\to np$ (right).
\label{Rescattering}}
\end{figure}

The measured annihilation rates are equal within errors and are given in the bottom rows of Table \ref{tab:Pontecorvo}. The contribution from rescattering in Figure \ref{Rescattering} (left) has been calculated to be about $10^{-7}$ for the $\Lambda$ mode and $10^{-9}$ for the $\Sigma^0$ mode \cite{Ko89,He89}, the former being larger because of the much stronger coupling $KN\Lambda$ than $KN\Sigma^0$ ($\Lambda(1405)$ resonance). Both predicted branching ratios are much smaller than experimental data. 

The fireball model gives a better estimate: In antiproton-nucleon annihilation the rates into $\pi^+\pi^-$ and $K^+K^-$ are compatible with the normalized phase space weights \cite{Cu84,Cu89}. Hence the annihilation rate of protons on two nucleons is written
as the probability $P_f$ to form a fireball (on the first nucleon then propagating to the second one), times the normalized final state phase space factor. One gets from the measured $\pbard\to\pi^-p$ branching ratio\footnote{Reference \cite{Cu89} uses  for $\pbard\to \pi^-p$ a preliminary branching ratio twice as large from an earlier experiment .} in Table \ref{tab:Pontecorvo} with the  phase space factor $4.7 \times 10^{-4}$ \cite{Cu89} the probability to form a fireball on two nucleons: $P_f = 3\%$.

Following \cite{Cu89}, for the  $\Lambda K^0$  and $\Sigma^0 K^0$ reactions we expect rates of  $P_f\times$ the phase space factors which are almost equal for $\Lambda K^0$  and $\Sigma^0 K^0$ ($\simeq 10^{-4}$). Hence one arrives at the prediction of about $3\times 10^{-6}$ for both channels, in very good agreement with data.

The fireball model could be checked by studying annihilation on three nucleons (fireball propagating to the third nucleon) such as $\pbar^3$He$\to np$ or $\pbar^3$H$\to nn$ (or even the similar reactions with low energy antineutrons) for which no data exist yet. The rates are expected to be of the order of $10^{-6}$ \cite{Cu89} much in contrast to the $10^{-8}$ to $10^{-7}$ predicted for the rescattering graph in Figure \ref{Rescattering} (right) \cite{Ko88}.

\subsection*{8. Antiproton annihilation on nuclei}
Let me complete this review with a few remarks on annihilation on nuclei from a non-expert, which are relevant to this workshop, and mention a few striking features. Comprehensive reviews have been written, see in particular \cite{Eg87,Gu89,Be94}.

\begin{figure}[htb]
\parbox{85mm}{\mbox{
\includegraphics[width=50mm]{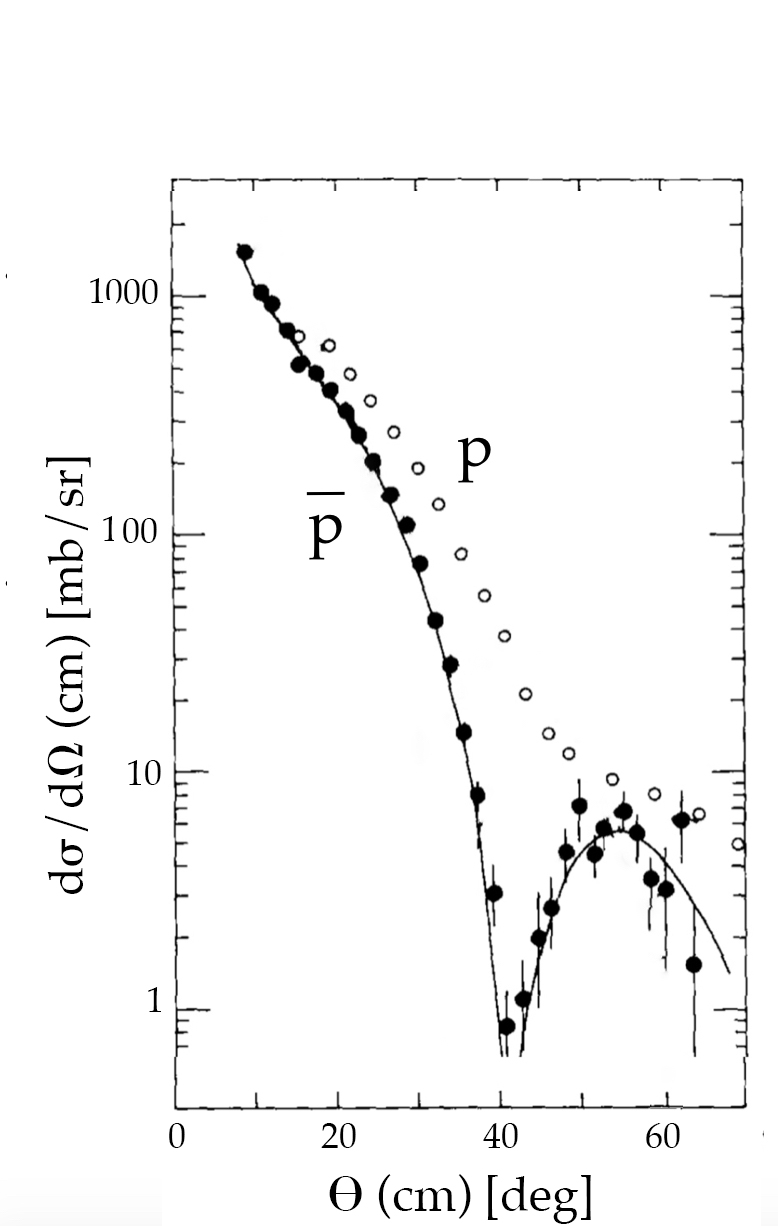}
}\centering}\hfill
\parbox{85mm}{\mbox{
\includegraphics[width=76mm]{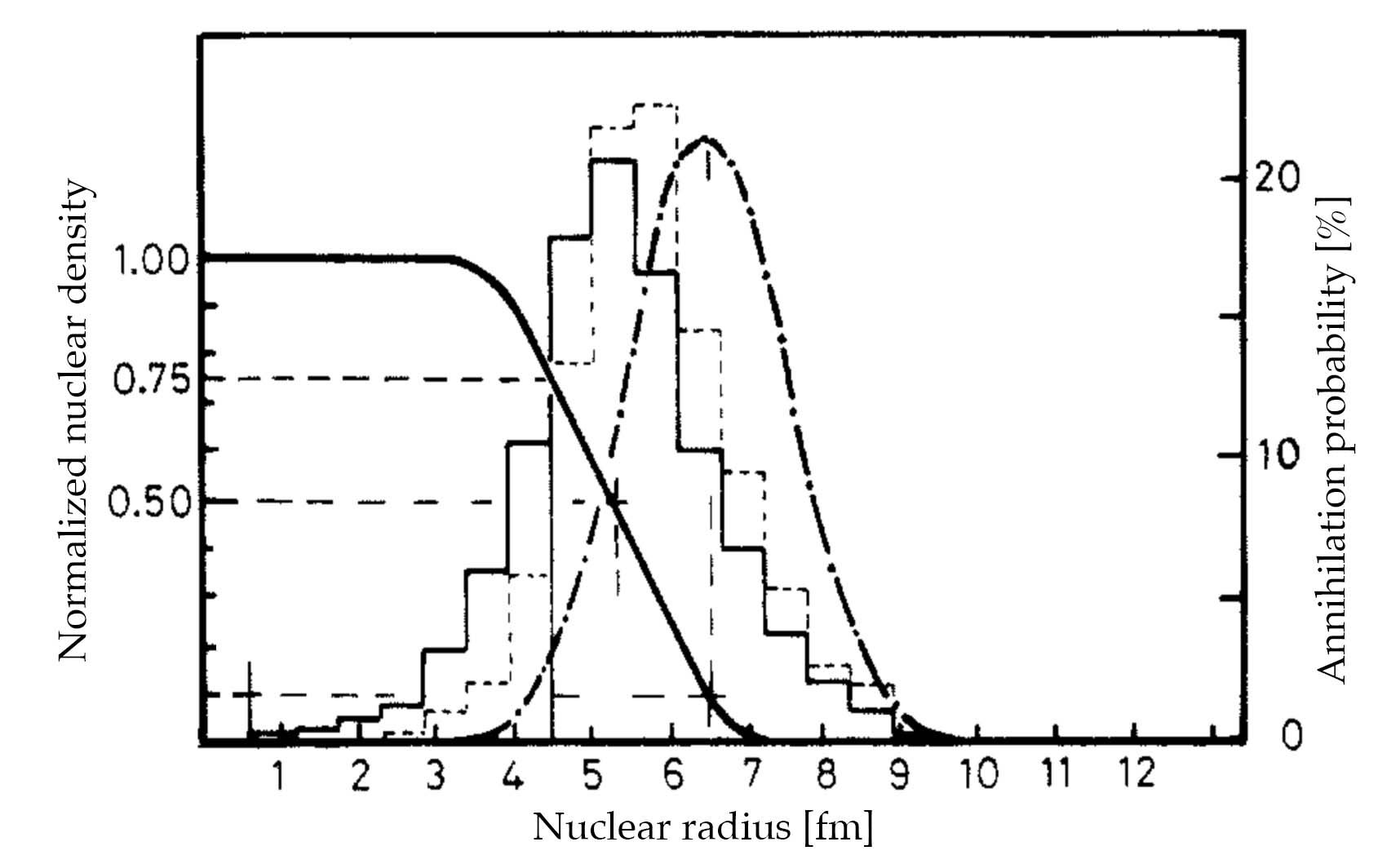}
}\centering} \caption[]{Left: elastic cross section at 300 MeV/c for  protons (open circles) and antiprotons on carbon (full circles), after \cite{Ga83}. Right: annihilation probability calculated with 590 MeV/c antiprotons on uranium (see the text).
The  dashed histogram corresponds to a vanishing $\pbar$-nucleus potential, the full histogram to 250 MeV. The full curve is the Woods-Saxon density distribution (after \cite{Ba86,Cl82}).
\label{Black}}
\end{figure}

The nucleus is a black sphere for antiprotons: Figure \ref{Black} (left) shows the familar diffraction pattern in the elastic cross section measured by the magnetic spectrometer of PS184 at LEAR with 300 MeV/c antiprotons on carbon, compared to the featureless cross section measured with protons of the same momentum. The annihilation is successfully described by the Intranuclear Cascade Model (INC) \cite{INC}. Annihilation occurs on one nucleon on the nuclear surface at about 10\% of the nuclear density. The pion multiplicity is low and the average pion momentum is 350 MeV/c, as in Figure \ref{Inclusive}. The pions emitted towards the interior of the nucleus have a short range due to the excitation of the $\Delta$ resonance, so that the energy deposit is strongly localized and leads  to high multiplicities ($\pi$, $p$, $d$, $\alpha$ and nuclear fragments). PS186 and PS203 have measured the emission of $p$, $d$, $^3$H, $^3$He,$^4$He and heavier fragments from lithium up to uranium with stopping antiprotons and using a multi-purpose charged particle spectrometer \cite{Pl93,Po95}. Not fully described by the INC is the larger neutron emission than expected from the $N/Z$ ratio \cite{Po95}. 

Evidence for a neutron halo around nuclei was reported earlier in Brookhaven with stopping antiprotons. The fraction of odd (even) charge multiplicity from $\pbarn$ ($\pbarp$) annihilation was measured with various targets immersed in a bubble chamber. The ratio of annihilations on neutron/protons was found to rise faster than $N/Z$ from carbon to lead targets \cite{Bu73}.

Most annihilations indeed occur on the nuclear surface: PS179 investigated the annihilation on the surface of Ag and Br nuclei by annihilating antiprotons on nuclear emulsions \cite{Ba86}. About 25\% of the events have a high multiplicity ($>10$) with antiproton momenta between 300 and 500 MeV/c. Figure \ref{Black} (right) shows a calculation of the annihilation probability as a function of nuclear radius for 590 MeV/c $\pbar$ on uranium. The 25\% of high multiplicity events would originate from the nuclear interior below a radius of 4.5 fm, where the nuclear density reaches 75\% of its maximum value. 

Nothing unusual in the production of strangeness as been observed (such as  enhancements due to the formation of quark-gluon plasma). The fraction of kaons at rest and between 400 and  900 MeV/c is 6.2\% (measured at ITEP in a xenon bubble chamber \cite{Dol91}), while the INC  predicts 6.25\% at 650 MeV/c. Hyperons are produced by kaons, e.g. via $K^-N\to\Lambda\pi$.

\begin{figure}[htb]
\parbox{85mm}{\mbox{
\includegraphics[width=60mm]{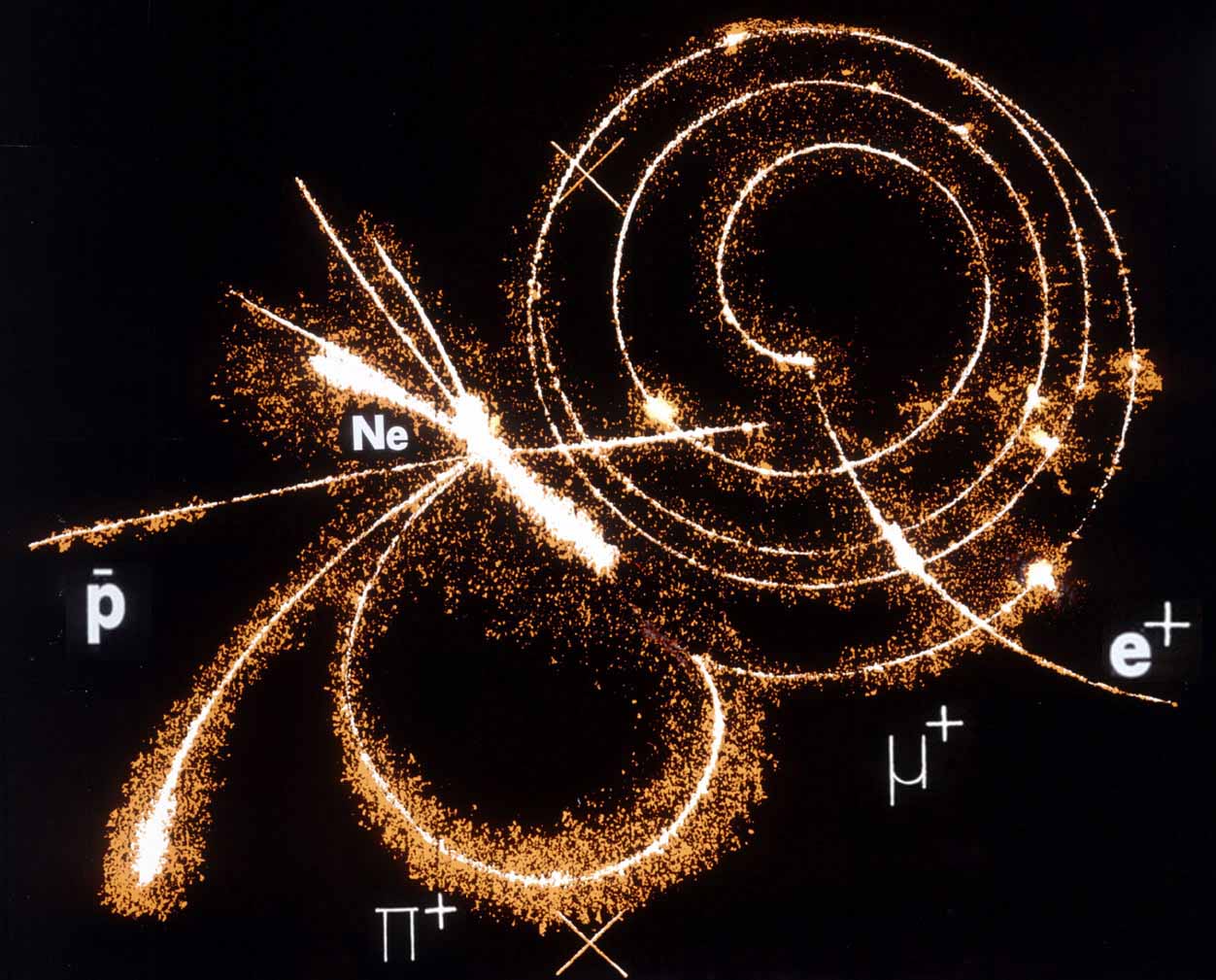}
}\centering}\hfill
\parbox{85mm}{\mbox{
\includegraphics[width=85mm]{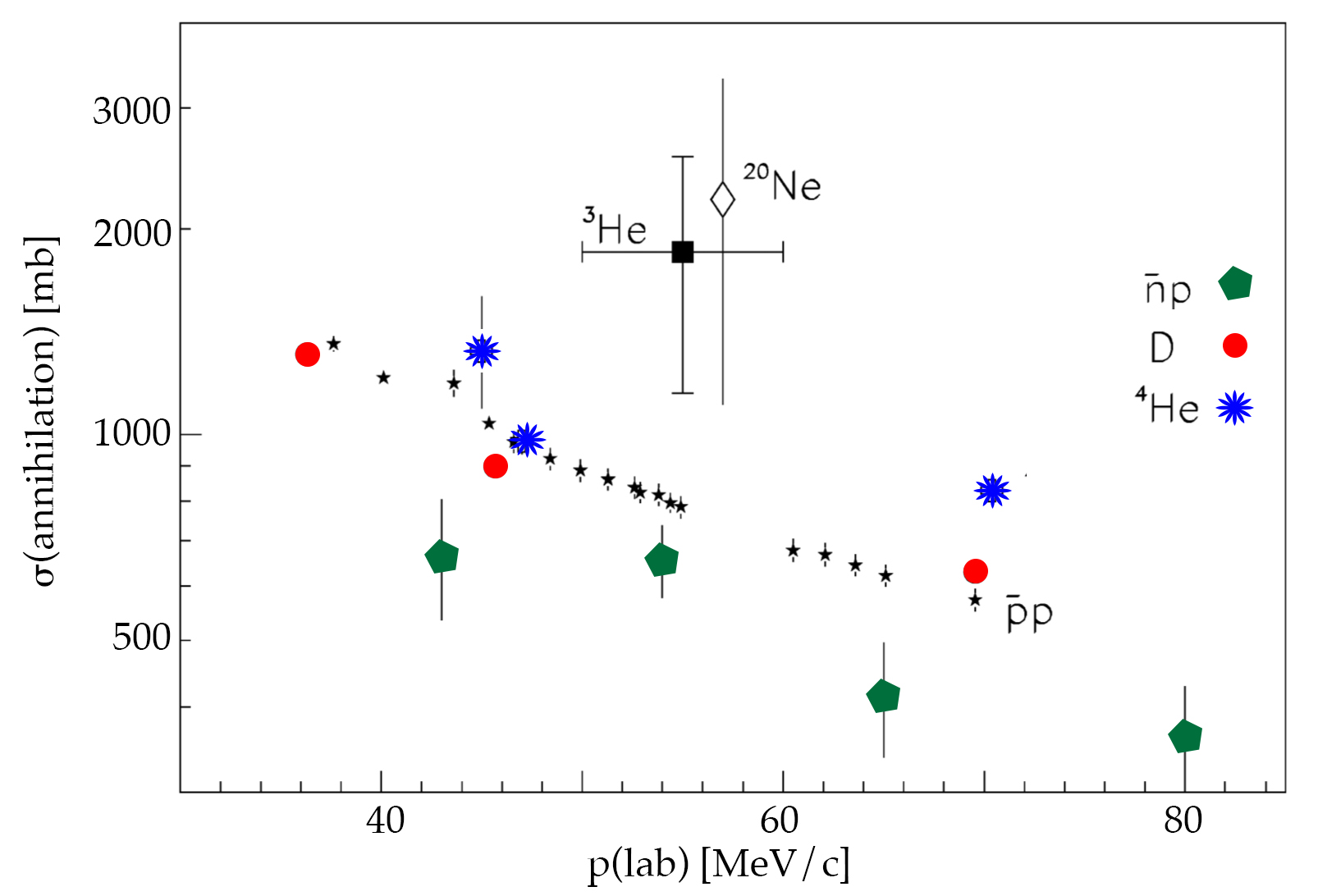}
}\centering} \caption[]{Left: $\pbar$Ne annihilation observed by the PS179 streamer chamber (photo CERN). Right: annihilation cross section for $\pbar^2$H, $\pbar^3$He, $\pbar^4$He, $\pbar^{20}$Ne, $\pbarp$ and $\nbarp$ \cite{Bi00}. 
\label{Streamer}}
\end{figure}

The PS179 collaboration studied antiproton annihilation in helium and neon. Figure \ref{Streamer} (left) shows a 600 MeV/c antiproton annihilating in the neon gas of the streamer chamber plunged in a magnetic field of 0.4 T. Figure \ref{Streamer} (right) shows the measured annihilation cross section below 80 MeV/c 
on deuterium, $^3$He, $^4$He and $^{20}$Ne, compared to $\pbarp$ and $\nbarp$ \cite{Bi00}. At low momenta the cross section for $\pbard$ is smaller than for $\pbarp$ due to the shadowing of the proton by the neutron ($\sigma(\pbarn) = \sigma(\nbarp) < \sigma(\pbarp)$ at  very low momenta, see Figure \ref{Sigmas}). The $\pbar^3$He cross section is surprisingly larger than the $\pbar^4$He one. The nearly equal cross sections on $^3$He and $^{20}$Ne are ascribed to Coulomb focusing of the antiproton on one of the nucleons.

\subsection*{9. Summary}
LEAR, which ran between 1983 and 1996, was a worldwide unique facility supplying intense pure antiproton beams between 60  and 1940 MeV/c, with which light meson spectroscopy was performed. Several new mesons were discovered with antiprotons annihilating at rest in hydrogen and with antineutron beams,  thanks to the slow extraction of the low energy antiproton beam and its narrow momentum bite.  Some of these mesons are candidates for tetraquark, baryonium, and glueball states. 

Annihilation at rest from $P$-waves was  found to contribute about 10\% in liquid hydrogen and about 50\% in NTP gas.  Hadron spectroscopy and annihilation dynamics were studied from $S$- and $P$-waves in $\pbarp$ and $\nbarp$ interactions. Statistical models describing multiplicity distributions are in rough agreement with data, but details are not well understood (such as the $\KKbar$ suppression from $P$-wave or the $\rho\pi$ suppression from the singlet $S$ state). Direct evidence for  quarks dynamics has been obtained, e.g. derived from measurements of the  pseudoscalar mixing angle. However, the exact mechanism involving quarks is still unclear. The physical nature of the surprisingly strong violation of the OZI rule is not clear either, but possibly due to sea (anti)quarks in the (anti)nucleon. The rare unusal annihilations involving more than one nucleon are presently best described by the global process of fireball formation. 

In the future, annihilation at rest could be further studied with the new ELENA ring (e.g. Pontecorvo $\pbar 3N$ annihilations such as $\pbar^3$He or $\pbar^3$H) by slowly extracting antiprotons from storage traps. The ASACUSA collaboration (AD3) traps presently some $10^6$ antiprotons/AD cycle (120s), which could be slowly released  between two AD cycles with an intensity of $\simeq$10$^4$ s$^{-1}$,  comparable to that in CRYSTAL BARREL at LEAR. This would require some post-acceleration to overcome target entrance windows or, alternatively, cryogenic targets such as solid hydrogen, jet targets or ion traps. Annihilation or cross section studies in the region 100 -- 200 MeV/c or below, in particular with  antineutron beams which require at least 100 MeV/c antiprotons, are unfortunately out of reach at the current AD.

\end{document}